\documentstyle[preprint,tighten,epsf,eqsecnum,aps,floats]{revtex}

\begin{document}
\def\be{\begin{equation}}
 \def \ee{\end{equation}}   
\def\bea{\begin{eqnarray}}\def\eea{\end{eqnarray}}
\preprint{NT@UW-00-04}
\title{Light Front Quantization--A Technique for Relativistic and Realistic
    Nuclear Physics}

\author{Gerald A. Miller}
                                
\address{Department of Physics\\
  University of Washington\\
  Seattle, Washington 98195-1560}
\maketitle

\begin{abstract}
Applications of relativistic
light front dynamics to computing the nucleonic and mesonic
components of nuclear wave functions are reviewed. In this approach
the fields are quantized at equal values of $\tau=x^0+x^3$. Our method is to
 use  a Lagrangian, and its associated  energy-momentum
tensor $T^{^+\mu}$ to define the total momentum operators
$P^\mu$ with $P^+$ as the plus-momentum and $P^-$ the $\tau$-development
operator.
The motivation for  unusual treatment of  nuclear physics is the desire
to use wave functions, expressed in terms  of plus-momentum variables,
 which are used to analyze high energy experiments  such
as deep inelastic scattering, Drell-Yan production, (e,e') and (p,p')
reactions. 
This motivation is discussed, and a simple overview of light front dynamics
is presented.
Some examples of ordinary quantum mechanics are solved to show that the
formalism is tractable. The necessary quantization
is reviewed and applied
to a series of problems: infinite nuclear matter within the mean field
approximation; a simple static source theory; finite nuclei using the mean
field approximation; low-energy pion-nucleon scattering using a chiral
Lagrangian; nucleon-nucleon scattering, within the one boson exchange
approximation; and, infinite nuclear matter including the effects of
two-nucleon correlations. Standard good results for nuclear
saturation properties are obtained, with a possible improvement in the lowered
 value, 180 MeV, of the  computed nuclear compressibility. A complicating
 feature of our
 light front dynamics is that manifest
 rotational invariance  is not used as an aid in doing
 calculations. But for each
 of the examples reviewed here,
manifest rotational invariance  emerges in
the results of the  calculations.
Thus    nuclear physics can be done in
a manner in which modern nuclear dynamics is respected, boost
invariance in the $z$-direction is preserved, and in which the
rotational invariance so necessary to understanding the basic features
of nuclei is maintained.  A salient feature is that $\omega,\sigma$
and $\pi$ mesons are important constituents of nuclei. It seems
possible to find
Lagrangians that yield reasonable descriptions of nuclear
deep inelastic scattering
and Drell-Yan reactions.   Furthermore, the presence of the
$\sigma$ and $\omega$ mesons could provide a nuclear enhancement of the ratio
of  the cross sections for longitudinally and transversely polarized virtual
photons in accord with recent measurements by the HERMES collaboration.
 
\end{abstract}
\newpage
\section*{Table of Contents}
1. INTRODUCTION AND MOTIVATION      3

A. {An Example}                     3

B. An Apparent Puzzle   5

C. More Motivation      5

D. Outline and Scope    6

2. WHAT IS LIGHT  FRONT DYNAMICS?  8

3. SIMPLE PROBLEMS                  9

A. Three-dimensional harmonic oscillator  9

B. Light Front Hulth\'en wave function    14

4. LIGHT FRONT QUANTIZATION   14 

A. Free Scalar Field          15

B. Free Dirac Field        17

C. Free Vector Meson       18

D. We need a Lagrangian no matter how bad 18

5. INFINITE NUCLEAR MATTER MEAN FIELD THEORY 22

A. Nuclear Momentum  Content  23
  
6. LIGHT FRONT STATIC SOURCE MODEL FOR NUCLEAR MESONS  26

A. Scalar meson distribution 27

B. Vector Meson Distribution 29

7. MEAN FIELD THEORY FOR FINITE SIZED NUCLEI 33

A. Nucleon Mode Equation     34

B. Dynamical meson fields   35

C. Finite Nucleus Solutions and Results 
and  Interpretation of the $p^-_n$ eigenvalues     37

D. Nuclear momentum content and  Lepton-nucleus deep inelastic scattering
 40

 8. CHIRAL SYMMETRY AND  PION NUCLEON SCATTERING   42

 9. NUCLEON-NUCLEON  SCATTERING ON THE LIGHT FRONT  46 

  A. Realistic one boson exchange potential 51

  B. Results for the two nucleon system   52

  10.  CORRELATED NUCLEONS IN  INFINITE NUCLEAR MATTER 54

  A. Nucleon truncation of the many-body problem 54

  B. Light Front Brueckner Hartree Fock Approximation 56

  C. Results for Energy versus density 59

  D. Meson degrees of freedom  60

  E. Momentum distributions 62

  F. Apparent Puzzle  Resolved 64

11. COMING ATTRACTIONS 64

A. Quark-composite nucleons in the nucleus 64

B. Few Body Problems 65

C. Possible Experimental signature-the HERMES effect 66

12. SUMMARY  67

ACKNOWLEDGMENTS 68

REFERENCES   69
\newpage 
\section{Introduction and Motivation}
The need for a relativistic methodology that is broadly applicable to nuclear
physics has never been more apparent. 
One of our  most important sets of problems involves understanding
the transition between the hadronic (baryon, meson)
and the underlying (quark, gluon) degrees of freedom.
Using a relativistic
formulation of the hadronic degrees of freedom is necessary to avoid a 
misinterpretation of a kinematic effect as a signal for the transition.
In particular, 
the goal of understanding future high energy and momentum transfer studies
of nuclear targets using 
 exclusive, nearly exclusive or inclusive processes can only be met through
using 
relativistic techniques. The light front approach of Dirac\cite{Di 49}
 in which the
time variable is taken as $t+z$ and the spatial variables 
are  $t-z,x,y$
\cite{fs}-\cite{hari}. 
is one of the promising approaches because
the momentum canonically conjugate to $t-z$, $p^+\equiv p^0+p^3$, is directly
related to the observables. 

To be specific, consider the lepton-nucleus deep inelastic scattering
observed by the 
EMC experiment \cite{emc,emcrevs}. This  showed that there is a significant
difference between the parton distributions of free nucleons and
nucleons in a nucleus. This difference can interpreted as a shift in 
the momentum distribution of valence quarks towards smaller values of
the Bjorken variable $x$. In the parton model, $x$ is the  ratio
of the plus-momentum $k^+=k^0+k^3$ of a quark to that of the target.
If one uses $k^+$ as a momentum
variable, the corresponding canonical spatial variable is $x^-=x^0-x^3$
and the time variable is $x^+=x^0 +x^3=\tau$\cite{blp}.
The $\tau$-development operator, which
plays the role of the light front Hamiltonian is $P^-=P^0-P^3,$ in which
$P^\mu$ is the total momentum operator. In equal time dynamics, the eigenstates
of
the Hamiltonian $P^0$ are complete. In light front dynamics the eigenstates of
$P^-$ are complete:
\bea
P^-|n\rangle \;&=&\;p^-_n|n\rangle\\
I\;&=&\;\sum_n |n\rangle\langle n|
\label{complete}.\eea
To do calculations
in this framework is
to use light front formalism or light front dynamics.
We have been attempting to derive the properties of nuclei using this light
front formalism. 

Light front dynamics applies to nucleons
within the nucleus as well as to partons of the nucleons, and 
this is a useful approach whenever  the momentum of
initial or final state nucleons is large compared to their
mass \cite{fs}. For example,
this technique  can be used for $(e,e'p)$ and $(p,2p)$
reactions at sufficiently high energies. It is important to realize that
the mere  use of
 light-front variables for nucleons in a nucleus is not sufficient to guarantee
 a reasonable result. Shouting relativity at a problem will not make it
 go away.
It is also necessary to include all the relevant features of realistic
 conventional nuclear dynamics. Combining the two aspects of relativity and
 realistic physics 
 provides
the technical challenge  addressed here.
\subsection{An Example}

A simple  example involving quasi-elastic scattering  of
high energy electrons from nucleons in nuclei  shows how  using the
light-front approach leads to important simplifications.
To be specific,
use coordinates in which  the four-momentum $q$ of the
exchanged virtual photon is
 $\left(\nu,0,0,-\sqrt{Q^2+\nu^2}\right)$. Here $Q^2=-q^2$, 
 and $\nu^2$ are both very large, but $Q^2/\nu$ is finite (the
Bjorken limit). Use the light-cone
variables, so that 
  $q^+\approx Q^2/2\nu\equiv Mx$
 ($M$ is the mass
of a nucleon),  
$q^-\approx 2\nu -Q^2/2\nu$, and $q^-\gg q^+$.

The  scattering cross
section for $e+A\to e' +(A-1)_f +p$, where $i,f$ represents the initial, final
nuclear eigenstate of $P^-$, and $p$ the four-momentum of the final
proton, takes the form
\begin{equation}
d\sigma\sim \sum_f \int
{d^3p_f\over E_f}\int d^4p\, \delta (p^2-M^2)\delta^{(4)}(q+p_i-p_f-p)|
\langle p,f\mid J(q)\mid i\rangle\mid^2,
\end{equation}
with  the operator $J(q)$  as a schematic representation of the
electromagnetic current. Kinematic factors and terms involving the
cross section are ignored here in order to bring out the
salient features. Performing the four-dimensional integral over
$p$ leads to the expression
\begin{equation}
d\sigma\sim \sum_f \int {d^2p_fdp_f^+\over p^+_f}
\delta\left((p_i-p_f+q)^2-M^2\right)\mid \langle p=p_i-p_f+q,f\mid J(q)\mid
i\rangle\mid^2 \label{int}.
\end{equation}
Equation~(\ref{int}) can be evaluated most easily by using a variable
\be\bbox{k}\equiv
\bbox{p_i}-\bbox{p_f},\ee
with $\bbox{k}\equiv(k^+,\bbox{k_\perp}),$
the momentum that the struck nucleon  had in the
initial state. The simplification emerges from the 
argument of the delta function, under the Bjorken limit. One finds:
\be(p_i-p_f+q)^2-M^2 \approx
-Q^2+q^-(p_i-p_f)^+=-Q^2+q^-k^+.\ee
Using this approximation, and changing the  variables of integration
to  $\bbox{k}$ gives:
\begin{equation}
d\sigma\sim \sum_f \int {d^2k\;dk^+\over k^+}
\delta\left(-Q^2+q^-k^+\right)\mid \langle p=k+q,f\mid J(q)\mid
i\rangle\mid^2 \label{int1}.
\end{equation}
For large values of $\nu$, one may use the single-nucleon approximation:
\be \langle p=k+q,f\mid J(q)
i\rangle=\langle f\mid b_{\bbox k}\mid
i\rangle, \ee
where $b_{\bbox k}$ is a nucleon destruction operator. Thus we find
\begin{equation}
d\sigma\sim \sum_f \int {d^2k\;dk^+\over k^+}
\delta\left(-Q^2+q^-k^+\right)\mid \langle f\mid b_{\bbox k}\mid
i\rangle\mid^2 \label{int2}.
\end{equation}
The important advantage is that  
  $p_f^-$ does not appear in the
argument of the delta function, or anywhere else in Eq.~(\ref{int2}). Thus the
sum over intermediate states can be replaced by unity, using
Eq.~(\ref{complete}).
In the usual
equal-time representation,  the argument of the delta function
is  $-Q^2+2\nu(E_i-E_f)$. The energy of the final state appears, and
one can not do the sum over states.
Using light front dynamics,  we re-write 
Eq.~(\ref{int2}) as
\be
d\sigma\sim \int {d^2k\;dk^+\over k^+}\delta(-Q^2+q^-k^+) \langle i\vert
 b^\dagger_{\bbox k} b_{\bbox k}\vert i\rangle
.\ee
Only a ground state matrix element is required; this is the simplification
we seek. Then integration over $k^+$ and noting that the operator
$b^\dagger_{\bbox k} b_{\bbox k}$ is  a number operator
\begin{equation}
d\sigma\sim
{ d^2{k}_\perp} \rho(M x,\bbox{k}_\perp),
\label{ds2}
\end{equation}
where $ \rho(M x,\bbox{k}_\perp)= \langle i\vert
 b^\dagger_{k^+=Mx,\bbox k_\perp} b_{k^+=Mx,\bbox k_\perp}\vert i\rangle$
 is the probability for a nucleon in
the ground state to have a momentum $(M x,{k}_\perp)$. Integration in
Eq.~(\ref{ds2}) leads to
\begin{equation}
\sigma \sim\int d^2k_\perp\, \rho(M x,{k}_\perp)\equiv f(Mx),
\end{equation}
with $f(Mx)$ as the probability for a nucleon in the ground state to
have a plus momentum of $Mx$, or the nucleon distribution function. 
The use of light-front dynamics to compute nuclear wave functions
should allow first-principles calculations of  $f(Mx)$.  
Using light-front dynamics incorporates the
experimentally relevant kinematics from the beginning, and therefore is
the most efficient way to compute the cross sections for nuclear deep
inelastic scattering and nuclear quasi-elastic scattering. 
\subsection{An Apparent  Puzzle}
Since much of this
work is
motivated by the desire to understand nuclear deep inelastic scattering
and related experiments,
it is worthwhile to review some of the features of the EMC
effect \cite{emc,emcrevs,fs2}. One key experimental result is the
suppression of the structure function for $x\sim 0.5$. This means that
the valence quarks of bound nucleons carry less plus-momentum than
those of free nucleons. This may be understood by
postulating that mesons carry a larger fraction of the plus-momentum in
the nucleus than in free space\cite{chls,et}.
While such a model explains the shift
in the valence distribution, one obtains at the same time a meson (i.e.
anti-quark) distribution in the nucleus, which is strongly enhanced
compared to free nucleons and which should be observable in Drell-Yan
experiments \cite{dyth}. However, no such enhancement has been observed
experimentally \cite{dyexp}, and  this has been termed as a severe crisis
for nuclear theory in
Ref.~\cite{missing}.

\subsection{More Motivation}
The use of light-front dynamics should allow us to compute the necessary
nuclear meson distribution functions using variables which are
experimentally relevant. The need for a computation of such functions
in a manner consistent with generally known properties of nuclei led
me to try to compute nuclear properties using light front dynamics.
There are  other motivations for using the light front formalism that have
been emphasized in many reviews\cite{lcrevs}.
One key feature is the belief (or hope) that the vacuum of the
theory is trivial because it can not create pairs.  Another is
that the theory is a Hamiltonian theory and the many-body techniques of 
equal time theory can be used here too.
I also quote the review by Geesaman
et al.\cite{emcrevs}
``In light front dynamics LFD,
the
particles are on mass-shell, and there are no off-shell
ambiguities. However, ... we have little or
no experience in
calculating the wave function of a  realistic nucleus in LFD''.
The aim
here is to provide such wave functions. 

I also like to say: Ask not what the light front can do for nuclear physics;
instead
ask
what nuclear physics can do for the light front. This is to provide a set of
non-trivial four dimensional examples with real physics content.

There are have been  several serious efforts devoted  to solving QCD using
light front
dynamics\cite{brodsky,pauli,osu}.
This is proving be a formidable task.  Much of the technical
difficulties are associated with the vanishing gluon mass and the nearly
vanishing values of the current quark masses.
In equal time dynamics, the vacuum is known to contain
both quark and gluon condensates, and recovering  the physics of
the complicated vacuum using light front dynamics requires a
careful treatment 
of the constraint equations and the related zero-modes. This is discussed
nicely in the conference proceeding~\cite{conf}. See also the most recent
review by Brodsky et al.\cite{lcrevs}.

As an example of the possible
difficulties one faces in solving a gauge theory using light front
quantization,  recall the work of Casher\cite{casher}. As explained by
Paston et al\cite{paston},
Casher  found that  the use of the gauge $A^+=0$ is
necessary. In that gauge the Feynman propagator has an extra power of
the plus-momentum in the denominator, which strengthens the infrared
singularities and introduces the need for a 
 special  regularization. One may 
"cut  out"  the  infrared regime 
momentum space, but this procedure   breaks 
the Lorentz invariance until  the  regularization  is  removed.
  In
principle, one  can preserve the gauge invariance  by using a discrete light
cone quantization
\cite{lcrevs}, but one also needs  to keep the zero modes.
This procedure typically introduces enormous complications.
But not all theories are so complicated. In the case of Yukawa theory,
which is very similar to our Lagrangians, the difficulty with the infrared
regime as already been solved in an convincing  manner Ref.~\cite{blb}.

The essential simplification encountered here is in our use of hadronic
Lagrangians,
which are not gauge theories. For example, consider that,
in the theories  we consider,  
 all of the particles are massive. No  one believes
that
the vacuum is a condensate of nucleon-anti-nucleon pairs. The vacuum really
should be trivial.
The implication is that the complications associated with zero modes can
legitimately be ignored.
Furthermore, the meson masses are
much larger ($\ge 135$ MeV ) than typical energy scales ($\sim$ 10 MeV) of
nuclear physics.
This means that truncation of the Fock space to terms
involving only 0,1 or 2 mesons should be a reliable approximation. Thus
many of the technical difficulties of using light front dynamics are not
present
for hadronic Lagrangians. There is no reason to prevent
light front dynamics from becoming
a standard tool of relativistic nuclear physics.
\subsection{Outline and Scope}
The  general goal is to provide a series of 
examples showing  that  light front dynamics  can be
used for high energy realistic and relativistic nuclear physics.
The goal is to present the physics in as simplified form as possible.

We shall 
begin with a very brief overview of  light front dynamics. Then
some simple examples\cite{toy}, requiring little technology are solved.
The formal procedures of light front quantization of a hadronic Lagrangian
${\cal L}$
are  discussed next. Our  first application of this quantization
is a study of 
infinite nuclear matter within the mean field approximation\cite{jerry}.
The distribution
functions $f(y)$ for nucleons and mesons are  computed.
The results are startling --the vector mesons are found to carry a substantial
fraction of the nuclear plus-momentum. This finding  is analyzed using a simple
static model\cite{bm98} in the following section.
Then finite nuclei\cite{bbm99} are studied
using the  mean field approximation. Here one confronts  the problem that
the use of $x^-=t-z$ as a spatial variable violates manifest
 rotational
invariance.  We found\cite{bbm99} 
that rotational invariance re-emerges in the results,
after one does the appropriate
dynamical calculation. The desire for a realistic approach means that
nucleon-nucleon correlations must be included and one must do physics 
 beyond the mean field approximation. Chiral symmetry and a simple example
 involving pion-nucleon scattering is then discussed\cite{jerry}.
 This is necessary
  preparation for the development of a new light-front one
  boson exchange potential\cite{rmgm98}.
 The influence of 
nucleon-nucleon correlations on the properties of nuclear matter is studied by
making the necessary light front calculations. Applications are to compute the
nuclear pionic content and to nuclear deep inelastic scattering and 
Drell-Yan processes. This study\cite{rmgm98}
indicates that the apparent puzzle is not a
real puzzle. Then some plans for future research are discussed. The impact of
the calculations discussed here is that  
 light front dynamics  can be
used for high energy realistic and relativistic nuclear physics. Reasonable
results for standard properties are obtained. There is a strong desire for an
experimental measurement of an observable  that has an importance  more
readily appreciated and that is more  easily computable using
light
front dynamics. As discussed, in Sect.~11,
the discovery of the HERMES effect\cite{hermes} that the ratio
$R\equiv \sigma_L/\sigma_T$ is significantly enhanced in nuclei may provide
such an observable. 

The emphasis of this review is on the influence of light front quantization on
nuclear physics. Hadronic Lagrangians are used. There are several excellent
reviews of light front techniques\cite{lcrevs,hari}, but these are mainly
devoted to applications in
particle physics\cite{brodsky,pauli,osu}. I am concerned here mainly with
heavy nuclei.
Applications of light front techniques
to nuclear reactions have been pioneered by Frankfurt and
Strikman\cite{fs,fs2}.
The present work reviews the first attempts, using realistic inputs,
to apply light front techniques to
the structure of heavy nuclei. This is the salient feature of the present
review. However, it is worth mentioning the  differences between
the present and previous uses of light front techniques. Apart from 
Ref.~\cite{georgia}, almost
all of the early work\cite{fs,fs2,coester,Ke 91,Ka 88,Fu 91} 
is devoted to the two-nucleon system, which is of tangential 
importance here.  But this is only one difference.
There are also differences in philosophy.
Our approach is to choose a Lagrangian, and then determine
 its
consequences for experimental observables. If the consequences are not
palatable, we change the Lagrangian. The approach of \cite{coester,Ke 91,Ka 88}
relies on symmetries which are used to write down the
allowed form of operators. This is not sufficient to determine uniquely, for
example,
all of
the components of the
electromagnetic current operator $J^\mu$. The approaches of
\cite{Ka 88,Fu 91}
try to restore manifest rotational invariance by introducing a general
direction $\widehat{n}$ to replace the three-direction.  We expect, and show,
that  the final calculations of observables do respect this invariance.
Our approach is closest to that
of Refs.~\cite{fs,fs2}, which is based on Feynman diagrams. The 
specification of a Lagrangian, should allow us to go further because
all of the relevant diagrams can be defined.

\section{ What is light front dynamics? }

This is a relativistic treatment
of dynamics in which the fields are quantized at a fixed ``time''$\tau =t+z
=x^0+x^3\equiv x^+$. This means that the independent
spatial variable must be 
$x^-\equiv t-z$ so that the canonical momentum is $ p^0+p^3\equiv p^+$. The
remainder of the spatial variables are given by
$ \vec{x}_\perp,\vec{p}_\perp$.

The consequence of using $\tau$ as a 
 ``time'' variable is that the canonical energy is $p^-=p^0-p^3$.
Our notation for four-vectors is to use:
\begin{equation}
A^\pm\equiv A^0\pm A^3,
\end{equation}
with
\begin{equation}
A\cdot B =A^\mu B_\mu={1\over2}\left(A^+B^- +A^-B^+\right)
-\vec{A}_\perp\cdot\vec{B}_\perp
.\end{equation}

The key reason for using such unusual coordinates is phenomenological.
 For a particle with  $\vec v\approx c\hat{e_3}$, the quantity   $p^+$
is BIG. Thus experiments tend to measure quantities associated with
$p^+$.

Another important feature is the relativistic dispersion relation
$p^\mu p_\mu =m^2$, which
in light front dynamics takes the form:
 \begin{equation}
 p^-={p_\perp^2+m^2\over p^+}
 \label{ONE}.\end{equation}
Thus one has a form of relativistic kinematics that avoids using a
square root. Eq.~(\ref{ONE}) is perhaps the most important equation in  light
front physics. Its consequences appear again and again in this review. If
a  reader, new to the light front,   decides to learn only one of the
many equations of this review--this should be the equation.

The main formal consequence of using light front dynamics is that the minus
component of the total momentum,
$P^-$, is used as a Hamiltonian operator, and the plus component 
 $P^+$ is used as a momentum  operator. These
 are obtained by using the energy-momentum tensor
\begin{equation}
T^{\mu\nu}=-g^{\mu\nu}{\cal L} +{\partial{\cal L}
  \over\partial
  (\partial_\mu\phi_i)}\partial^\nu\phi_i, 
\end {equation}
in which the degrees of freedom are labeled as $\phi_i$.
The term $T^{+\mu}$  is the density for the operator $P^\mu$, with 
\begin{equation}
P^\mu={1\over 2}\int d^2x_\perp dx^- T^{+\mu}.
\end{equation}
The use of a Lagrangian to provide these operators is the salient feature of
our
approach.
 The procedures to obtain these
 operators for the different Lagrangians are discussed below.

\section{Simple problems}
We introduce light front techniques by solving the wave equations for  
the three dimensional (scalar) harmonic oscillator, and the three
dimensional Hulth\'en
potential.
We consider a situation in which a single nucleon moves under the influence of
an external potential. This external potential arises from interactions with 
residual  $(A-1)$--particle nucleus.
The physics of the entire $A$--nucleon system is Lorentz invariant.  In the
present example the external potential is most simply expressed in the rest
frame of the nucleus, for which it independent of time.  
For simplicity, we neglect the effects of
spin and use the Klein-Gordon equation.

\subsection{Three Dimensional Harmonic Oscillator}

In ordinary coordinates, the Klein-Gordon  equation for a 3 dimensional
relativistic oscillator
\be
E^2 \phi  = \left[{\vec p}^2 + m^2 + \kappa {\vec x}^2 \right] \phi
\ee
closely resembles the Schroedinger equation for a non-relativistic harmonic 
oscillator, and the  energy eigenvalues are
\be
E_n^2 = m^2 + \omega \left( n+ \frac{3}{2} \right),
\label{eig}\ee
where $\omega^2 = 4 \kappa$.

The light front version of the Helmholtz equation is obtained by changing
variables to $x^\pm=x^0\pm x^3$, with
\be \partial^\pm =2{\partial\over \partial x^\mp}.\ee
Then \be
\left(\partial^-\partial^+-\nabla^2_\perp+m^2+\kappa
(x^2_\perp+(x^+-x^-)^2/4)\right)\phi=0.\label{lfs}\ee
A   potential that is static in the equal time formulation corresponds to
a LF-``time'', i.e. $x^+$, dependent
 potential in light-front coordinates.
A static source in a 
rest-frame corresponds to a uniformly moving source on the 
light-front because the line $z=z_0$ corresponds to $x^\pm=t\pm z_0$ and
$\Delta x^-/\Delta x^+=1$. Thus the time dependence  is
only due to a uniform translation, and  it should be easy to transform 
Eq.~(\ref{lfs})   into
a form which contains a potential that does not depend on  $x^+$. 
 For this purpose, we consider the equation of motion 
satisfied by  fields  obtained by an $x^+$ (LF-time) 
dependent translation 
\begin{eqnarray}
\phi  ({\vec x}_\perp,x^-,x^+)
&\equiv& e^{-ix^+P^+/2}{ \chi} ({\vec x}_\perp,
x^-,x^+) 
\label{eq:tilde}
\end{eqnarray}
Using $P^+=-i2{\partial\over \partial x^-}$, we find
\begin{eqnarray}
e^{ix^+ P^+/2} f\left({x^+-x^-\over2}\right) e^{-ix^+ P^+/2} &=&
f\left({-x^-\over 2}\right)
\nonumber\\
e^{ix^+ P^+/2} \partial^-e^{-ix^+ P^+/2} &=& 
\partial^- - \partial^+ \label{trans}.
\end{eqnarray}

The use of Eqs.~(\ref{eq:tilde})    and (\ref{trans})
in Eq.~(\ref{lfs}) then leads to the result
\be 
\left((\partial^--\partial^+)\partial^+-\nabla^2_\perp+m^2+\kappa (x^2_\perp+
({x^-\over 2})^2)\right)\chi=0 .\label{lfst}\ee
Note that the potential does not depend on $x^+$ and is instead a function
of $x^2_\perp+({x^-\over 2})^2$. The term $({x^-\over 2})^2$ can be thought of
as playing the role of $z^2$, so that the potential is actually spherically
symmetric. Indeed, if we simply set $x^+=0$, then $x^0=-x^3$ and
$x^-=-2x^3=-2z$.   

It is instructive to write the result (\ref{lfst}) in an operator form:
\be i\partial^-\chi=\left[i\partial^+ + {-\nabla^2_\perp+m^2\over i\partial^+}
  +{1\over \sqrt{i\partial^+}}\;V\;{1\over
    \sqrt{i\partial^+}}\right]\chi,\label{recoil} \ee
      where $V\equiv \kappa (x^2_\perp+({x^-\over 2})^2).$ This looks like an
      operator version of Eq.~(\ref{ONE}), but including an interaction and
      an 
  additional $p^+$ 
operator which
accounts for the recoil of the $A-1$-nucleon  residual nucleus,
see Sect. VII. The factors ${1\over \sqrt{i\partial^+}}$ are boson
normalization factors; see Sect.~IV.

The  potential appearing in Eq.~(\ref{lfst}) does not depend on $x^+$ so that
there will be 
solutions of the  form:
\be
\chi(x)=e^{-ip_n^-x^+/2}\chi_n(x^-,\vec {x}_\perp).\ee
Using this and  completing the square,  leads  to  the result
\be
\left( -(\partial^++ip^-_n/2)^2-\nabla^2_\perp+m^2+\kappa (x^2_\perp+
({x^-\over 2})^2)\right)\chi_n=(p^-_n/2)^2\chi_n .\ee
One converts the operator $ (\partial^++ip^-_n/2)^2$ to $(\partial^+)^2$ using
yet  another transformation:
\be
\chi(x^-,\vec{x}_\perp)=e^{-ip_n^-x^-/4}F(x^-,\vec{x}_\perp)\ee
to find
\be \left( -(\partial^+)^2-\nabla^2_\perp+m^2
  +\kappa\left[x^2_\perp+({x^-\over 2})^2\right]\right)F=(p_n^-/2)^2 F.\label{312}\ee
This is the same form as 
the equation in the equal time coordinates, and $p^-_n/2$
takes on the values of $E_n$:
\be
(p_n^-/2)^2=m^2+\omega(n+3/2).\ee

The light front version gives the same results for the eigenvalues
as   the ET development. So one 
might wonder why we are doing the light front at all. The point is that we are 
      able to compute the light front
wave functions that depend on $x^-$, or in  momentum space depend on $p^+$.
The wave function of the ground  state of Eq.~(\ref{lfs}) is
     given by
     \be
  \chi_0(x^-,\vec{x}_\perp)=e^{ip_0^-x^-/4}N_0 \exp
  {\left(-{1\over2}\sqrt{\kappa}
    (x_\perp^2+{{x^-}^2\over
          4})\right)}.
      \label{wave0}\ee

      The number density $n_0(p_\perp,p^+)$  is  defined as the square
      of the momentum space version of $\chi_0$.
This quantity is accessible in high energy proton and electron nuclear
quasi-elastic reactions.
      It is useful to define
      the light front variable \be\alpha\equiv p^+/({p_0^-\over2}).
      \label{lfv}\ee For a shell-model nuclear target $p_0^-$ is a definite
      fraction of the total minus-momentum. In the rest frame,
      the minus-momentum is the same as the plus-momentum. Thus the variable
      $\alpha$ is a ratio of plus-momenta and  $n_0(p_\perp,\alpha)$ is
      independent of frame. 
      
The Fourier transform of Eq.~(\ref{wave0}) is obtained by multiplying
by $e^{ i\bbox{p}_\perp\cdot\bbox{x_\perp}}
e^{-i{1\over 2}p^+x^-}$ and doing the
  integral over all $\bbox{x_\perp}$ and $x^-/2.$
 Then one  determines that
      \be n_0(p_\perp,\alpha)=\tilde{N}_0e^{-{p_\perp^2\over\sqrt{\kappa}}}
      e^{{-(p_0^-(\alpha-1))^2\over4\sqrt{\kappa}}}.
      \label{nho}\ee
Note that one finds the same $p_\perp$ distribution for each value of the
variable $\alpha$.  This is not a general 
feature of light-front wave functions, as we show in the next sub-section.

In an exact calculation, the wave functions and
number density should vanish for values of
a plus-momentum fraction variable  that are not between 0 and 1.
This is referred to as having proper support or satisfying the spectrum
condition.
This property is not evident here because the system consists of one particle
that interacts with an external source which supplies the external potential
$V$ and is also responsible for the term $i\partial^+$ on the right hand side
of Eq.~(\ref{recoil}). To better understand the support and the origin of the
term $i\partial^+$, consider a two-particle problem in which a light particle
(nucleon)
of mass $m$ interacts with a heavy particle (residual nucleus) of mass $M$. The
total four-momentum of the system is denoted as $P^\mu$. For
a nucleus with
$A$ nucleons, $M
\approx (A-1)m$ and $M\gg m$. This is a weak binding limit (binding energy per
particle much less than the mass)  appropriate for
nuclear physics. The light particle  is labeled as $1$ and the
heavy particle as $2$. The plus-momentum fraction carried by the light particle
is denoted as $\beta$ with
\be \beta={p^+_1\over P^+},\ee
so that
\be {\beta\over \alpha}={p_0^-\over2P^+}. \ee 
  Using the given
masses $m,M$ it will always be true  that $\beta\ll1$, and that
$\alpha\sim 1$.
The relative momentum is given by
\be
\bbox{p}_\perp=(1-\beta)\bbox{p_1}_\perp-\beta \bbox{p_2}_\perp,\ee
with \be\bbox{p}_\perp\approx\bbox{p_1}_\perp.\ee
The use of these variables allows  leads to a simple expression for the kinetic
energy \bea p^-_1+p^-_2&=&{p_\perp^2\over \beta(1-\beta)P^+}+{m^2\over \beta P^+}
+{M^2\over(1- \beta) P^+}\label{good1}\\
&  \approx & {p_\perp^2+m^2\over p^+}+p^++{M^2\over P^+},\label{heavy}\eea
in which the weak binding
approximation ${M\over P^+}^2\to 1$ is used. The right hand side
of Eq.~(\ref{heavy}) is the kinetic part of the  right hand side of
Eq.~(\ref{recoil}). Thus the term $i\partial^+=p^+$ represents the momentum of
the heavy particle which supplies the external potential. The use of the
kinetic energy operator of Eq.~(\ref{good1}) instead of that of
Eq.~(\ref{recoil})
would lead to obtaining exact support properties.
  
The approximation does not cause much trouble. 
For large values of the particles mass 
$m$, ($m\gg\kappa^{1/4})$ (which corresponds to
the situation of
nuclear physics in which the 
product of the nucleon mass and the nuclear radius is a very large number)  
the value of $\alpha$ must always be close to unity.
In nuclear physics ${m^2\over \sqrt{\kappa}}$ is
of the order of $(5R_A/ \rm{Fm})^2$ which taking $R_A=4$ Fm, yields
$e^{- 400}$ at $x=0$. In that case, there is no problem with the support.

To provide an example, we consider the 
 electromagnetic form factor of the ground
state of a heavy nucleus. This  has been    measured, for many nuclei,
in elastic electron scattering,
and
the sizes of nuclei have been determined as one of the classic achievements of
nuclear physics. Interest in this topic has been revised because of a
recent proposal to Jefferson Laboratory to use parity-violating electron
scattering to measure the neutron radius\cite{happex}.
Determining this to  high precision is needed, and
can be obtained provided one knows the proton distribution. 
Therefore it is useful 
 to examine the influence of  presumably
small effects such as  relativistic
corrections.
One works using a reference frame, the Drell-Yan frame
\cite {dyframe} in which the plus component of the four
vector $q^\mu$ of the virtual photon vanishes, so that $Q^2=-q^2=q^2_\perp$.
Now consider the various momenta involved. Suppose initially,
$\bbox{P_\perp}=0$.
If the light particle absorbs the virtual photon the final momenta are
$\bbox{p_\perp+q_\perp} $
for particle 1 and  $\bbox{-p_\perp}$ for particle 2. The relative
momentum  is then $\bbox{p_\perp}+(1-\beta)\bbox{q_\perp}$
In this case the form factor $F(Q^2)$
(matrix element of the plus component of the
electromagnetic current operator) is given by
\be F(Q^2)=\int d^2p_\perp d\beta\; \beta \chi_0(\beta,\bbox{p}_\perp)
 \chi_0(\beta,\bbox{p}_\perp+(1-\beta) \bbox{q}_\perp),
 \ee
in which the influence of relativity appears in the integral over $\beta$
and the factor $\beta$. For the harmonic oscillator ground state we
find:
\be F(Q^2)= N\int d^2p_\perp d\alpha\;  \alpha e^{-\left(
    {p_\perp^2\over \sqrt{\kappa}}
    +{Q^2(1-\beta)^2\over 4\sqrt{\kappa}}\right)}
e^{-{{p_0^-}^2\over 4\sqrt{\kappa}}(\alpha-1)^2}\label{form}
,\ee
where \be {{p_0^-}^2\over 4}=m^2+3\sqrt{\kappa}.\ee

Our purpose here is the study of nuclear physics, so we are interested in
the non-relativistic limit and the corrections to it.
To this end, we define a variable $p_z$ using
\be \alpha=1+{p_z\over m}.\ee The non-relativistic limit of (\ref{form}) is
obtained by letting $m$ approach infinity, but with $m/M\approx 1/(A-1)$, and
\be \beta={m\over M}\alpha.\ee
Then $ {{p_0^-}^2\over 4}=m^2$ and
we find
\bea F_{NR}(Q^2)&=& N_{NR}\int {d^3p\over m} e^{-\left(
    {p^2\over \sqrt{\kappa}}+{Q^2(1-m/M)^2\over
      4\sqrt{\kappa}}\right)}\nonumber\\
&=& e^{-{Q^2(1-m/M)^2\over 4\sqrt{\kappa}}}
.\eea
The mean square radius  $-6{d F(Q^2)\over d Q^2}\mid_{Q^2=0}$, is given by
\be R^2_{NR}={3\over 2} {(1-m/M)^2\over\sqrt{\kappa}}.\ee
The experimental value of the nuclear radius is given approximately by
\be R^{\rm exp}=1.1 A^{1/3}\; {\rm fm}.\label {expr}\ee 
The leading correction to this comes from the  product of
the factor $p_z\over m$ in the
integrand of Eq.~(\ref{form})
with a $p_z\over M$ which comes from a term in the exponent.
This is correction is  of order
$p_z^2/mM\sim \sqrt{\kappa}/mM$. We define a semi-relativistic limit $SR$ via
the use of (\ref{form}) and keeping the leading correction terms. Performing
the straightforward evaluations, using $R_{NR}=R^{\rm exp}$,  leads to the result:
\be
\delta\equiv{R_{SR}^2-R_{NR}^2\over R_{NR}^2}= 
{\sqrt{\kappa}\over mM},
\ee
or using Eq.~(\ref{expr})
\be \delta\approx{0.055\; 
  \over  A^{2/3}(A-1)}.\label{relre}\ee
This corresponds to very small ($8\times\;10^{-6}$)
effects for large nuclei $A\sim 200.$

\subsection {Light Front Hulth\'en Wave Function}
\begin{figure}
\unitlength1cm
\begin{picture}(10,8)(-3,2.5)
\includegraphics{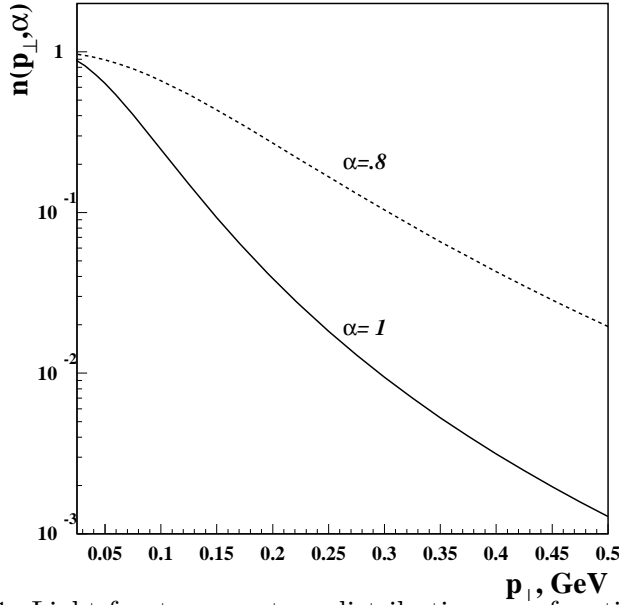}
\end{picture}
\caption{Light front momentum distribution as a function
  of $\alpha$ and $p_\perp$.
 }
\label{fig:hulth}
\end{figure}

Any static potential of the form $V(\vec{x}^2=x_\perp^2+z^2)$
 can be solved on the light front. The transformation
 (\ref{eq:tilde}) corresponds to including the $p^+$ term in the
 $x^+$ development operator and 
a  simple prescription of replacing $z$ by $-x^-/2$ in $V$.
We present here  the solution for the Hulth\'en potential. This allows us to
demonstrate an
interesting contrast between the implications of different forms of 
potentials.
In the equal time formulation we have the wave equation:
\be
E^2 \phi  = \left[{\vec p}^2 + m^2 + V^H ({\vec x}^2) \right] \phi,
\ee
in which $V^H$ is chosen so that the lowest energy solution is 
\be
\phi(r)=N(e^{-ar}-e^{-br}),
\ee
where $b>a$. The eigenenergy is given by
\be
E=\sqrt{m^2-a^2}.
\label{EH}\ee
The light front version of the wave equation is obtained in the same manner as
Eq.~(\ref{312}), with the result
 \be
    {{p_n^-}^2\over 4}\chi(x^-,\vec{x}_\perp)=\left[\left(-2i{\partial\over
          \partial x^-}\right)^2+p_\perp^2+V^H\left(x_\perp^2+{{x^-}^2\over
          4}\right)\right]\chi(x^-,\vec{x}_\perp)
     . \ee The lowest value of $p_n^-/2$ is clearly the
     same as $E$ of Eq.~(\ref{EH}), and the
ground state      wave function is
     given by
     \be
  \chi_0(x^-,\vec{x}_\perp)=e^{ip_n^-x^-/4}N_0^H
  \left[\exp{\left(-a\sqrt{x_\perp^2+{{x^-}^2\over 4}}\right)}
    -\exp{\left(-b\sqrt{x_\perp^2+{{x^-}^2\over 4}}\right)}\right]
  .
      \ee
The momentum distribution $n^H_0(p_\perp,p^+)$ obtained here provides an
interesting contrast with that of the harmonic oscillator (\ref{nho}).
We find
\be
n^H_0(p_\perp,\alpha)
=\tilde{N}_0^H\left[{1\over \left( a^2+p_\perp^2+(p^-_n/2)^2
    (\alpha-1)^2 \right)^2}
-{1\over \left( b^2+p_\perp^2+(p^-_n/2)^2
    (\alpha-1)^2\right)^2}\right]
.\ee
There is  a different $p_\perp$ distribution
for each value of
$\alpha$, with broader functions of $p_\perp$ obtained for smaller values
of $\alpha$; see Fig.~\ref{fig:hulth}. The results in the figure
are obtained using $m$=.94 GeV, $E$=.932 GeV and $b=5 a$.
An experimental hint of such a behavior has been found
recently\cite{ep}.

\section{Light Front Quantization}

I discuss the basic aspects in an  informal way as possible.
For more details, see the reviews and the references. It's easiest to
 start by considering
one free field at a time. These are  the scalar meson $\phi$, the Dirac
fermion $\psi$ and  the massive vector meson $V^\mu$. 
\subsection{Free Scalar field}
Consider the Lagrangian
\begin{eqnarray}
{\cal L}_\phi 
= {1\over 2} (\partial^+\phi \partial^-\phi -\bbox{\nabla}_\perp\phi
\cdot\bbox{\nabla}_\perp\phi-m_s^2\phi^2).\label{lagphi}
\end{eqnarray}
The notation is such that
$
  \partial^\pm=\partial^0\pm\partial^3=2 {\partial\over \partial x^\mp}$.
The Euler-Lagrange equation
leads to the wave equation
\begin{eqnarray}
  i\partial^-\phi={-\nabla^2_\perp+m_s^2\over i\partial^+}\;\phi.
\end{eqnarray}
The most general solution is a superposition of plane waves:
\begin{equation}
\phi(x)=
\int{ d^2k_\perp dk^+ \theta(k^+)\over (2\pi)^{3/2}\sqrt{2k^+}}\left[
a(\bbox{k})e^{-ik\cdot x}
+a^\dagger(\bbox{k})e^{ik\cdot x}\right],
\label{expp}\end {equation} 
where 
$k\cdot x={1\over2}(k^-x^++k^+x^-)-\bbox{k_\perp\cdot x}_\perp$ with
 $k^-={k_\perp^2+m_s^2\over k^+}$, and $\bbox{k}\equiv(k^+,\bbox{k}_\perp)$.
The $\theta$ function restricts $k^+$ to positive values.
Note that
\begin{equation}
  i\partial^+e^{-ik\cdot x}=k^+e^{-ik\cdot x}
      .\end{equation}
    The value of $x^+$ that appears in Eq.~(\ref{expp}) can be set to zero, but
    only after taking necessary derivatives.

    Deriving the equal $x^+$ commutation relations for the fields
    is a somewhat obscure
    procedure \cite{yan12,lks70,r71},
  but the result can be stated in terms of  familiar
commutation relations:
\begin{equation}
[a(\bbox{k}),a^\dagger(\bbox{k}')]=
\delta(\bbox{k}_\perp-\bbox{k}'_\perp)
\delta(k^+-k'^+)\label{comm}
\end {equation}
with $[a(\bbox{k}),a(\bbox{k}')]=0$.

It is interesting\cite{hari} to consider the commutation relations between
fields at the same value of light front time:
\be
[\phi(x),\phi(y)]_{x^+=y^+}={-i\over 4}\epsilon
(x^--y^-)\delta^{(2)}(\bbox{x}_\perp-\bbox{y}_\perp)
,\ee
where $\epsilon (x)=\theta(x)-\theta(-x)$. This commutator, evaluated for
$x^0=y^0$, vanishes because the separation between $x$ and $y$ would
necessarily be space-like. However, for $x^+=y^+$ the separation is
$-(\bbox{x}_\perp-\bbox{y}_\perp)^2$ which vanishes for
$\bbox{x}_\perp=\bbox{y}_\perp$.  

The next step is compute the Hamiltonian $P^-$ for this system.
The conserved  energy-momentum tensor is given in terms of the Lagrangian:
\begin{equation}
T^{\mu\nu}_\phi=-g^{\mu\nu}{\cal L_\phi} +{\partial{\cal L}_\phi
  \over\partial
  (\partial_\mu\phi)}\partial^\nu\phi.\label{tmunup}
.\end {equation}
This brings us to the question of what is $g^{\mu\nu}$?
This is straightforward, although the results (viewed for the first time)
can be surprising:
\begin{eqnarray}
  g^{+\nu}=g^{0\nu}+g^{3\nu},
  \end{eqnarray}
  which implies
 \begin{eqnarray} g^{++}&=&g^{00}+g^{03}+g^{30}+g^{33}=1+0+0-1=0\nonumber\\
 g^{ij}&=&-\delta_{i,j} (i=1,2,j=1,2);\quad
g^{+-}=g^{-+}=2.
\end{eqnarray}
Then one finds that
\begin{equation}
T^{+-}_\phi=
\bbox{\nabla}_\perp \phi \cdot \bbox{\nabla}_\perp\phi+
m^2_s \phi^2.
\end {equation}

In general, the term $T^{+-}$  is the density for the operator $P^-$:
\begin{equation}
P^-={1\over 2}\int d^2x_\perp dx^- T^{+-}.
\end{equation}
The use of the field expansion (\ref{expp}), along with normal ordering
and  integration leads to the result:
\begin{equation}
P^-_\phi=\int d^2k_\perp dk^+
\theta(k^+)a^\dagger(\bbox{k})a(\bbox{k}){k_\perp^2+m_s^2\over k^+}.
\end{equation}
One defines a vacuum state $\mid0\rangle$ such that
$
  a(\bbox{p})\mid0\rangle=0.
$
Then the creation operators acting on the vacuum give the usual single particle
states:
\begin{eqnarray}
  P^-_\phi a^\dagger(\bbox{p})\mid0\rangle={p_\perp^2+m_s^2\over p^+}
  a^\dagger(\bbox{p})\mid0\rangle.\end{eqnarray}
The momentum operator $P^+$ is constructed by integrating $T^{++}$:
\begin{equation}
P^+_\phi=\int d^2k_\perp dk^+
\theta(k^+)a^\dagger(\bbox{k})a(\bbox{k}) k^+.
\end {equation}

It is interesting to consider how the results for $P^-$ are changed if
interactions are included. If we add an interaction term  of the form
${1\over 2}\phi V\phi$ one would get an equation for $P^-$ of the form of
Eq.~(\ref{recoil}), but without the term $i\partial^+$ on the right hand side.

Another consideration involves the vacuum.
Suppose we take the Lagrangian
\begin{eqnarray}
{\cal L}={1\over 2} (\partial_\mu \phi \partial^\mu
\phi-m_s^2\phi^2)+\lambda \phi^4.
\end{eqnarray}
The operator $\phi$ creates or destroys a particles of plus-momenta
$k^+>0$. Thus  possible terms in which $\lambda \phi^4$ term converts the
vacuum $\mid0\rangle$ into a four particle state vanishes by virtue of the
conservation of plus-momentum. The vacuum of $p^+=0$ can not be connected
to four particles, each having a positive $k^+$.
 This vanishing simplifies  Hamiltonian  ($x^+$-ordered perturbation)
 calculations. However, one must be careful about physics involving
 $k^+=0$. For example, Burkardt\cite{mb:adv} constructs non-vanishing
 perturbative corrections to the self energy of
 the
 scalar meson which do involve the spontaneous creation 
of four scalar mesons from the vacuum. For  a detailed  discussions, see
Ref.~\cite{conf}.
  \subsection{Free Dirac Field}
Consider the Lagrangian
\begin{eqnarray}
{\cal L}_\psi=
\bar{\psi}(\gamma^\mu
{i\over 2}\stackrel{\leftrightarrow}{\partial}_\mu
-M)\psi, \label{lagpsi}
\end{eqnarray}
and its equation of motion:
\be
\left(i \gamma^\mu\partial_\mu-M\right)\psi=0
\label{dirac0}  .\ee
A
fermion has spin 1/2, so there can only be two independent degrees of
freedom. The standard Dirac spinor has four components, so two of these must
represent dependent degrees of freedom. In the light front formalism one
separates the independent and dependent degrees of freedom by using projection
operators:
$
\Lambda_\pm\equiv {1\over 2} \gamma^0\gamma^\pm
.$ Then the independent field is $\psi_+=\Lambda_+\psi$ and the
dependent one is  $\psi_-=\Lambda_-\psi$

The Dirac equation (\ref{dirac0}) is re-written as
\be
\left({i\over 2}\gamma^+\partial^- +{i\over2}\gamma^-\partial^+
  +i\bbox{\gamma}_\perp\cdot\bbox{\nabla}_\perp-M\right)\psi=0
\label{dirac1}.\ee
Equations for $\psi_\pm$ can be obtained by multiplying
Eq.~(\ref{dirac1}) on the left by $\Lambda_\pm$: 
\bea
i\partial^- \psi_+=(\bbox{\alpha}_\perp\cdot{ \bbox{\nabla}_\perp\over i}
+\beta M)\psi_-\nonumber\\
i\partial^+ \psi_-=(\bbox{\alpha}_\perp\cdot{ \bbox{\nabla}_\perp\over i}
+\beta M)\psi_+,
\eea
so that the equation of motion of $\psi_+$ becomes
\be
i\partial^- \psi_+=(\bbox{\alpha}_\perp\cdot{ \bbox{\nabla}_\perp\over i}+\beta
M)
{1\over i\partial^+}(\bbox{\alpha}_\perp\cdot{ \bbox{\nabla}_\perp\over
  i}+\beta M)\psi_+
.\ee

One can make the field expansion and determine the momenta in a manner similar
to the previous section.
The key results are
\bea
T^{+-}_\psi
&=&\psi^\dagger _+\left(\alpha_\perp\cdot{\bbox{\nabla}_\perp\over i}
  +\beta M\right){1\over
  i\partial^+} \left(\alpha_\perp\cdot{\bbox{\nabla}_\perp\over i}
  +\beta M\right)\psi_+,
\\
P^-_\psi
&=&\sum_\lambda\int d^2p_\perp dp^+\theta(p^+){p_\perp^2+M^2\over  p^+}
\left[b^\dagger(\bbox{p},\lambda)b(\bbox{p},\lambda)+
  d^\dagger(\bbox{p},\lambda)
  d(\bbox{p},\lambda)\right],
\eea
where  $b(\bbox{p},\lambda),d(\bbox{p},\lambda)$ are nucleon and
anti-nucleon destruction operators.
\subsection{Free Vector Meson}
The formalism for massive vector mesons was worked out by Soper\cite{des71}
  and later
by Yan\cite{yan34}
using a  different formulation. I generally follow Yan's
approach. The  formalism in the references
is  lengthy and detailed in the references, so
 I try to state the
minimum. There are  three independent degrees of freedom, 
even though the Lagrangian depends on $V^\mu$ and
$V^{\mu\nu}=
\partial ^\mu V^\nu-\partial^\nu V^\mu$. 
These are chosen to be $V^+$ and $V^{+i}$. 
The other terms $ V^-,V^i,V^{-i}$ and $V^{ij}$ can be 
written in terms of $V^+$ and $V^{+i}$.

The expression for the vector meson field operator is
\begin{equation}
V^\mu(x)=
\int{ d^2k_\perp dk^+ \theta(k^+)\over (2\pi)^{3/2}\,\sqrt{2k^+}}
\sum_{\omega=1,3}\epsilon^\mu(\bbox{k},\omega)\left[
a(\bbox{k},\omega)e^{-i\bbox{k}\cdot\bbox{x}}
+a^\dagger(\bbox{k},\omega)e^{i\bbox{k}\cdot\bbox{x}}\right],\label{vfield}
\end {equation} 
where the polarization vectors are the usual ones:
\begin{eqnarray}
k^\mu\epsilon_\mu(\bbox{k},\omega)=0,\qquad \epsilon^\mu(\bbox{k},\omega)
\epsilon_\mu(\bbox{k},\omega')=-\delta_{\omega\omega'},\nonumber\\
\sum_{\omega=1,3}\epsilon^\mu(\bbox{k},\omega)
\epsilon^\nu(\bbox{k},\omega)=-(g^{\mu\nu}-{k^\mu k^\nu\over m_v^2}).
\label{p1}
\end{eqnarray}
Once again the four momenta are on-shell, with 
$k^-={k_\perp^2+m_v^2\over k^+}.$ The commutation relations are
\begin{equation}
[a(\bbox{k},\omega),a^\dagger(\bbox{k}',\omega')]=\delta_{\omega\omega'}
\delta^{(2,+)}(\bbox{k}-\bbox{k}'), 
\end{equation}
with $[a(\bbox{k},\omega),a(\bbox{k}',\omega')]=0$,
and lead to commutation relations amongst the
field operators that are the 
same as in Ref.~\cite{yan34}.

\subsection{We need a Lagrangian no matter how bad} 

It seems to me that one can not do complete dynamical calculations
using the light front formalism without specifying some Lagrangian.
One starts\cite{jerry} with $\cal L$
and derives field equations. These are used to express
the dependent degrees of freedom in terms of independent ones. One also uses
 $\cal L$ to derive $T^{\mu\nu}$ (as a function of independent degrees of
 freedom) which is used to obtain the total momentum
 operators $P^\pm$. It is $P^-$ that acts as a Hamiltonian operator in the
 light front $x^+$-ordered perturbation theory.

 We start with a Lagrangian containing scalar and vector mesons and nucleons
 $\psi'$.
 This is the minimal Lagrangian for obtaining a caricature of nuclear physics
 because the exchange of scalar mesons provides a medium range attraction
 which can bind the nucleons and the exchange of vector mesons provides the
 short-range repulsion which prevents a collapse. It is also useful, for
 phenomenological purposes, to include scalar meson self-coupling terms.
 Thus we take 
\begin{eqnarray}
{\cal L} &=&{1\over 2} (\partial_\mu \phi \partial^\mu \phi-m_s^2\phi^2)
-{\kappa\over 3!}\phi^3-{\lambda\over 4!} \phi^4
-{1\over  4} V^{\mu\nu}V_{\mu\nu} +{m_v^2\over 2}V^\mu V_\mu \nonumber\\
&+&\bar{\psi}^\prime\left(\gamma^\mu
({i\over 2}\stackrel{\leftrightarrow}{\partial}_\mu
-g_v\;V_\mu) -
M -g_s\phi\right)\psi', \label{lag} 
\end{eqnarray}
with the effects of other mesons included elsewhere and below.
The equations of motion are
\begin{eqnarray}
\partial_\mu V^{\mu\nu}+m_v^2 V^\nu&=&g_v\bar \psi'\gamma^\nu\psi'
\label{vmeson}\\
\partial_\mu\partial^\mu \phi+m_s^2\phi+{\kappa\over2}\phi^2+{\lambda\over
  6}\phi^3 
&=&-g_s\bar\psi'\psi',
\label{smeson}\\
(i\partial^--g_vV^-)\psi'_+&=&(\bbox{\alpha}_\perp\cdot
(\bbox{p}_\perp-g_v\bbox{V}_\perp)+\beta (M +g_s\phi))\psi'_-\label{nfg0}\\
(i\partial^+-g_vV^+)\psi'_-&=&(\bbox{\alpha}_\perp\cdot
(\bbox{p}_\perp-g_v\bbox{V}_\perp)+\beta (M +g_s\phi))\psi'_+. 
\label{nfg}
\end{eqnarray}
The presence of the interaction term $V^+$ on the left-hand  side of
Eq.~(\ref{nfg}) presents a problem because one can not easily solve for
$\psi_-$ in terms of $\psi_+$.
This difficulty is handled by using the Soper-Yan
transformation:
\be\psi'=e^{-ig_v\Lambda(x)}\psi ,\qquad
\partial^+ \Lambda=V^+. \label{sytran}\ee
 Using this in Eqs.~(\ref{nfg0})-(\ref{nfg}) leads to the more usable form
\begin{eqnarray}
(i\partial^--g_v \bar V^-)\psi_+=(\bbox{\alpha}_\perp\cdot 
(\bbox{p}_\perp-g_v\bbox{\bar V}_\perp)+\beta(M+g_s\phi))\psi_-\nonumber\\
i\partial^+\psi_-=(\bbox{\alpha}_\perp\cdot 
(\bbox{p}_\perp-g_v\bbox{\bar V}_\perp)+\beta(M+g_s\phi))\psi_+. 
\label{yan}
\end{eqnarray}
The cost of the transformation is that one gets new terms resulting from
taking derivatives of $\Lambda(x)$. One uses  $\bar V^\mu$ with 
\be
 \bar V^\mu=V^\mu-{1\over\partial^+}\partial^\mu V^+,
\label{vbar}\ee
and 
  $\bar V^\mu$
enters in the nucleon field equations, but  
 $V^\mu$  enters in the meson field equations.

We also need the eigenmode expansion for $\bar V^\mu$. This is given by
\begin{equation}
\bar V^\mu(x)=
\int{ d^2k_\perp dk^+ \theta(k^+)\over (2\pi)^{3/2}\,\sqrt{2k^+}}
\sum_{\omega=1,3}\bar\epsilon^\mu(\bbox{k},\omega)\left[
a(\bbox{k},\omega)e^{-ik\cdot x}
+a^\dagger(\bbox{k},\omega)e^{ik\cdot x}\right],\label{nvfield}
\end {equation} 
where, using Eqs.(\ref{vbar}) and (\ref{vfield}),
the polarization vectors $\bar\epsilon^\mu(\bbox{k},\omega)$ are
\begin{equation}
\bar\epsilon^\mu(\bbox{k},\omega)=
\epsilon^\mu(\bbox{k},\omega)-{k^\mu\over k^+}\epsilon^+(\bbox{k},\omega).
\label{p2}
\end{equation}
Note that
\begin{equation}
\sum_{\omega=1,3}\bar\epsilon^\mu(\bbox{k},\omega)
\bar\epsilon^\nu(\bbox{k},\omega)=-(g^{\mu\nu}-g^{+\mu}{k^\nu\over k^+}
-g^{+\nu}{k^\mu\over k^+}).
\label{ebar}
\end{equation}

Then we may construct the total four-momentum operator from 
\begin{equation}
P^\mu={1\over2}\int dx^-d^2x_\perp\,T^{+\mu}(x^+=0,x^-,\bbox{x}_\perp),
\end{equation}
with
\begin{equation}
T^{\mu\nu}=-g^{\mu\nu}{\cal L} +\sum_r{\partial{\cal L}\over\partial
(\partial_\mu\phi_r)}\partial^\nu\phi_r\, ,\label{tmunu}
\end {equation}
in which the degrees of freedom are labeled by $\phi_r$. We need
$T^{++}$ and $T^{+-}$, which are given by
\begin{equation}
T^{++}=\partial^+\phi\partial^+\phi+V^{ik}V^{ik}
+m_v^2V^+ V^+
+2\psi^\dagger_+ i\partial^+ \psi_+,
\label{tpp}
\end{equation}
and 
\begin{eqnarray}
T^{+-}&=&\bbox{\nabla}_\perp\phi\cdot\bbox{\nabla}_\perp\phi +m_\phi^2\phi^2
+2\left({\kappa\over 3!}\phi^3+{\lambda\over 4!} \phi^4\right)
+{1\over 4}(V^{+-})^2+{1\over 2}V^{kl}V^{kl} +m^2_vV^kV^k\nonumber\\
&&+\bar\psi
\left(\bbox{\gamma}_\perp\cdot(\bbox{p_\perp}-g_v\bbox{\bar{V}^-})
+M+g_s\phi\right)\psi.
\label{tpm2}
\end{eqnarray}

This form is still not useful for calculations because the constrained
field $\psi_-$ contains interactions. We follow Refs.~ \cite{des71,mpsw}
in expressing $\psi_-$ as a sum of terms, one $\xi_-$ whose relation
with $\psi_+$ is free of interactions, the other $\eta_-$ containing
the interactions. That is, rewrite the second of Eq.~(\ref{yan}) as
\begin{eqnarray}
\xi_-&=&{1\over i\partial^+}(\bbox{\alpha}_\perp\cdot 
\bbox{p}_\perp+\beta M)\psi_+\nonumber\\
\eta_-&=&{1\over i\partial^+}(-\bbox{\alpha}_\perp\cdot 
g_v\bbox{\bar V}_\perp+\beta g_s\phi)\psi_+. \label{yan1}
\end{eqnarray}
Furthermore, define $\xi_+(x)\equiv\psi_+(x)$, so that 
\begin{equation}
\psi(x)=\xi(x)+ \eta_-(x), \label{fcon}
\end{equation}
where $\xi(x)\equiv \xi_-(x)+\xi_+(x)$. This separates the dependent
and independent parts of $\psi$.

One needs to make a similar treatment for the vector meson fields. The
operator $V^{+-}$, is determined by
\begin{equation}
V^{-+}={2\over\partial^+}\left[g_v\;J^+-m^2_vV^+-\partial_iV^{i+}\right].
\label{vpm}
\end{equation} 
Part of this operator is determined by a constraint equation, because
the independent variables are $V^+$ and $V^{i+}$. To see this examine
Eq~(\ref{vpm}), and make a definition
\begin{equation}
V^{+-}=v^{+-}+\omega^{+-}, \label{vcon}
\end{equation}
where 
\begin{equation}
\omega^{+-}=-{2\over \partial^+}J^+.
\end{equation}
The sum of the last term of Eq\.~(\ref{tpm2}) and the terms involving
$\omega^{+-}$ is the interaction density. Then one may use
Eqs.~(\ref{tpm2}), (\ref{fcon}), and (\ref{vcon}) to rewrite the $P^-$
as a sum of different terms, with 
\begin{equation}
P_{0N}^-={1\over 2}\int d^2x_\perp dx^-\,
\bar\xi\left(\bbox{\gamma}_\perp\cdot\bbox{p}_\perp +M\right)\xi,
\label{freef}
\end{equation}
and the interactions
\begin{equation}
P^-_I=v_1+v_2+v_3, \label{defv}
\end{equation}
with 
\begin{equation}
v_1=\int d^2x_\perp dx^-\,\bar\xi\left(g_v\gamma\cdot\bar V+g_s\phi
\right)\xi,\label{v1}
\end{equation}
\begin{equation}
v_2=\int d^2x_\perp dx^-\,\bar\xi\left(-g_v\gamma\cdot\bar V
+g_s\phi\right)\;{\gamma^+\over 2i\partial^+}\;\left(-g_v\gamma\cdot\bar V
+g_s\phi
\right)\xi, \label{v2}
\end{equation}
and 
\begin{eqnarray}
v_3&=&{g_v^2\over8}\int d^2x_\perp dx^-\int dy^-_1\,
\epsilon(x^--y^-_1)\,
\xi^\dagger_+(y^-_1,\bbox{x}_\perp)\gamma^+\xi_+(y^-_1,\bbox{x}_\perp)
\nonumber\\
&&\times \int dy^-_2\,\epsilon(x^--y^-_2)
\xi^\dagger_+
(y^-_2,\bbox{x}_\perp)\gamma^+\xi_+(y^-_2,\bbox{x}_\perp),\label{v3}
\end{eqnarray}
where $\epsilon(x)\equiv \theta(x)-\theta(-x)$. The term $v_1$ accounts for
the emission or absorption of a single vector or scalar meson. The term
$v_2$ includes contact terms in which there is propagation of an
instantaneous fermion. The term $v_3$ accounts for the propagation of
an instantaneous vector meson.

Our variational procedure will involve the independent fields $\psi_+$,
so we need to express the interactions $P_{0N}^-$ and $ v_{1,2}$ in terms of
$\xi_+$. A bit of Dirac algebra shows that
\begin{eqnarray}
P^-_N&\equiv& P^-_{0N}+v_1+v_2\nonumber\\
&=&
\int d^2x_\perp {dx^-\over 2}\,\xi^\dagger_+\Bigl[2g_v\bar{V}^-
\nonumber\\
&+& \left(\bbox{\alpha}_\perp
\cdot(\bbox{p}_\perp-g_v\bbox{\bar V}_\perp)+\beta(M+g_s\phi)\right)
{1\over i\partial^+}
\left(\bbox{\alpha}_\perp
\cdot(\bbox{p}_\perp-g_v\bbox{\bar V}_\perp)
+\beta(M+g_s\phi)\right)\Bigr]\xi_+.
\label{good}
\end{eqnarray}

It is worthwhile to define the contributions to $P^\pm$ arising from
the mesonic terms as $P_s^\pm$ and $P_v^\pm$. Then one may use
Eqs.~(\ref{tpm2}) and (\ref{tpp}) along with the field expansions to
obtain
\begin{eqnarray}
P_s^-&=&{1\over2}\int d^2x_\perp dx^-\,
\left(\bbox{\nabla}_\perp\phi\cdot\bbox{\nabla}_\perp\phi +m_s^2\phi^2
\right)\nonumber \\
&=&\int d^2k_\perp dk^+\,\theta(k^+)a^\dagger
(\bbox{k})a(\bbox{k}){k_\perp^2+m_s^2\over k^+},
\label{pmph}
\end{eqnarray}
\begin{equation}
P_s^+=\int d^2k_\perp dk^+\,\theta(k^+)a^\dagger
(\bbox{k})a(\bbox{k})k^+,
\label{ppph}
\end{equation}
\begin{equation}
P_v^-=\sum_{\omega=1,3}\int{ d^2k_\perp dk^+ \,
\theta(k^+){k_\perp^2+m_v^2\over k^+}} 
a^\dagger(\bbox{k},\omega) a(\bbox{k},\omega)\;+v_3,\label{pvm}
\end{equation}
and
\begin{equation}
P_v^+=\sum_{\omega=1,3}\int d^2k_\perp dk^+ \,\label{ppm}
\theta(k^+)k^+
a^\dagger(\bbox{k},\omega) a(\bbox{k},\omega).
\end {equation}
The term $v_3$ is the vector-meson instantaneous term, and we include
it together with the purely mesonic contribution to $P_v^-$ because it
is canceled by part of that contribution.

Thus, our result for the total minus-momentum operator is
\begin{equation}
P^-=P_N^- +P_s^- + P_v^-,
\label{bigm}
\end{equation}
and for the plus-momentum
\begin{equation}
P^+=P_N^+ +P_s^+ + P_v^+
,\label{bigp}
\end{equation}
where from Eq.~(\ref{tpp})
\begin{equation}
P_N^+\equiv{1\over 2} \int d^2x_\perp dx^-\, 2\xi_+^\dagger i\partial^+\xi_+.
\end{equation}

\section{ Infinite   Nuclear Matter Mean Field Theory}

The aim of our approach is to do the realistic, relativistic physics of large
nuclei using light front dynamics. Since this is a difficult project, it is
very worthwhile
to consider first the simplest possible example. This is to consider the
nucleus of infinite size. One considers a limit in which
the radius and nucleon number are each taken to be
infinity,
but the baryon density, or number of nucleons per unit volume, is taken as
finite.
Furthermore, a useful first step is to consider the mean field approximation to
to the dynamics in which nucleons are treated as  moving
under the influence of self-consistent potentials. Field-theoretic versions of
relativistic mean field theory were pioneered by Walecka and exploited heavily
by his students and many other workers\cite{bsjdw}. Until our work,
the only light front version of relativistic field theory applied to infinite
nuclear matter was that of Glazek and Shakin\cite{gs} who treated the nucleus
as bound under the influence of scalar mesons.

The philosophy\cite{bsjdw} behind field-theoretic
mean field theory  is that the nucleonic densities which
are  mesonic sources
are large enough to generate a  large number of
 mesons to enable a classical treatment (replacing the  operators which
 represent the sources  by 
 expectation values).  In infinite nuclear matter, the volume is taken
 as infinity so that all positions are equivalent. Then, the nucleon mode
 functions are plane waves and the nuclear matter ground state is assumed
 to be a normal Fermi gas, of Fermi momentum $k_F$ and of large volume
 $\Omega$. Under these
 conditions one finds solutions of  Eqs.~(\ref{vmeson}) and
 (\ref{smeson}) in which the meson fields are constants:
\begin{eqnarray}
\phi=-{g_s\over m_s^2} \langle \bar \psi \psi\rangle\equiv -{g_s\over m_s^2}\rho_S
\label{sphi}\\
V^\mu={g_v\over m_v^2} \langle \bar \psi
\gamma^\mu\psi\rangle=\delta^{0,\mu}{g_v\rho_B\over m_v^2},\label{sv}
\end{eqnarray}
where  the expectation values refer to a ground state expectation value, and
$\rho_B=2k_F^3/3\pi^2$.
This result that $V^\mu$ is a constant.
The notion that there is no special direction in space is used.
 Eq.~(\ref{vbar}),  
 tells us that the  
only non-vanishing component of $\bar V$ is $\bar {V}^-=
V^0$.
Since the potentials entering the light-front Dirac equation (\ref{yan}) are
constant, the 
nucleon modes are plane waves $\psi \sim e^{ik\cdot x}$,
and the many-body system is a  
Fermi gas. The solutions of Eq.~(\ref{yan}) are
\be i\partial^- \psi_+=g_v\bar{V}^-\psi_++{k_\perp^2+(M+g_s\phi)^2\over
k^+}\psi_+.\label{sd}\ee
Solving
the equations (\ref{sphi}),(\ref{sv}) and (\ref{sd}) yields
a  self-consistent
solution. 
The light front eigenenergy $(i\partial^-\equiv k^-)$
is the sum of a kinetic energy term in which the mass is shifted by the
presence of the scalar field, and an energy arising from the vector field.
Comparing 
this equation with the one for free nucleons shows that   the nucleons 
have a  mass $M+g_s\phi$ and  move
in plane wave states. The nucleon 
field operator is constructed using the solutions of 
Eq.~(\ref{sd}) as the plane wave basis states. This means that 
the nuclear matter ground state, defined by operators that create and 
destroy baryons in eigenstates of Eq.~(\ref{sd}), is the correct 
wave function and that Equations~ (\ref{sphi}), (\ref{sv}) and (\ref{sd})
represent the solution of the approximate
field equations, and the diagonalization of the  Hamiltonian.

\subsection{ Nuclear Momentum Content}
The expectation value of $T^{+\mu}$ is used to obtain the total momentum:
\be P^\mu=
{1\over 2}\int d^2x_\perp dx^-
\langle T^{+\mu} \rangle. \ee The expectation value is constant so that
the volume $\Omega={1\over 2}\int d^2x_\perp dx^- $ will enter as a factor.
A straightforward evaluation leads to the results
\bea
{P^-\over\Omega}&=&m_s^2\phi^2+2\left({\kappa\over 3!}\phi^3 +{\lambda\over
  4!}\phi^4\right)+ {4\over
(2\pi)^3}\int_F d^2k_\perp dk^+\;{k_\perp^2+(M+g_s\phi)^2\over k^+}\\
{P^+\over\Omega}&=&m_v^2(V^-)^2+{4\over
(2\pi)^3}\int_F d^2k_\perp dk^+\;k^+.\label{pplus}\eea

To proceed further one needs to define the Fermi surface $F$. The use of a
transformation
\be  k^+\equiv \sqrt{(M+g_s\phi)^2+\vec{k}^2} +k^3\equiv E(k)+k^3\label{dkp}\ee
to define a new variable
$ k^3$ enables one to simplify the integrals. One replaces the integral over
$k^+$ by one over $k^3$, including the Jacobian factor \be{\partial k^+\over
  \partial k^3}={k^+\over E}.\ee 
Then one computes 
the nuclear energy $E$ as the average of $P^+$ and $P^-$:
$
E\equiv{1\over 2}\left(P^-+P^+\right)$. The results are 
the very same expressions as in the original Walecka models. This provides a
useful check on the algebra, which is important because this model has been
solved in a manifestly covariant fashion.

There is a potential problem:
for nuclear matter in its rest frame we need to have $P^+=P^-=M_A$. If one
looks at the expressions for $P^\pm$ this result does not seem likely. 
However, the value of the Fermi momentum has not yet been determined. There is
one more condition to be satisfied:
\be\left({\partial (E/A)\over\partial k_F}\right)_\Omega=0.\ee
Satisfying this equation determines $k_F$ and for the value so obtained the
values of $P^+$ and $P^-$ turn are the same.

Thus we see that
our light front procedure reproduces standard results for energy and  density.
We discuss two sets of results. The first involves the original Walecka model,
in which $\lambda=\kappa=0$. Then
we may use the parameters of Chin and Walecka\cite{cw} $g_v^2M^2/m_v^2=195.9$ and
$g_s^2M^2/m_s^2=267.1$ to obtain first numerical results. In this case, 
$k_F=1.42  \quad$fm$^{-1}$, the binding energy per nucleon is 15.75 MeV and
$M+g_s\phi=0.56
M$. The last number
corresponds to a huge attraction that is nearly canceled by the
huge repulsion. Then one may use Eq.~(\ref{pplus}) to obtain the separate
contributions of the vector mesons and nucleons, with spectacular  results.
The use of Eq.~(\ref{pplus}) leads immediately to the unusual result that 
nucleons carry only 65\% of the plus-momentum. Thus is much less than
 the  90\% needed to explain the EMC effect for infinite nuclear
 matter\cite{sdm}. According to Eq.~(\ref{pplus}) 
the vector mesons must
carry the remaining 35\% of the plus-momentum, an amazingly large
number. 

It is instructive to evaluate the vector $U_V\equiv g_V\;V^-$ and scalar $U_S$
potentials that the nucleon
feels with the present parameters of QHD1. These are given by 
\bea
U_V={g_V^2\over m_V^2}\rho_B=330\; \rm{MeV}\\
U_S=-{g_S^2\over m_S^2}\rho_S=-420\; \rm{MeV}
.\eea
These very large potentials, obtained with $m_V=783 $ MeV and $m_S=550$ MeV,
are the distinctive features of the Walecka
model\cite{bsjdw}. The vector and scalar meson fields are given according to
Eqs.~(\ref{sphi}) and (\ref{sv}) as
\bea
V^0=U_V/g_V=28.3 \;\rm{MeV}\\
\phi=-43.9\;\rm{MeV}.\eea
  
\begin{figure}
\unitlength1cm
\begin{picture}(10,8)(-1,2.7)
\includegraphics{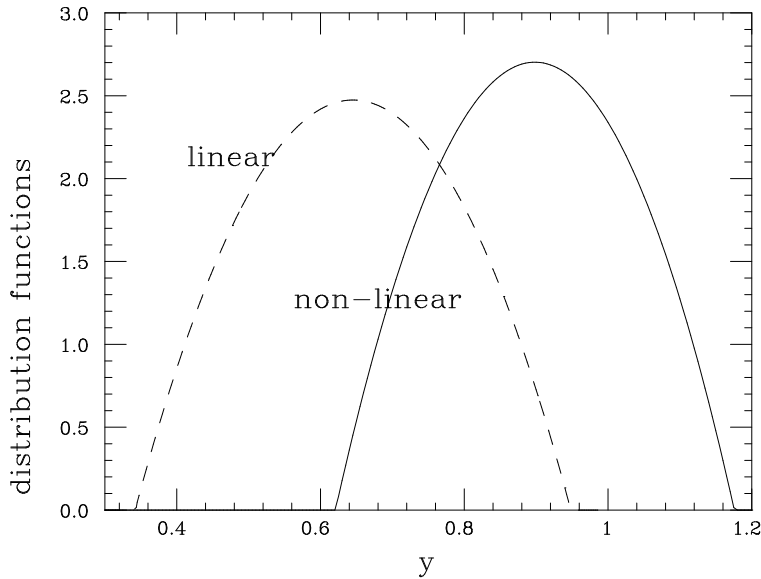}
\end{picture}
\caption{Light front momentum distribution as a function
  of $y$ for linear and non-linear models. 
 }
\label{fig:fnynm} 
\end{figure}

The nucleonic momentum distribution  
is the input to calculations of the nuclear structure functions.
This distribution function
can be 
computed from the integrand of Eq.(\ref{pplus}). The probability that
a nucleon has plus momentum $k^+$ is determined from the condition that
the plus momentum carried by  nucleons, $P^+_N$, is given by 
$P^+_N/A=\int
dk^+\;k^+ f(k^+)$, where $A=\rho_B\Omega$.
 It is convenient to 
use the  dimensionless variable $y\equiv {k^+\over \bar{M}}$
with $\bar{M}=M-15.75 $ MeV. Then we find the result: 
\begin{equation}
f(y)={3\over 4} {\bar{M}^3\over k_F^3}\theta(y^+-y)\theta(y-y^-)\left[
{k_f^2\over \bar{M}^2}-({E_f\over \bar{M}}-y)^2\right],\label{ndist}
\end{equation}
where
$y^\pm\equiv {E_F\pm k_F\over \bar{M}}$ and
$E_F\equiv\sqrt{k_F^2+(M+g_s\phi)^2}$. This function is displayed in
Fig.~\ref{fig:fnynm}. 
The average value of $y$,$\langle y\rangle$
can be computed from this distribution:
\be
\langle y\rangle=\int_0^\infty dy\; yf(y)= {E_F\over \bar{M}}.\label{ybar}\ee

The  relation to experiments is obtained by
 recalling that the nuclear structure function $F_{2A}$ 
can be obtained from the light front distribution function
$f(y)$ (which gives the probability for a nucleon to have 
a plus momentum fraction $y$) and the nucleon structure function
$F_{2N}$ using  the relation:
\begin{equation}
{F_{2A}(x)\over A}=\int dy f(y) F_{2N}(x/y), \label{deep}
\end{equation}
where $y$ is $A$ times the fraction of the 
nuclear  plus-momentum carried by the nucleon, and
$x$ is the Bjorken variable computed using the
nuclear mass minus the binding energy.
This formula  is the expression of the usual convolution model, with
validity determined by a number of assumptions.
Eq.~(\ref{deep}) is essentially Eq. (5.2) of 
Frankfurt and Strikman, \cite{fs2}, with a correspondence
between our $A\;f(y)$ 
and their $\int d^2p_\perp \rho^N_A(\alpha,p_\perp)/\alpha$. The use of the
light front
formalism enables us to calculate
the function $f(y)$ from the integrand of 
Eq.(\ref{pplus}).

The distribution of the 
vector meson plus-momentum is also an interesting quantity. The
mean fields $\phi, V^\mu$  are constants in space and time. Thus 
 $V^-$ has support only for $k^+=0$. The physical interpretation of this
 is that an infinite 
 number of mesons carry a vanishingly small $\epsilon$ of the
plus-momentum, but the product is  35\%. One can also show\cite{bm98} that
\be k^+f_v(k^+)=0.35 M \delta(k^+).\label{v35}\ee
There is an important phenomenological
consequence the value 
$k^+=0$ corresponds to $ x_{Bj}=0$
which can not be  reached in experiments. This means one can't use the
momentum sum rule as a phenomenological tool to analyze deep inelastic
scattering data to determine the different
contributions to the plus-momentum.

Of course these noteworthy  results are
caused by solving a simple model for a simple
system with a simple mean field approximation. It is necessary to ask if any
of the qualitative features of the present results will persist in more
detailed treatments.
A simple first step is to include the effects of scalar meson self-coupling.
We examine  the parameters of Ref.~\cite{fpw87}, and find one set which seems
consistent with deep inelastic scattering data. This is $\kappa=2500 $ MeV,
$\lambda=$200, $m_S=368.6$ MeV.
Here $k_F=1.3$ fm$^{-1}$ and the binding energy per nucleon is
again 15.75 MeV. The other parameters are $g_s^2=27.96,\;g_v^2=52.44$.
The value of $M+g_s\phi=0.84M$. This corresponds to nucleons carrying 90\% of
the
plus momentum, and
\be k^+f_v^{\rm non-linear}(k^+)=0.10 M \delta(k^+).\label{v10}\ee
which is roughly consistent with the data\cite{sdm}.
The nucleon momentum distribution is also shown in Fig.~\ref{fig:fnynm}. 
The vector 
and scalar 
fields  are now given by 
\bea
V^-={g_V\over m_V^2}\rho_B=97\; \rm{MeV}\\
\phi=-29  \rm{MeV}
.\eea

\section{light front static source  model for nuclear Mesons }

The previously discussed calculations of 
Ref.~\cite{jerry}, using a Lagrangian in which Dirac nucleons
are coupled to massive scalar and vector mesons\cite{bsjdw},
treated the example of infinite 
nuclear matter within the mean field approximation. In this case, 
the scalar and vector
meson fields are constants in both space and time, but only the vector mesons
carry a non-zero plus-momentum. The constant nature  means that the 
momentum distribution has support only at $k^+=0$. Such a distribution would
not be accessible experimentally, so that the suppression of the plus-momentum
of valence quarks would not imply the existence of a corresponding testable
enhancement of anti-quarks. However, 
it is necessary to ask if the result is only a artifact of the infinite nuclear
size and of the mean field approximation.

In section, I review  a previous investigation\cite{bm98} of
the dependence on nuclear size. The technique was to  
 construct  the scalar and vector meson states for
a model in which the nucleus is represented as a static source of radius
$R$ and 
mass $M_A$. In this case,  $M_A$ is very large
for large $A$, and the nucleus acts as an extended static 
source. In the mean field approximation, the nucleus 
acts as a static source if it is infinitely heavy. Then a coherent state is 
the ground state of the
Hamiltonian.

\subsection{Scalar  Meson Distribution}

We take 
the  scalar mesons, as coupled to a large static 
nucleus represented by a scalar  source $J({\vec r})$.
Such a system is described by the Lagrangian density
\begin{equation}
{\cal L}= \frac{1}{2}\partial_\mu \phi \partial^\mu \phi
-\frac{m^2_S}{2}\phi^2 + J \phi .
\label{eq:lager}
\end{equation} 
To be 
specific,  consider a spherical nucleus
of radius $R$ with constant density, so that 
\begin{equation}
J({\vec r}) = J_0 \Theta (R-|{\vec r}|).
\label{eq:j}
\end{equation}
The quantity $J_0$ can be thought of as arising from the product
of a scalar meson coupling constant $g_S$ and an appropriate scalar density
$\rho_S$ with 
\begin{equation}
J_0=g_S\rho_S.
\end{equation}
This is a static source in the rest frame, expressed in the usual
$(t,\bbox{r})$
coordinates.

The essential observation for the LF formulation of systems
coupled to static sources is that static sources in a rest-frame
correspond to uniformly 
moving 
sources in a LF framework \cite{zako}.
In particular, 
a static charge distribution in a rest-frame
$J_{RF}({\vec x}_\perp,x^3)$ corresponds to a uniformly
moving charge distribution on the LF, where at $x^+=0$
\begin{equation}
J_{LF}({\vec x}_\perp,x^-)= J_{RF}({\vec x}_\perp,z=-x^-/2) .
\label{eq:jj}
\end{equation}

Using the usual technique to construct $T^{\mu\nu}$, one finds
\begin{eqnarray}
P^-&=&\sum_{k^+,{\vec k}_\perp}
\left[ \left(\frac{m_S^2+{\vec k}_\perp^2}{k^+}
+{k^+}\right) a_{k^+{\vec k}_\perp}^\dagger
a_{k^+{\vec k}_\perp}\right. 
-2 \left.\frac{1}{ \sqrt{2 k^+}}\left(\tilde{J}_{RF}(
{\vec k}_\perp,k^+)a_{k^+{\vec k}_\perp}^\dagger
+ h.c. \right)\right] ,
\label{eq:hlf}
\end{eqnarray}
where 
\bea \tilde{J}_{RF}({\vec k}_\perp,k^+)
&=&{1\over (2\pi)^{3/2}}\int \;d^2x_\perp\;{dx^-\over
2}\;
e^{-i\bbox{k_\perp}\cdot \bbox{x_\perp}} e^{{ik^+\;x^-\over2}}\;
g_S\rho_S\Theta(R-\sqrt{x_\perp^2+(x^-/2)^2}\;),\\
&=& {1\over (2\pi)^{3/2}} {4\pi\;R^3\over 3}{3j_1(kR)\over k R},
\eea    with
    \be k\equiv \sqrt{k_\perp^2+{k^+}^2}.\ee
The interaction depends only on a combination of momenta that corresponds to the
magnitude of a vector $\bbox{k}$ with a positive $z$ component, $k^+$. 
The form factor ${3j_1(kR)\over k R}$ can be well approximated\cite{cbm} by the
expression $e^{-k^2R^2/10}$ so that the typical momenta are bounded by
$\sqrt{10}/R$.    The term proportional to $k^+$  arises from the plus momentum
of the heavy source and its inclusion is necessary to  maintain the
rotational invariance of the light-front approach,  
see Sect.~III. 
 
The LF Hamiltonian $P^-$  [Eq. (\ref{eq:hlf})] 
is quadratic  in the scalar fields and its ground state  is a coherent state
\begin{equation}
|\psi_0 \rangle_{LF} \propto \left[ 
\prod_{k^+,{\vec k}_\perp}
\exp \left( \frac{2\tilde{J}_{RF}^*({\vec k}_\perp,k^+)
a_{k^+{\vec k}_\perp}^\dagger }
{\sqrt{2k^+}\left(\frac{m^2+{\vec k}_\perp^2}{k^+}+
 k^+\right) 
}\right)
\right]|0\rangle.
\label{eq:psi0lf}
\end{equation}
    The action of Eq. (\ref{eq:hlf})  on $|\psi_0 \rangle_{LF} $ yields
    again the
same
state with eigenvalue
\begin{eqnarray}
P^-_0= -2\int_0^
\infty \frac{dk^+}{k^+}\int d^2k_\perp
\frac{|\tilde{J}_{RF}({\vec k}_\perp,k^+)|^2}
{\frac{m_S^2+{\vec k}_\perp^2}{k^+}+k^+ }.
\label{eq:e0lf}
\end{eqnarray}
We  replace the integral $\int_0^\infty dk^+$ by ${1\over
  2}\int_{-\infty}^\infty dk_z$ so that
\begin{eqnarray}
P^-_0= -(g_S\rho_S)^2 \int {d^3k\over (2\pi)^3}
\;\left({4\pi\;R^3\over 3}{3j_1(kR)\over k R}\right)^2
{1\over m_S^2+k^2}.
\end{eqnarray}
Integration of the above yields the result that
\begin{eqnarray}
{\rm lim}\;_{R\to\infty}\;\;P^-_0=-(g_S\rho_S)^2 {4\pi\over 3}{R^3\over m_S^2}
.\end{eqnarray}

The LF-momentum distribution for the scalar mesons
\begin{equation}
\rho_S({\vec k}_\perp,k^+)
 =\langle\psi_0\mid 
a_{k^+{\vec k}_\perp}^\dagger
a_{k^+{\vec k}_\perp}\mid \psi_0\rangle
\end{equation}
can be calculated using 
Eq.~(\ref{eq:psi0lf}), with the result
\begin{equation}
\rho_S ({\vec k}_\perp,k^+)
= \frac{2k^+\left|\tilde{J}_{RF}({\vec k}_\perp,k^+)\right|^2}
{\left[m_S^2+k^2 
\right]^2 
} .
\label{eq:lfmom}
\end{equation}
Since $\tilde{J}_{RF}({\vec k}_\perp,k^+)$ is strongly
peaked for $k \sim 1/R$, the momentum distribution is also
trivially peaked near small momenta (for $R \gg 1/m_S$). 
However, $\rho_S$ vanishes at $k^+=0$, so that it necessarily is very small.
Note that the  light frame momentum distribution function and
the related plus-momentum distribution can only be obtained using the light front
formulation.

The plus-momentum carried
by the meson field is of great interest here. We compute this using 
 Eq. (\ref{eq:lfmom}), to obtain 
\begin{eqnarray}
\langle k^+ \rangle 
&=& \frac{1}{3}\int d^3k 
\frac{{\vec k}^2\left|\tilde{J}_{RF}({\vec k})\right|^2}
{\left[m_S^2+{\vec k}^2 
\right]^2 } .
\label{eq:kplus}
\end{eqnarray}
where we used rotational invariance of the source in the rest-frame.
It is interesting to express the quantity $\langle k^+ \rangle $
in terms of a coordinate space integral. We find:
\begin{equation}
\langle k^+ \rangle  = 
\int\!\!\! d^3rd^3r' J(\vec {r})\!\!\left[\left(
1-{m_S\over2}|\vec{r}-\vec{r}\;'|\right){e^{-m_S|\vec{r}-\vec{r}\;'|}
\over 12\pi |\vec{r}-\vec{r}\;'|}\right]\!\! J(\vec {r}\;').
\label{eq:kpr}
\end{equation}
The quantity in brackets has a volume integral of 0 (with the
integration variable as $|\vec{r}-\vec{r}\;'|$). 
Thus the integral receives non-vanishing contributions only from 
regions near the nuclear surface, as is expected from the notion
that 
 the scalar meson field would be constant for a nucleus of infinite
 size
and so  receives nonzero Fourier components only from
the regions near the surface of the nucleus.
This means that $\langle k^+ \rangle  \propto R^2$ which gives a far
smaller
magnitude than the $R^3$ behavior of the binding energy. In
particular,
\be {\langle k^+ \rangle \over  P^-_0}\sim
{1\over R},\ee
which vanishes in the limit
${R\to\infty},$
 in accord with Ref.~\cite{jerry} which shows that
$\langle k^+ \rangle$ vanishes for the case of infinite nuclear
matter. 

A detailed investigation of the scalar meson momentum distribution,$f_S(x)$,
probability
that
a scalar meson 
carries a momentum
fraction
\begin{equation} x\equiv {k^+\over M_N} \label{xdef}\end{equation}
is given in Ref.~\cite{bm98}.

\subsection{Vector Meson Distribution} 
The calculation of vector meson distributions is based on the
formalism in Ref. \cite{jerry}, which used the light  front quantization
procedure
of Ref.~\cite{des71,yan34}. 
The model we consider is defined by  taking the vector mesons
$V^\mu$ to be coupled to a large nuclear source of baryon current
$J^\mu$. Thus the relevant Lagrangian density ${\cal L}_V$ is given by
\begin{eqnarray}
{\cal L}_V =
-{1\over  4} V^{\mu\nu}V_{\mu\nu} +{m_V^2\over 2}V^\mu V_\mu 
-J^\mu \bar{V}_\mu\label{lagv}
\end{eqnarray}
where $
\bar{V}^\mu$ of Eq.~(\ref{vbar}) enters in the interaction term,
with
$\bar{V}^+=0$.

Note that 
\begin{equation}
\tilde{J}^\mu_{LF} ({\vec k}_\perp, k^+) = \delta(\mu,0)
\tilde{J}_{RF}^V({\vec k}_\perp, k^+),
\label{eq:jplus}
\end{equation}
where 
\begin{equation}
J_{RF}^V=g_V\rho_B \Theta(R-r),
\end{equation}
because there is no special direction in space. The nucleus is in its rest frame.

The main difference 
between vector mesons and scalar mesons is the appearance
of the polarization vector 
${\epsilon}^\mu$ in the coupling of vector
mesons to the nucleon current (\ref{nvfield}).
The  transverse components of the nucleon current vanish in
our model, and  $\bar\epsilon^+=0$, so that the only non-zero 
component of $\bar\epsilon^\mu$ is $\mu=-$. 

The coherent state  takes the form
\begin{equation}
|\psi_0^V \rangle \propto \left[ 
\prod_{k^+,{\vec k}_\perp}
\exp \left( \frac{2\tilde{J}_{RF}^\mu({\vec k}_\perp,k^+)^*
\bar{\epsilon}_\mu
a(\bbox{k},\omega)}
{\sqrt{2k^+}\left(\frac{m^2_V+{\vec k}_\perp^2}{k^+}+
{k^+}\right) 
}\right)
\right]|0\rangle,
\label{eq:psi0lfV}
\end{equation}
where
\begin{eqnarray}
\tilde{J}_{RF}^V( {\vec k}_\perp,k^+)=
g_V\rho_B\sqrt{2\over\pi}R^3{j_1(kR)\over kR},
\end{eqnarray}
with 
$ 
k\equiv\sqrt{\vec {k}_\perp^2 +{k^+}^2}.
$
The expressions for the energy and momentum distribution are obtained by
taking expectation values using this wave function.
 The  expressions obtained are similar to those for the
scalar mesons except that a factor of 
$(\bar{\epsilon}^-)^2 $ summed over all polarization
states is present. 
The polarization sum is given by Eq.~(\ref{p2}), with $\mu=-$, and
$k^-={k_\perp^2+m_V^2\over k^+}$, so that 
\begin{equation}
\sum_{\omega=1,3} \bar\epsilon^-({\vec k},\omega)\bar\epsilon^-
({\vec k},\omega) =4 \frac{m_V^2+{\vec k}_\perp^2}{{k^+}^2}.
\end{equation}
We shall see that
the denominator factor ${k^+}^2$ plays a large role in causing the vector meson
momentum distribution function to be large for small values of $k^+$.

The light front Hamiltonian $P^-$ consists of three terms: the 
kinetic energy (Eq. (2.24) of Ref.~\cite{jerry}); the linear coupling and, the
effects of the instantaneous vector meson exchange. Thus  the light-front
energy of the 
nucleus due to coupling to the vector meson field consists of
 a contribution which arises from physical vector
meson intermediate 
states, and  an instantaneous 
interaction  which exactly cancels the most infrared
singular  terms of the term due to dynamical mesons. 
This term is obtained by canonical light front quantization and e.g.
is
included in Eq. (2.48) of Ref.~\cite{jerry}.
One takes the matrix element of the light front Hamiltonian to 
obtain the ground state energy of the vector meson field
coupled to the fixed source:

\begin{equation}
P^-_0(V)= 2\int_0^\infty \!\frac{dk^+}{k^+}
\int \!\!d^2k_\perp
\frac{\left|\tilde{J}_{RF}^V({\vec k}_\perp,k^+)\right|^2}
{\frac{m_V^2+k^2}{k^+}+k^+}, 
\label{eq:p0lf}
\end{equation}
which is  related to the vector meson contribution to the binding
energy $P^-_0$,  which except for the  sign (reflecting the fact that
scalar meson give rise to attraction, while vector mesons give
rise to repulsion between nucleons) is 
of the same form as the scalar result Eq. (\ref{eq:e0lf}).

The result of taking matrix elements of the plus-momentum operator in the 
coherent state leads to the result:
\begin{eqnarray}
\rho_V({\vec k}_\perp,k^+) 
&=& \left| \tilde{J}_{RF}^V \right|^2 \left\{
\frac{2}{k^+} \frac{1}{m^2_V+k^2 } 
\right.
\left.-
\frac{2k^+}{\left[
m^2_V+ + k^2\right]^2}
\right\} .
\label{eq:rhov}
\end{eqnarray}

The second term on the right hand side  of Eq. (\ref{eq:rhov}) is (except for
the  sign and a
differing mass) 
identical to the momentum distribution of scalar mesons. 
 The contribution of this second
term to the momentum (per nucleon)
carried by the vector mesons vanishes in the nuclear matter limit and 
explicit numerical evaluation shows that for finite nuclei
it is negligible compared with the
the first term.
This singular term
\begin{equation}
\rho_V^{\rm sing}({\vec k}_\perp,k^+) \equiv
\frac{2\left| \tilde{J}_{RF}^V \right|^2 }{k^+} 
\frac{1}{m_V^2+k^2} 
\label{sing}
\end{equation}
is more interesting since it diverges as $k^+ \rightarrow 0$.
By direct comparison one can verify that the contribution
from this term to the total momentum carried by the vector mesons
is identical to the contribution 
of the vector mesons to the rest-frame energy of the nucleus
\begin{eqnarray}
\int_0^\infty \!\!\!dk^+ \!\!\!\int \!d^2k_\perp \rho_V^{\rm sing} k^+
&=& \int_0^\infty \!\!\!dk^+ \!\!\int \!d^2k_\perp 
\frac{2\left| \tilde{J}_{RF} \right|^2  
}{m_V^2+k^2} 
= P^-_0(V) .
\end{eqnarray} 
This means that  in the infinite nuclear
matter limit the momentum carried by the vector mesons is given by
\begin{equation}
\langle k^+_V \rangle = P_0^-( V ). 
\end{equation}
 This result 
that the momentum carried by vector mesons equals the $P^-$    due to vector 
mesons holds regardless whether or not there is also a scalar interaction
present. Both scalar and vector interaction are rather large in
nuclei (of opposite sign, such that their net effect is small), so
vector mesons may carry a substantial fraction of the nucleus'
momentum. In the previous section, we have shown that scalar mesons
carry only a small fraction so the net plus-momentum carried by the mesons
is essentially the potentially very large
plus-momentum carried by the vector mesons.

The  result that vector mesons carry a large part of the nuclear
plus-momentum, which was obtained  in Ref. \cite{jerry}
using more general arguments,
is at first surprising since the vector meson field (very much
like the scalar meson field) is constant in space and time for nuclear
matter in the mean field approximation. Therefore, one would expect
that the vector meson field
for nuclear matter contains only quanta with vanishing $+$ momentum.
It thus seems paradoxical that vector mesons nevertheless
carry a finite fraction of the nucleus' $+$ momentum in this limit.
To resolve this apparent paradox, we study the momentum distribution
arising from the crucial singular piece
$\rho_V^{\rm sing}$ of Eq.~(\ref{sing}).

A $+$-momentum distribution function per nucleon
$(A={4\pi\over 3}R^3\rho_B)$)is defined  using the variable of
Eq.~(\ref{xdef}). Then, the use of Eq.~(\ref{sing}) leads to the result 
\begin{equation}
f_V(x)\equiv {3M_N\over 4\pi R^3\rho_B}
\int d^2k_\perp \rho_V^{\rm sing}({\vec k}_\perp,k^+),
\end{equation} or
\begin{equation}
xf_V(x)=R^3{6\over \pi}{g_V^2\rho_B}
 \int_{xM_NR}^\infty {dy\over y}  {1\over y^2
+m^2_V R^2}j_1^2(y). \label{exact}
\end{equation}
A qualitative understanding may be gained by noting that the quantity 
$j^2_1(y)/y$ peaks at $y\approx1.6$, and that $m_V\;R$ and
$M_N\;R$ are   very large numbers. Thus, in the limit $R\to\infty$
the integral  vanishes unless
$x=0$.  In the limit of infinite  $R$
there is a sharp                     
distinction between the results for $x=0$ and for non-zero values,
no matter how small. If $x=0$, the integral can be done. The net result is that 
\begin{equation}
\lim_{R\to\infty} xf_V(x) = ({g_V\over m_V})^2{\rho_B\over M_N} \delta(x), 
\label{delta}
\end{equation}
in accord with the 
result expected from earlier work.
The present meaning of the function $\delta(x)$ is that integrals over $x$
including this delta function are non-vanishing provided the lower limit is
infinitesimally close to zero.
We follow Ref.~\cite{jerry} and again use the parameters of 
of Chin and Walecka\cite{cw}. Then the use of  Eq.~(\ref{delta}) leads to 
 $\langle x_V \rangle =0.348,$
which agrees with the result of Ref.~\cite{jerry}.

\begin{figure}
\begin{Large}
\unitlength1.cm
\begin{picture}(15,6.3)(1.4,-8.0)
\includegraphics{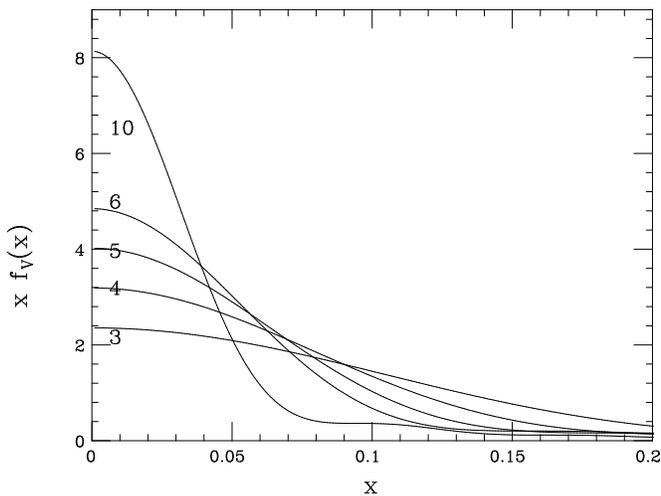}
\end{picture}
\end{Large}
\caption{The quantity $x f_V(x)$ vs. $x$ 
 for nuclei of radii 3,4,5,6, 10   fm. }
\label{fig:vectorf}
\end{figure}

The distribution function $f_V(x)$ of 
Eq.(\ref{exact}) for finite values of $R$ is  
shown in Fig.~(\ref{fig:vectorf}).
These  results 
 show the typical behavior of distributions that approach 
delta functions.  The approach to the delta function limit is slow. All of the
results show a significant spread and there is  support for 
values of $x$ such that $x$ greater than about 0.1 for $0<R<6$ fm.
The key physics  question  to be addressed is whether or not the 
distribution is non-zero for values of $x$ that are too small to be
observable.  This is discussed in Sect.~11.

The scalar and vector meson states
 for a mean field
model of large nuclei have been constructed using coherent states,
and  the 
meson distribution functions have been obtained.
The light-front
momentum distribution for scalar mesons in the mean field approximation
is localized near $k^+\rightarrow 0$, and  the total
momentum carried by the scalar mesons in the mean field approximation
is vanishingly small for nuclear matter.
This calculation thus confirms the results of Ref.~\cite{jerry}:
even though scalar mesons contribute to the nuclear binding in nuclear
matter, they carry only a vanishing fraction of the momentum in
the mean field calculation. 

Also in accord with Ref.~\cite{jerry} is the present result that
vector mesons do contribute to the plus-momentum of
the nucleus in the same limit of infinite nuclear matter. 
The momentum fraction carried by vector mesons in mean field approximation
and in the nuclear matter limit is given by the ratio between the 
the vector meson contribution  to the potential energy and
the nuclear mass.
This result is obtained even though the
vector meson distribution functions are zero for
non-zero $x$ for infinitely large nuclei. 
This is because   $xf_V(x)\propto \delta(x)$.

For nuclear radii $R$ corresponding to realistic large nuclei,
the vector meson distributions, while strongly peaked at low values of $x$,
are wide enough so that some fraction of the vector mesons could be observable.
Thus the limitation of the support to $k^+=0$ found in  Ref.~\cite{jerry}, does
not occur. However, the
predictive power of the momentum sum rule is  still vitiated because
a significant fraction of the mesons are hidden at small values of $x_{Bj}$.

\section{Mean Field Theory for finite-sized nuclei}
It is important to make calculations for finite nuclei because all laboratory
experiments are
done for such targets or projectiles. The most remarkable  feature of
all of nuclear physics is that the shell model is able to explain the magic
numbers. Rotational invariance causes the $2j+1$
degeneracy of the single particle orbitals, and full occupation leads to
increased  binding. But light front dynamics does not
make rotational invariance
manifest because the different components $x^-,\bbox{x}_\perp$
of the spatial variable are
treated differently. There is only an explicit  invariance for rotations of the
$xy$ plane about the $z$ axis. However, any set of correct  final results must
respect rotational invariance. The challenge of making  successful
calculations of the properties of finite nuclei is important to us.

In the following we review the work of Ref.~\cite{bbm99}
Let's start with a schematic  
 discussion of  how it is that we will be 
able to find spectra which have the correct number of degenerate
states.
The concern here is  with systems with a large number of nucleons, so that applying
the usual light
front Fock state expansion  is too difficult. Instead, a variational
calculation, a light front version of Hartree-Fock theory, is needed.   
The light front 
Hamiltonian, is $P^-$,  but  minimizing
the expectation value of that operator
would lead to nonsensical results because $P^-=M_A^2/P^+$.
One can  reach
zero energy by letting $P^+$ be infinite.
This difficulty is avoided by  
 performing a constrained variation, in which
the total light front  momentum is fixed by including a Lagrange multiplier term
proportional to the total momentum in the light front  Hamiltonian.
We minimize the expectation value of $P^+$ subject to the condition that the
expectation values of $P^-$ and $P^+$ are equal. This is
the same as minimizing
the expectation value of the average of $P^-$ and $P^+$.

 We use a simple example to demonstrate the need to
include the plus-momentum along with the minus momentum.
Consider a nucleus of $A$ nucleons of momentum $P_A^+=M_A$,
${\bbox{P}_A}_\perp=0$, which consists of a nucleon of momentum
$(p^+,\bbox{p}_\perp)$, and a residual $(A-1)$ nucleon system which
must have momentum $(P^+_A-p^+,-\bbox{p}_\perp)$. The kinetic energy
$K$ is given by the expression
\begin{equation}
K={p_\perp^2+M^2\over p^+}+{p_\perp^2+M_{A-1}^2\over P^+_A- p^+}.
\end{equation}
In the second expression, one is tempted to neglect the term $p^+$ in
comparison with $ P^+_A\approx M_A$. This would be a mistake. Instead
expand
\begin{eqnarray}
K&\approx&{p_\perp^2+M^2\over p^+}+{M_{A-1}^2\over P^+_A}\left(1+ {p^+\over
 P_A^+}\right)\nonumber\\ 
&\approx&{p_\perp^2+M^2\over p^+}+p^+ +M_{A-1}, 
\end{eqnarray}
using  $M^2_{A-1}/P_A^2\approx 1$ for large $A$.
The second term is the plus-momentum operator  mentioned above.
For free particles,
of ordinary three momentum $\bbox{p}$ one has $E^2(p)=\bbox{p}^2+m^2$
and $p^+=E(p)+p^3$, so that
\begin{equation}
K\approx {\left(E^2(p)-(p^3)^2\right)\over E(p)+p^3}
+E(p)+p^3+M_{A-1}=2E(p)+M_{A-1},
\end{equation}
We see that $K$ depends only on the magnitude of a three-momentum and
rotational invariance is restored. The physical mechanism of this
restoration is the inclusion of the recoil kinetic energy of the
residual nucleus.

\subsection{ Nucleon Mode Equation}

The key feature of 
 Ref.~\cite{bbm99} was the derivation of the equation that defines the nucleon
 single-particle modes. This was done by minimizing the value of
 $P^-$ subject to the constraint that the expectation values of $P^+$ and
 $P^-$ were equal. These equations are given in Eq.~(4) of
 the short paper, and (3.28) of the long paper. Here we present a more
 intuitive derivation\cite{toy} of the nucleon single particle wave equation.
The first step  is to
develop the light front  
formalism for potentials that are static in the nuclear  rest frame. Let us
start from the Dirac equation for a static potential in the usual $(t,
\bbox{r})$  (ET)  coordinates
\begin{equation}
\left[i\gamma^\mu\partial_\mu - m - g_SV_S({\vec r}) - g_V
\gamma^0V_0 ({\vec r})  \right] \psi^{ET} =0.
\label{eq:dirac_et}
\end{equation}
Since $\gamma^0= (\gamma^++\gamma^-)/{2}$ and since couplings
using  the ``bad'' component $\gamma^-$ are difficult to handle in
the light front  framework, we perform the transformation 
\begin{equation}
\psi^{ET} = e^{ig_V \Gamma} {\psi}^\prime,\label{tg}
\end{equation}
where
\begin{equation}
\partial_3 \Gamma = V_0 ,
\end{equation}
and   $\Gamma$ does not depend on time.
Using the transformation (\ref{tg}),
one  finds
\bea
0 &=& \left[i\gamma^\mu\partial_\mu
+i{\vec \gamma}_\perp\cdot ( {\vec \partial}_\perp\Gamma)
  - m - g_SV_S({\vec r}) - g_V
\left(\gamma^0+\gamma^3\right) V_0 ({\vec r})  \right]{\psi}^\prime ,
\label{psip}\eea
which may be expressed in terms of 
light front  coordinates ($\vec{x}_\perp=\vec{r}_\perp$)  as 
\begin{eqnarray}
& &\left[{1\over2}\left(i\gamma^+\partial^- + i\gamma^-\partial^+\right)
+i{\vec \gamma}_\perp\cdot\left( {\vec \partial}_\perp
  + g_v {\vec \partial}_\perp
  \Gamma({\vec x}_\perp, \frac{x^+-x^-}{{2}}) \right)- m \right.
\label{eq:dirac_lf1}
\\
& &\quad \left.- 
V_S({\vec x}_\perp, \frac{x^+-x^-}{{2}} )- 
\gamma^+V_0 ({\vec x}_\perp, \frac{x^+-x^-}{{2}} )
\right] {\psi}^\prime =0. \nonumber\end{eqnarray}

Even though the potential is static in the equal time formulation,
the Dirac equation for the same potential in light-front coordinates
is light front -``time'', i.e. $x^+$, dependent.
This is because  a static source in a 
rest-frame corresponds to a uniformly moving source on the 
light-front. Given that the time dependence of the external fields is
only due to a uniform translation, we transform 
Eq. (\ref{eq:dirac_lf1}) into
a form which contains only static (with respect to $x^+$)
potentials. For this purpose, we consider the equation of motion 
satisfied by Dirac fields which are obtained by an $x^+$ (light front -time) 
dependent translation 
\begin{eqnarray}
{\psi}^\prime  ({\vec x}_\perp,x^-,x^+)
&\equiv& e^{-ix^+P^+/2}{ \psi} ({\vec x}_\perp,
x^-,x^+) 
\label{eq:tilde1}
\end{eqnarray}

The techniques of Sect.~III, in particular
 Eq.~(\ref{trans}), are used to show that 
the equation of motion for $\psi$ takes the form:
\begin{eqnarray}
& &[{1\over 2}i\gamma^+(\partial^--\partial^+) +{1\over2}
  i\gamma^-\partial^+
+i{\vec \gamma}_\perp\cdot
({\vec \partial}_\perp+ig_v\Gamma({\vec x}_\perp,-\frac{x^-}{{2}}))-m 
\label{eq:dirac_lf2}
\\
& &\quad - 
g_SV_S({\vec x}_\perp, -\frac{x^-}{{2}} )- 
g_V\gamma^+V^- ({\vec x}_\perp, -\frac{x^-}{2} )
] {\psi} =0, \nonumber\end{eqnarray}
with $V^-=V_0$.
The translated fields satisfy an equation of motion 
with  potentials that do not depend on $x^+$. Moreover, the 
static potentials evaluated at ${\vec r}$ correspond to light front potentials
evaluated at $({\vec x}_\perp,-\frac{x^-}{{2}})$.
An even  simpler
derivation of this can be obtained from evaluating $z=(x^+-x^-)/2$
at $x^+=0$ which says   $z=-x^-/2$.

That the  result (\ref{eq:dirac_lf2}) is  the same as the equations for
$\psi_\pm$ in Ref.~\cite{bbm99}.  
 can be seen by making a decomposition  
 into a dynamical and a constraint equation.
Multiplication of Eq. (\ref{eq:dirac_lf2}) by  $\gamma^+$ from the left
yields
a constraint equation 
\begin{equation}
i \partial^+ \psi_-
=  \left[
i{\vec \alpha}_\perp\cdot ({\vec \partial}_\perp+ig_V
({\vec \partial}_\perp\Gamma))+\beta m  +V_S
\right] \psi_+ 
\label{eq:tconstr}
\end{equation}
where  $\psi_\pm \equiv \frac{1}{2} \gamma^0
\gamma^\pm \psi$.
Multiplication of Eq. (\ref{eq:dirac_lf2}) by  $\gamma^-$ from the left
yields
an equation for $\psi_+$:\begin{equation}
i (\partial^--\partial^+-ig_vV^-) \psi_+
=  \left[
i{\vec \alpha}_\perp\cdot ({\vec \partial}_\perp+ig_V
({\vec \partial}_\perp\Gamma))+\beta m  +V_S
\right] \psi_-.
\label{eq:dyn}
\end{equation}
One may use 
 the constraint equation (\ref{eq:tconstr}) to eliminate 
$\psi_{-}$ in Eq. (\ref{eq:dyn}) to obtain the  
 equation of motion for 
for the dynamical degrees of freedoms. The results (\ref{eq:tconstr})
and (\ref{eq:dyn}) are the desired equations. The difference between these
and the original fermion field equations (\ref{yan}) is the appearance of the
term $-i\partial^+$ which appears on the left side of the equation,

\subsection{Dynamical Meson Fields} 

The light-front Schroedinger equation for the complete nuclear
ground-state wave function $\mid\Psi\rangle$ is
\begin{equation}
 P^-\mid\Psi\rangle = M_A\mid\Psi\rangle.
\end{equation}
We choose to work in the nuclear rest frame so that we also need
\begin{equation}
 P^+\mid\Psi\rangle = M_A\mid\Psi\rangle.
\end{equation}
 As explained above, one must minimize the expectation value of $P^-$ subject
to the condition that the expectation value of $P^+$ is equal to the
expectation value of $P^-$. This is the same as minimizing the average
of $P^-$ and $P^+$, which is the rest-frame energy of the entire
system. To this end we define a light-front Hamiltonian
\begin{equation}
 H_{LF}\equiv {1\over 2}\left(P^++P^-\right).
\end{equation} 
We stress that $H_{LF}$ is not usual the Hamiltonian, because the
light-front quantization is used to define all of the operators that
enter. 

The wave function $\mid\Psi\rangle$ consists of a Slater determinant of
nucleon fields $\mid\Phi\rangle$ times a mesonic portion
\begin{equation}
\mid\Psi\rangle= \mid\Phi\rangle\otimes\mid \rm{mesons}\rangle,
\end{equation}
and the mean field approximation is characterized by the replacements
\begin{eqnarray}
\phi&\to& \langle\Psi\mid \phi\mid\Psi\rangle\nonumber\\
V^\mu&\to& \langle\Psi\mid V^\mu\mid\Psi\rangle. \label{replace1}
\end{eqnarray}

The essential difference of our approach and the procedure to handle the
original Walecka model  is that the fields are treated as dynamical objects.
 The  equations for the meson fields are derived  in terms of
 expectation values of
creation and destruction operators. This means that other expectation
values involving meson fields, such as the plus-momentum density, also
have non-zero values.

We detail 
 the treatment of  the expectation
value of $\phi(x)$. Meson self-coupling terms are ignored here.
Consider the quantity $H_{LF}
a(\bbox{k})\mid \Psi\rangle$, and use commutators to obtain
\begin{equation}
H_{LF}a(\bbox{k})\mid \Psi\rangle=[ H_{LF},a(\bbox{k}) ]\mid \Psi\rangle
+M_A a(\bbox{k})\mid \Psi\rangle.\label{ha}
\end{equation}
The operators $P_s^\pm$ of Eqs.~(\ref{pmph}) and (\ref{ppph})
and the standard commutation relations allow one to obtain
\begin{equation}
[ H_{LF},a(\bbox{k}) ]=-{k_\perp^2+{k^+}^2+m_s^2\over 2k^+}
a(\bbox{k}) +
{J(\bbox{k})\over (2\pi)^{3/2}\sqrt{2k^+}},
\end{equation}
where 
\begin{equation}
{J(\bbox{k})\over (2\pi)^{3/2}\sqrt{2k^+}}={1\over 2}
[P^-_I,a((\bbox{k}) ],
\label{jdef}\end{equation} 
and $P^-_I$ is given in Eq.~(\ref{defv}).

We use Eqs.~(\ref{v1})-(\ref{v3}), and take the commutator of the
interactions $v_i$ with $a(\bbox{k})$. Then re-express the results in
terms of $\xi$ to obtain
\begin{equation}
J(\bbox{k}) =-{1\over 2}g_s\int
d^2x_\perp dx^-\, e^{i\bbox{k}\cdot \bbox{x}}
\bar\xi(\bbox{x})\xi(\bbox{x}) . \label{jk}
\end{equation}
Take the overlap of Eq.~(\ref{ha}) with $\langle \Psi\mid$ to find
\begin{equation}
\langle \Psi\mid a(\bbox{k})\mid \Psi\rangle
{k_\perp^2+{k^+}^2+m_s^2\over 2k^+}
={ \langle \Psi\mid J(\bbox{k})\mid \Psi\rangle 
\over (2\pi)^{3/2}\sqrt{2k^+}}\;. \label{pap}
\end{equation}
Multiply  Eq.~(\ref{pap}) by a factor ${\sqrt{2k^+}\over
(2\pi)^{3/2}}e^{-i\bbox{k}\cdot \bbox{x}}$ and  add the result
to its complex conjugate. The integral of the resulting
equation over all $\bbox{k}_\perp$ and 
$k^+>0$ 
leads to the result
\begin{eqnarray}
\lefteqn{
\left(-\nabla_\perp^2-
\left(2{\partial \over \partial x^-}\right)^2+m_s^2 
\right)
\langle \Psi\mid \phi(x)\mid \Psi\rangle =}
\qquad\qquad&&\nonumber \\
&&\langle \Psi\mid 
\int {d^2k_\perp dk^+\theta(k^+)\over (2\pi)^3}\,\left(J(\bbox{k})
e^{-i\bbox{k}\cdot \bbox{x}} +J^\dagger(\bbox{k})
e^{+i\bbox{k}\cdot \bbox{x}}\right)\mid \Psi\rangle .
\label{ph1}
\end{eqnarray}

The evaluation of the right-hand-side of Eq.~(\ref{ph1}) proceeds by
using Eq.~(\ref{jk}) and its complex conjugate. The combination of
those two terms allows one to remove the factor $\theta(k^+)$ and
obtain a delta function from the momentum integral. That ${1\over2}k^+x^-$
appears in the exponential leads to the removal of the factor
${1\over2}$ of Eq.~(\ref{jk}). One can also change variables using
\begin{equation}
z\equiv{-x^-\over 2},\qquad \bbox{x}\equiv (z,\bbox{x}_\perp).
\label{relate}
\end{equation}
The minus sign enters to remove the minus sign between the two terms of
the factor $k\cdot x$.  Then one may use a simple definition 
\be -\nabla_\perp^2- \left(2{\partial \over \partial
x^-}\right)^2\equiv -\nabla^2.\ee
Note Ref.~\cite{bm98} obtained the  relation (\ref{relate})
 by examining the space-time diagram for a static 
(independent of $x^0$) source. The net result is that
\begin{equation}
\left(-\nabla^2+m_s^2 \right)
\langle \Psi\mid \phi(\bbox{x})
\mid \Psi\rangle =-g_s \langle \Psi\mid
\bar{\xi}
(\bbox{x})\xi(\bbox{x})\mid \Psi\rangle ,\label{phieq}
\end{equation}
which has the same form as the equation in the usual equal-time
formulation. Note that the right hand side of Eq.~(\ref{phieq}) should
be a function of $|\bbox{x}|$ for the spherical nuclei of our
present concern. Our formalism for the nucleon fields uses
$\bbox{x}_\perp$ and $x^-$ as independent variables, so that obtaining
numerically scalar and vector nucleon densities that depend only
$x_\perp^2+(x^-/2)^2$ will provide a central, vital test of our
procedures and mean field theory. This
does occur\cite{bbm99}, so the scalar field $\langle\Psi\mid
\phi(\bbox{x})\mid\Psi\rangle$  depends only $|\bbox{x}|$
according to (\ref{phieq}).

We stress that the use of Eq.~(\ref{relate}) is merely a convenient way
to simplify the calculation --- using it allows us to treat the $\perp$
and minus spatial variables on the same footing, and to maintain
explicit rotational invariance. We will obtain the mesonic
plus-momentum distributions from the ground state expectation value of
 operators expressed in terms of light front coordinates.

The procedure of Eqs.~(\ref{ha}) to (\ref{phieq}) can also be applied
to the vector fields,\cite{bbm99}. After some manipulations mandated
by the appearance of the barred vector potential, one finds 
\begin{equation}
\left(-\nabla^2+m_v^2\right)
\langle \Psi\mid \bar{V}^\mu (\bbox{x})
\mid \Psi\rangle =g_v \langle \Psi\mid
\bar\xi(\bbox{x})\gamma^\mu\xi(\bbox{x})\mid \Psi\rangle
.\label{veq}
\end{equation}

\subsection{Finite Nucleus Solutions,  Results and
  interpretation of the eigenvalues}
The main content of the nucleon mode equation is 
that it is the usual  Dirac equation expressed in terms of light front
variables.
It should not be a surprise that the mode equation of the ET theory 
turns out to be the same as that of the light front theory. In the former,
one minimizes the energy of the chosen single determinant wave function. In the
light front one minimizes $P^-$ subject to the constraint that the
expectation value of $P^+$ is the same as that of $P^-$. For the correct
wave function, each of $P^\pm$ is one half the mass or energy of the nucleus.

The nuclear wave function $\mid\Psi\rangle$ consists of a
Slater determinant of nucleon
fields $\mid\Phi\rangle$ times a mesonic portion,
and the mean field approximation is characterized by the replacements:
$
  \phi\to \langle\Psi\mid \phi\mid\Psi\rangle,
V^\mu\to \langle\Psi\mid V^\mu\mid\Psi\rangle. 
$
We  quantize the
nucleon   fields using
\begin{equation}
\psi(x) =\sum_n \langle x^-,\bbox{x_\perp}\mid n\rangle e^{-i p_n^-x^+/2}\;b_n\;,
\label{quant}\end{equation}
in which the variable $x$ represents both the spatial variables
$\bbox{x}_\perp,x^-$ and the spin, isospin indices, and $b_n$ obeys the usual
anti-commutation relations. The meson fields are also treated as
functions of these variables.
 The Slater
determinant $\mid\Phi\rangle$ is defined by allowing $A$ nucleon states to be
occupied.

The use of Eq.~(\ref{quant}) in Eqs.~
(\ref{eq:tconstr})
and (\ref{eq:dyn})
leads to
 \begin{eqnarray} 
 p_n^- \mid n\rangle_+&=&\left(i\partial^++
2 g_v\bar{V}^-\right)
 \mid n \rangle_+ +
 (\bbox{\alpha}_\perp\cdot 
(\bbox{p}_\perp-g_v\bbox{\bar V}_\perp)+\beta(M+g_s\phi))\mid n \rangle_-\;,
\label{nnminus}\\
 i\partial^+\mid n\rangle_-&=& \left(\bbox{\alpha}_\perp\cdot 
(\bbox{p}_\perp-g_v\bbox{\bar V}_\perp)+\beta(M+g_s\phi)\right)\mid n
\rangle_+\;,\label{nnplus}\\  \mid n\rangle_\pm&=&\Lambda_\pm\mid n\rangle
,  
\end{eqnarray}
which together with Eqs.(\ref{phieq}) and (\ref{veq}) forms the self-consistent
  set of equations to be solved. The first step of the solution procedure
 is to use a representation\cite{harizhang}
in which $\mid n\rangle_\pm$ are each represented as Pauli spinors.
The light front quantization respects 
manifest rotational invariance for rotations 
about the $z$-axis. Thus each single-particle state has a good $J_z$. 
We use a momentum representation for the longitudinal variable
in which the values of $p^+$ take on those
of a discrete set:
$p_m^+=(2m+1)\pi/(2L)$ where $m\ge0$ and $L$ is a quantization
length. 
This means that the nucleon  wave functions
have support only for  $p^+\ge0$. This spectrum condition is
a requirement for exact solutions for  any theory.
The coordinate space representation
is used for the $\perp$ variables.
Then we 
have 
\begin{eqnarray}
\langle p_m^+,\bbox{x}_\perp\mid n\rangle_+=\left[\begin{array}{c}
U_m^{(n)}(\bbox{x_\perp})e^{i(J_z-1/2)\phi}\\
L_m^{(n)}(\bbox{x_\perp})e^{i(J_z+1/2)\phi}\end{array}
\right],\label{spinor}
\end{eqnarray}
in which  the upper (lower) entrees of the Pauli spinor correspond to 
$m_s=1/2(-1/2)$, and $\bbox{x}_\perp\equiv (x_\perp,\phi)$.
When Eq.~(\ref{spinor}) is used in Eq.~(\ref{eq:dyn}),
one finds that the equations 
do not depend on the magnitude of $J_z$;  solutions for $\pm$ 
a given magnitude of $J_z$ are degenerate.
The functions $U_m(x_\perp)$ and $L_m(x_\perp)$ are expanded in a basis
of B-splines of degree five in $x_\perp$ \cite{bspline}.

The technical aspects of the solution procedure are detailed
in  Refs.~\cite{bbm99}, so I  concentrate on summarizing the results.
If our  solutions  are to have
any relevance,  they should respect rotational invariance. The
success in achieving this is examined in Table I 
  which gives
our results for the spectra of $^{16}$O and $^{40}$Ca. 
Scalar and vector meson parameters are taken from Horowitz and
Serot\cite{hs}, and we have ignored electromagnetic effects.
Note that the
$J_z=\pm1/2$ spectrum contains the eigenvalues of all states, since all
states must have a $J_z=\pm1/2 $ component. Furthermore,
the expected degeneracies among states with different
values of $J_z$ are reproduced numerically. This is an essential success of our
procedure.

The obtained eigenvalues of the nucleon mode equation   are essentially
the same as the single particle energies of the ET
formalism, to within the expected numerical accuracy of our program.
The origin of this remarkable feature is the  large value of the product of the
nucleon mass and the nuclear radius. This is discussed next. 

\begin{table}[t]

\rule{0in}{2ex}
\begin{center}
\begin{tabular}{ldddd}
\multicolumn{2}{c}{ET} & \multicolumn{3}{c}{LF} \\
\cline{1-2} \cline{3-5}
State $n$ & $p^-_n/2-M_N$ (MeV) & $J_z=\pm1/2$ & $J_z=\pm3/2$ & $J_z=\pm5/2$\\
\cline{1-1} \cline{2-2} \cline{3-3} \cline{4-4} \cline{5-5}
0s$_{1/2}$ & $-$55.40 & $-$55.39 & & \\
0p$_{3/2}$ & $-$38.90 & $-$38.90 & $-$38.90 & \\
0p$_{1/2}$ & $-$33.18 & $-$33.18 & & \\
0d$_{5/2}$ & $-$22.75 & $-$22.75 & $-$22.75 & $-$22.74 \\
1s$_{1/2}$ & $-$14.39 & $-$14.36 & & \\
0d$_{3/2}$ & $-$13.87 & $-$13.87 & $-$13.88 & \\
\end{tabular}
\end{center}
\label{table1}
\caption{Comparison of the single particle spectra of $^{40}$Ca in the
equal time (ET) formalism ($\epsilon_n-M_N$) with the light front (LF)
method ($p_n^-/2-M_N$).  }
\end{table}


We may understand the 
near equality of single particle eigenvalues using an analytic argument,
which is essentially the inverse of the one used to begin this section.
First we use a representation in which $V^\mu$ appears in the equation
for the nucleon field. That is, define
$\langle x\mid n\rangle'\equiv e^{-ig_v\Lambda(x)}\langle x\mid n\rangle$,
with $\partial^+\Lambda=V^0$.
Then multiply the equation (\ref{nnminus}) for $\mid n\rangle'_+$ by 
 $\gamma^+$ and the  equation  (\ref{nnplus})
 for $\mid n\rangle'_-$  by $\gamma^-$. Adding the resulting two equations
  gives
\begin{equation}
(\gamma^0(p^-_n-2\gamma^0g_vV^0
-\gamma^3(2p^+-p^-_n/2))\mid {n}\rangle'
=2(\bbox{\gamma}_\perp\cdot  \bbox{p}_\perp+M+g_s\phi  
)\mid {n}\rangle'.\label{almost}
\end{equation}
Solving Eq.~(\ref{almost}) in coordinate space  is expected to lead to
solutions in which the 
spectrum condition is not respected exactly. Equation~(\ref{almost})
may be converted into a manifestly rotationally invariant
equation using the relation:
$x^-=- 2z,$ so
that $p^+=i\partial^+=
2i{\partial\over \partial x^-}\to-i{\partial\over \partial z}$.
The operator $p^+$ acts as a $p^3$ operator, and the
result (\ref{almost}) looks like the Dirac equation of the equal time (ET)
formulation (of eigenvalue $p^-_n/2$)
except for an offending term  $-p^-_n/2$
multiplying the $\gamma^3$. This term may be eliminated by including a phase 
factor: 
$\langle x\mid n\rangle'=e^{ip^-_n z/2}\langle x\mid n\rangle_{ET}.$
The result is that $\langle x\mid n\rangle_{ET}$ satisfies the standard
Dirac equation of the equal time formulation. This means that
the eigenvalues $\epsilon_n=p_n^-/2$ must be approximately the same in the two formulations.
The net result of these transformations is that if one were to include
$p^+<0$ in the LF calculation then the eigenvalues $\epsilon_n$
in the LF and ET calculations would be identical and the LF
wave function would be expressed as
\begin{equation}
\langle p^+,x_\perp\mid n\rangle={1\over\sqrt{2\pi}}\int_{-\infty}^\infty dz
e^{-i(p^+-p_n^-/2)z}e^{ig_v\Lambda(x_\perp,z)}
\langle z,x_\perp\mid n\rangle_{ET}. \label{emcv}
\end{equation}
The 
relation (\ref{emcv}) tells us that 
$\langle p^+,x_\perp\mid n\rangle$ peaks at $p^+\approx M-g_VV^0$, with a width
of the order of the inverse of the radius of the entire nucleus. 
Therefore, neglecting $p^+<0$ is only a minor approximation, which
is the reason why our above LF calculation, with $p^+<0$ 
gives $P_n^-/2$ that are 
approximately equal to the ET results.

Eq. (\ref{emcv})
shows  that
the influence of the vector 
potential is to remove plus-momentum from the nucleons. Furthermore, the large
value of the nuclear radius causes the region of support to be very narrow, so
that $\langle p^+,x_\perp\mid n\rangle$ is very small for negative values of
 $p^+$.
 
\subsection{Nuclear momentum content and lepton-nucleus deep inelastic scattering}
Table 7.2\cite{bbm99}
gives the contributions to the total $P^+$ momentum from the
nucleons, scalar mesons, and vector mesons for $^{16}$O, $^{40}$Ca, and
$^{80}$Zr, as well as the nuclear matter limit.
The vector mesons carry approximately 30\% of the nuclear
plus-momentum. The technical reason for the difference with the scalar
mesons (which have negligible effect) is that the evaluation of
$a^\dagger(\bbox{k},\omega)a(\bbox{k},\omega)$ counts vector mesons
``in the air''and the resulting expression contains polarization
vectors that give a factor of ${1\over k^+}$  which
enhances the distribution of vector mesons of low $k^+$. The results
for the nucleon and vector meson plus-momentum distributions
are shown in Figs.~7.1 and 7.2
\cite{bbm99}. 
As the size of the nucleus increases the enhancement of the
distribution at lower values of $k^+$ becomes more evident. 

\begin{figure}
\unitlength1.cm
\begin{picture}(15,9)(-14,1) 
\includegraphics{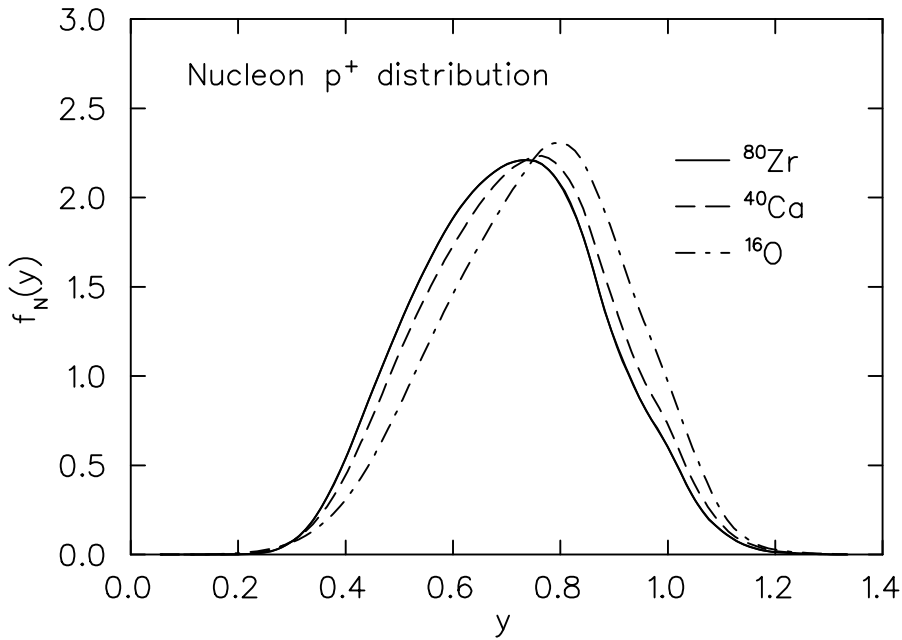}
\end{picture}
\caption{Nucleon plus-momentum distribution function}
\label{fig:fny}
\end{figure}
\begin{figure}
\unitlength1.cm
\begin{picture}(15,9)(-14,1)
\includegraphics{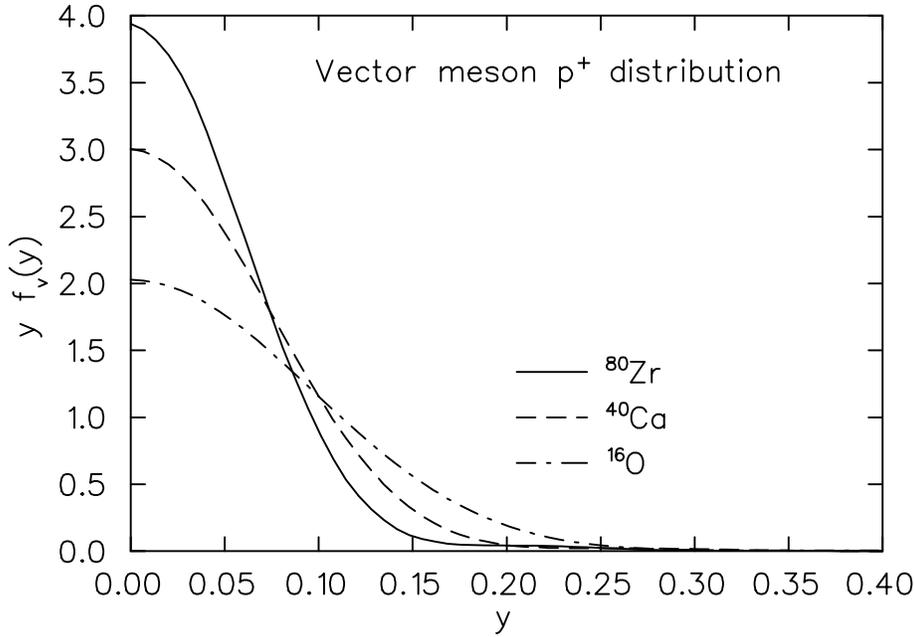}
\end{picture}
\caption{Vector meson plus-momentum distribution $y f_v(y)$, for
$^{16}$O, $^{40}$Ca, and $^{80}$Zr. }
\label{fig:fvy}
\end{figure} 

\begin{table}
\caption{Nuclear momentum content  $^{16}$O, $^{40}$Ca,
$^{80}$Zr, and nuclear matter (NM) in MeV. Contributions from nucleons,
$P_N^+/A$, vector mesons, $P_v^+/A$, and scalar mesons $P_v^+/A$ are displayed
along with the sum $P^+/A$ 
}
\rule{0in}{2ex}
\begin{center}
\begin{tabular}{ldddd}
Nucleus & $P_N^+/A$ & $P_s^+/A$ & $P_v^+/A$ & $P^+/A$\\
\hline
$^{16}$O & 704.7 & 6.4 & 221.8 & 932.9\\
$^{40}$Ca & 672.6 & 4.7 & 253.3 & 930.6\\
$^{80}$Zr & 655.2 & 3.6 & 270.2 & 929.0\\
NM & 569.0 & 0.0 & 354.2 & 923.2\\
\end{tabular}
\end{center}
\label{table3}
\end{table}


It is worthwhile to see how the present results are related to 
lepton-nucleus deep inelastic scattering experiments. We find that the
nucleons carry only about 70\% of the plus-momentum. The use of our
$f_N$ in standard convolution formulae lead to a reduction in the
nuclear structure function that is far too large ($\sim$95\% is
needed \cite{emcrevs}) to account for the reduction
observed \cite{emcrevs} in the vicinity of $x\sim 0.5$. The reason for
this is that the quantity $M +g_s\phi$ acts as a very small
nucleon effective mass
of about 670 MeV. While such a low value
is  needed to reproduce the nuclear
spin orbit force,
it causes difficulties, not only in deep inelastic scattering,
but also  in understanding the quasi-elastic 
$(e,e')$ reaction \cite{frank}.
The use of other
Lagrangians\cite{zim,qmc,qmcme} will lead to improved
results. Including  effects beyond the mean field
  lead to a significant effective tensor coupling of the iso-scalar
vector meson \cite{ls}, and to an increased value of the effective mass.
Such effects are incorporated in Brueckner theory, and a light-front
version \cite{rmgm98} could be applied to finite nuclei with better
success in reproducing the data.

A simple way to improve the phenomenology is to modify $\cal L$, by for example
including 
 scalar meson self coupling terms: $
\phi^3,\phi^4$. A wide variety of parameter sets reproduce the binding
energy and density of nuclear nuclear matter~\cite{fpw87}. For one set,  
nucleons  carry 90\% of $P^+,\;$
so that  vector mesons carry 10\%, see Sect. 5. This could be acceptable.  
There is a problem with this parameter set, the related nuclear   
spin-orbit splitting  is found to be too small~\cite{fu96a}.
This is not so bad,
since there are a variety of non-mean field mechanisms which can supply
a spin orbit force. Thus one finds a need to 
 go beyond mean field theory. This involves the introduction of  
light front Brueckner theory.

\section {Chiral Symmetry and Pion-Nucleon Scattering}
The essence of Brueckner theory is the use of an in-medium
nucleon-nucleon scattering matrix to derive the self-consistent fields. This
scattering matrix is derived from a realistic nucleon-nucleon potential.
This means that the influence of pions and approximate chiral symmetry
must be incorporated. 
In this section, we discuss the light -front quantization of a chiral
Lagrangian, and review its application to low energy pion-nucleon scattering.

The
starting point 
is the definition of a  non-linear chiral
model in which the 
nuclear constituents are nucleons $\psi$ (or $\psi')$, pions $\bbox{\pi}$
 scalar mesons $\phi$ and
 vector mesons
$V^\mu$. 
The  Lagrangian without the influence of chiral symmetry is given above in
Eq.~(\ref{lag}). 
The additional pieces which include pions and implement chiral symmetry is
${\cal L}_\chi$, with  
\begin{eqnarray}
{\cal L}_\chi =
+{1\over  4}f^2Tr (\partial_\mu\;U\;\partial^\mu\;
U^\dagger)+{1\over  4}m_\pi^2f^2\;
Tr(U +U^\dagger-2)
-\bar{\psi}^\prime\left( 
(M +g_s\phi)(U-1)\right)\psi'. \label{lagxhi}
\end{eqnarray}
 The unitary  matrix $U$ can be
chosen from amongst three forms $U_i$:
\begin{equation}
U_1\equiv e^{i  \gamma_5 \bbox{\tau\cdot\pi}/f},\quad
U_2\equiv{1+i\gamma_5\bbox{\tau}\cdot\bbox{\pi}/2f\over
1-i\gamma_5\bbox{\tau}\cdot\bbox{\pi}/2f},\quad
U_3=\sqrt{1-\pi^2/f^2}+i\gamma_5\bbox{\tau\cdot\pi}/f, \label{us}
\end{equation}
which correspond to different definitions of the fields.

The pion-nucleon  coupling here is chosen as that of  linear representations
of chiral symmetry used by  Gursey \cite{gursey}, with the 
the Lagrangian approximately $(m_\pi\ne 0$)
invariant under the chiral transformation
\begin{eqnarray}
\psi^\prime\to e^{i \gamma_5 \bbox{\tau}\cdot\bbox{a}}\psi^\prime\nonumber\\
U\to e^{-i \gamma_5 \bbox{\tau}\cdot\bbox{ a}} \;U\; 
e^{-i \gamma_5 \bbox{\tau}\cdot \bbox{a}}.
\label{chiral}
\end{eqnarray}
One may transform the fermion fields, by taking $U^{1/2}\psi'$ as the
nucleon field. One then gets Lagrangians of the non-linear
representation  \cite{weinberg}. In this case
the early soft pion theorems are manifest in the Lagrangian, and the
linear pion-fermion coupling is of the pseudovector type.  However,
the use of light front theory, requires that one find an easy way to
solve the constraint equation that governs the fermion field.
The constraint is  handled in a usual  fashion if 
Eq.~(\ref{lagxhi}) is used, and we'll show that this contains  the early soft
pion theorems.

The constant $M\over f$ plays the role of the bare 
pion-nucleon coupling constant.
If $f$ is chosen to be the pion decay constant, the Goldberger-Trieman
relation yields the result that the axial vector coupling constant
$g_A=1$, which would be  a problem for the Lagrangian, unless loop effects
can make up the needed 25\% effect. Corrections of that size 
are typical of order $({M\over f})^3$ effects found in the 
cloudy bag model\cite{cbm} for many observables, including $g_A$.

There are no explicit $\Delta's$ in the above Lagrangian. We note that
treating the higher order effects of the pion-nucleon inherent in this
Lagrangian is likely to lead to a resonance in the (3,3) channel of
pion-nucleon scattering.  Such resonance
effects can be included in the two-pion
exchange contribution to nucleon-nucleon scattering. However,
the $\Delta$ really is dominated by its 3-quark component\cite{cbm}
and carrying out 
 such computations 
 seems cumbersome. Thus it would be useful to incorporate the
 $\Delta$ in the Lagrangian.

The  scalar meson of Eq.~(\ref{lag}) is kept, even
though the effects of $\pi-\pi$ interactions, which could lead to
similar effects, are now included in the Lagrangian. We follow many
authors and  include a scalar meson to
simplify calculations and neglect the $\pi-\pi$ interactions
of Eq.~(\ref{lagxhi}).  Note that, in the present Lagrangian,
the scalar meson $\phi$ is not a chiral
partner of the pion- the chiral transformation is that of
Eq.~(\ref{chiral}).

It is now worthwhile to mention  a subtle feature regarding chiral symmetry in
light front formalisms.
Chiral invariance is defined as invariance under the transformation defined by
Eq.(\ref{chiral}) if the equal time formalism is used. Now the independent 
fermion variable is  $\psi_+$ and $\psi_-$ is a functional of this.
Thus chiral invariance is the invariance under the transformation
\begin{equation}
\psi_+\to e^{i\gamma_5\bbox{\tau}\cdot\bbox{a}}\psi_+,\label{newchiral}
\end{equation}
which is not the same as Eq.(\ref{chiral}) \cite{mustaki,osu}.
However,
the $T^{+-}$ (or  light front Hamiltonian density) derived from ${\cal L}+
{\cal L}_\chi$
 is invariant under the transformation
(\ref{newchiral}) if the pion mass is neglected\cite{jerry},
so the usual chiral 
properties are obtained in these light front dynamics.

We begin by  showing that, if one starts with a non-linear representation of
chiral symmetry,  the requirement of solving the constraint 
equation for the minus component of the fermion field leads one to a Lagrangian
of the Gursey-type  linear representation. 

The focus is on chiral properties and pion-nucleon scattering, so we
momentarily dispense
with the vector and non-chiral $\phi$ meson fields,  and
 examine  only the following fermion-pion term
 of a non-linear representation:
\begin{equation}
{\cal L}_{N\pi}=\bar{N}\left[\gamma_\mu i\partial^\mu-M+{1\over 1+(\pi/2f)^2}
\left({1\over 2f}\gamma^\mu\gamma_5\bbox{\tau\cdot}\partial^\mu\bbox{\pi}-
({1\over 2f})^2
\gamma^\mu\bbox{\tau\cdot \pi\times}\partial^\mu\bbox{\pi}\right)\right]N.
\end{equation}
The most general non-linear realization does not specify
the ratio of the two pion-nucleon couplings. 
Next obtain the fermion field equation and
make the usual decomposition:  $N_\pm\equiv
\Lambda_\pm N$ with
\begin{eqnarray}
\left(i\partial^--O^-\right)N_+=
[\bbox{\alpha}_\perp\cdot(\bbox{p}_\perp-\bbox{O}_\perp)
+\beta M
]N_-\label{nn}
\nonumber\\
\left(i\partial^+-O^+\right)N_-=[
\bbox{\alpha}_\perp\cdot\bbox{(p}_\perp-\bbox{O}_\perp)+\beta M]N_+,
\end{eqnarray}
in which  the operator $O^\mu$ is  defined as 
\begin{equation}
O^\mu\equiv {-1\over 1+(\pi/2f)^2}
\left({1\over 2f}\gamma_5\bbox{\tau\cdot}\partial^\mu\bbox{\pi}-
({1\over 2f})^2
\bbox{\tau\cdot \pi\times}\partial^\mu\bbox{\pi}\right).
\end{equation}
We need to remove the $O^+$ term from the left hand side of the equation
for $N_-$. This can be done by defining, in analogy with the Soper-Yan
transformation,  a unitary operator $F$ and fermion
field $\chi$ such
that 
\begin{equation}N=F \chi\label{defchi}\end{equation}
 with \begin{equation}i\partial^+F=O^+F.\label{feq}\end{equation}
 The identity\cite{gursey} 
\begin{equation}
U_2^{{1\over2}}\partial^\mu U_2^{-{1\over2}}=i O^\mu,
\end{equation}
where $U_2$ is given in Eq.~(\ref{us}),
helps a good deal. Its use in Eq.~(\ref{feq}), combined with the 
condition $\partial^\mu (U_2 U_2^{-1})=0$, leads to the result
\begin{equation}
F=U_2^{{1\over2}} \label{ff},
\end{equation}
so that 
using Eqs. (\ref{ff}) and (\ref{defchi}) in (\ref{nn}) yields
\begin{eqnarray}
i\partial^-\chi_+=\left[
\bbox{\alpha}_\perp\cdot \bbox{p}_\perp+\beta MU_2\right]\chi_-
\nonumber\\
i\partial^+\chi_-=\left[
\bbox{\alpha}_\perp\cdot\bbox {p}_\perp+\beta MU_2\right]\chi_+.
\end{eqnarray}
This is of the desired form in which no interactions appear on the 
left-hand-side of the equation for $\chi_-$.
Thus the use of light front quantization mandates that the 
pion-nucleon interactions be of the form of Eq.~(\ref{lagxhi}), in which the
possibility of using $U_1$ or $U_3$ arises from a field re-definition of that
Lagrangian.

The new chiral terms of ${\cal L}_\chi$ lead to additions to the
potentials
$v_1,v_2$ given above in Eqs.~(\ref{v1},\ref{v2}). These additions are:
\begin{equation}
v_1^\chi=\int d^2x_\perp dx^-\bar\xi(U-1)(M+g_s\phi)\xi,\label{cv1}
\end{equation}
and
\bea 
v_2^\chi&=&\int d^2x_\perp dx^-\;\bar\xi\left[
(U-1)(M+g_s\phi)\;{\gamma^+\over 2p^+}\;\left(-g_v\gamma\cdot\bar V
+(U-1)(M+g_s\phi)
\right)\right]\xi  \nonumber\\
&-&\int d^2x_\perp dx^-\;\xi\left[g_v\gamma\cdot\bar V\;{\gamma^+\over
    2p^+}\;(U-1)(M+g_s\phi)\right]\xi .
\label{cv2}
\eea 
The term $v_1^\chi$ accounts for  the emission or absorption
of any number of pions through the operator $U-1$. The term $v_2^\chi$ 
includes contact terms in which there is propagation of an instantaneous 
fermion.

The first test for any chiral formalism is to reproduce the early soft pion
theorems\cite{softpi}.
Here we concentrate on low energy pion-nucleon scattering
because of its relation to the nucleon-nucleon force. We work to second 
order in $1/f$ in this first application. In this case, each of the $U_i$
takes the same form:
\begin{equation}
U=1+i\gamma_5 {\bbox{\tau\cdot\pi}\over f} -{1\over 2f^2}\pi^2. \label{um1}
\end{equation}

This expression is to be used in the potentials 
$v_1^\chi$ and $v_2^\chi$ of Eqs.~(\ref{cv1}) and (\ref{cv2}).
Substitution leads to  the approximations
\begin{equation}
v_1^\chi\approx{M\over f}\int d^2x_\perp dx^-\bar\xi
\left[i\gamma_5\;\bbox{\tau\cdot\pi} -{1\over 2f}\pi^2\right]
\xi,\label{cv11}
\end{equation}
and
\bea 
v_2^\chi&\approx &{M^2\over f^2}\int d^2x_\perp dx^-\;\bar\xi\left[
i\gamma_5\;\bbox{\tau\cdot\pi}\;{\gamma^+\over
  2p^+}\;i\gamma_5\;\bbox{\tau\cdot\pi}\right]\xi
\label{cv22}
\eea 
The second-order scattering
graphs are of three types and are shown as time $x^+$ ordered 
perturbation theory diagrams in Fig.~8.1. 
The kinematics are such that
$\pi (q) N(k)\to \pi(q') N(k')$, with $P_i=q+k$ and $P_f=q'+k'$.
 The iteration
of $v_1$ to second order yields the direct and crossed graphs of
Fig.~8.1a.  
In this formalism  $v_1^\chi$ is proportional to the matrix element of $\gamma_5$ 
between $u$ spinors, so it is  
proportional to the momentum of the absorbed or  
emitted pion.
 Thus the terms of Fig.~8.1a vanish near threshold. The terms of Fig.~8.1b
 are
generated by the $\bar u \gamma_5 v$ terms of $v_1^\chi$. Using the   field 
expansions:
\begin{equation}
\bbox{\pi}(x)=
\int{ d^2k_\perp dk^+ \theta(k^+)\over (2\pi)^{3/2}\sqrt{2k^+}}\left[
\bbox{a}(\bbox{k})e^{-ik\cdot x}
+\bbox{a}^\dagger(\bbox{k})e^{ik\cdot x}\right],
\end {equation}
and
\begin{equation}
\xi(x)=\int{ d^2k_\perp dk^+ \theta(k^+)\over (2\pi)^{3/2}\sqrt{2k^+}}
\sum_{\lambda=+,-}\left[u(\bbox{k},\lambda)e^{-ik\cdot x}b(\bbox{k},\lambda)+
v(\bbox{k},\lambda)e^{+ik\cdot x}d^\dagger(\bbox{k},\lambda)\right],
\label{dq}
\end{equation}
in the expression (\ref{cv11})
for $v_1^\chi$ leads to the result that 
plus-momentum is conserved and the plus momentum of 
every  particle is greater than zero. This means that the first of Fig.~8.1b
vanishes identically and the second vanishes for values of the initial 
pion plus momentum that are less than twice the nucleon mass.
 The net result is that only the instantaneous term of $v_2^\chi$ and 
the $\pi^2$ 
term of $v_1^\chi$ (shown in Fig.~8.1c) 
remain to be evaluated.

Proceeding more formally, we evaluate the S-matrix given by
\begin{equation}
S=T_+e^{-{i\over2}\int^\infty_{-\infty} dx^+ \hat P_I^-(x^+)}, \label{smat1}
\end{equation}
where $T_+$ is the $x^+$ (light-front time) ordering operator and
$\hat P_I^-$ is the interaction representation light front Hamiltonian. Then
\begin{equation}
(S-1)_{fi}=-2\pi i\delta (P_i^--P_f^-)\langle f|T(P^-_i)|i\rangle,\label{smat2}
\end{equation}
with 
\begin{equation}
T(P^-_i)= P^-_I+ P^-_I{1\over P_i^--P^-_0}T(P^-_i)
\end{equation}
The evaluation proceeds by using the field expansions in the expressions
for $v_1^\chi$ and $v_2^\chi$. Integrating over $d^2x_\perp dx^+$ and evaluating the
result between the relevant initial and final pion-nucleon states
leads to the result that each contribution to the S-matrix
 is proportional to a common factor,
$${\delta^{(2,\perp)}(P_i-P_f)\over 2 (2\pi)^3 \sqrt{k'^+k^+q'^+q^+}},$$
which combines with the result of the required integration over the light cone
time ($x^+$) to provide the necessary momentum conservation and flux factors.
The remaining factor of each term is its 
contribution to the invariant amplitude ${\cal M}$. The result is
\begin{equation}
{\cal M}=\tau_i\tau_f {M^2\over f^2} {\bar u(k')\gamma^+u(k)\over 2(k^++q^+)} +
\tau_f\tau_i {M^2\over f^2} {\bar u(k')\gamma^+u(k)\over 2(k^+-q^+)} 
-\delta_{if}{M\over f^2}\bar u(k')u(k)
\end{equation}
where the three terms here correspond to the three terms of
Fig.~8.1c 
At threshold, $k'^+=k^+=M$ and $q'^+=q^+=m_\pi$, so the role
 of cancellations in the reduction of  the term proportional to $\delta_{if}$ 
is immediately  apparent.
In more  detail, one finds 
\begin{equation}
{\cal M}= 
\delta_{if}{2m_\pi^2\over f^2} +2i\epsilon_{fin}\tau_n{m_\pi M\over f^2}
\label{wtt}\end{equation}
to leading order in $m_\pi/M$. The weak nature 
of the $\delta_{if}$ term and the presence of the second Weinberg-Tomazowa term
is the hallmark of chiral symmetry\cite{softpi}.

The same results could be obtained using the linear sigma model, with
$\sigma$ exchange playing the role of  the $\pi^2$ term of 
Eq.(\ref{um1}).

\begin{figure}[t]
\noindent
\epsfysize=4.5in
\hspace{1.0in}
\epsffile{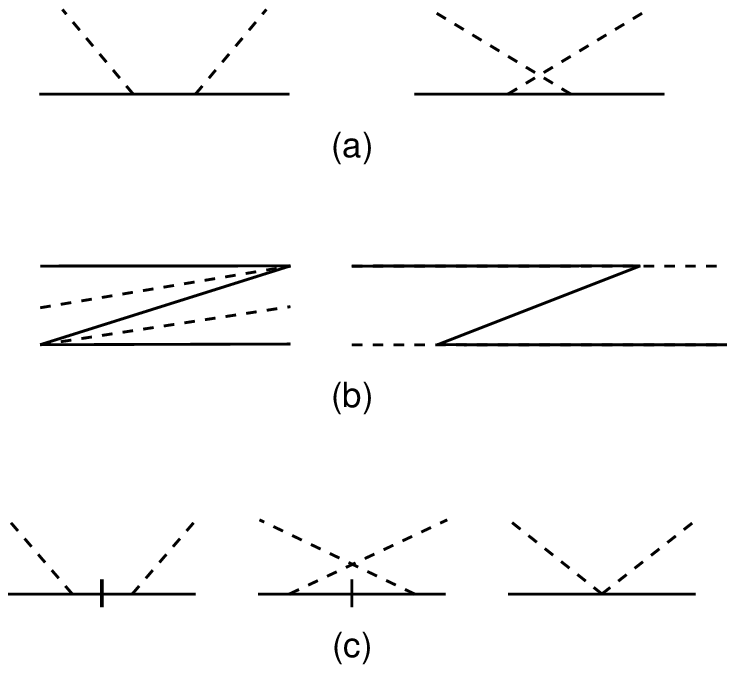}
\begin{center}
\caption {Low energy pion-nucleon scattering,
 $x^+$-ordered graphs, 
(a) Second-order effects of the $\bar u \gamma_5 u$ term $v_1^\chi$. 
(b) Second-order effects of the $\bar u \gamma_5 v$ and  $\bar v \gamma_5 u$
 terms of  $v_1^\chi$. (c) Effects of the instantaneous fermion propagation terms
of $v_2^\chi$, and of the $\pi^2$ term of $v_1^\chi$. The terms $v_i^\chi$ are defined
in Eqs.~(\ref{cv11},\ref{cv22})}.
\end{center}
\label{fig:pin}
\end{figure}

\section{Nucleon-Nucleon Scattering on the Light Front}

The correlations between nucleons are caused by the nucleon-nucleon
interaction. Thus a necessary first step towards a light-front
theory of nuclear
correlations is the derivation of a light-front theory of the nucleon-nucleon
interaction. Previous work \cite{jerry} showed that the light-front
version of the Lippmann-Schwinger equation, the Weinberg equation, can be
transformed (except that the retardation effect is kept)
into the Blankenbecler-Sugar equation\cite{BbS}.
Kinematic  invariance 
under boosts in the three-direction is maintained, and
a one-boson exchange potential,  is in reasonably good
agreement with the NN phase shifts, is obtained\cite{rmgm98}.

\begin{figure}[t]
\noindent
\epsfysize=4.5in
\hspace{1.0in}
\epsffile{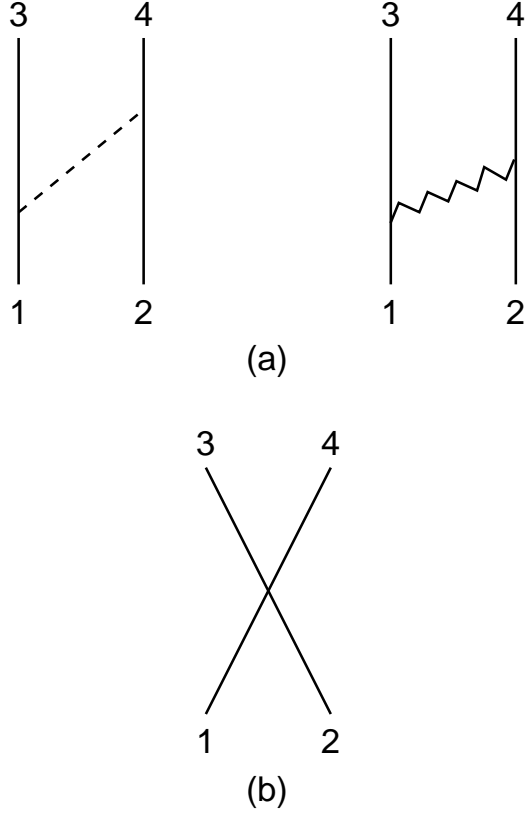}
\begin{center}
\caption{  $x^+$-ordered graphs  One boson exchange contributions 
to nucleon-nucleon scattering,$x^+$-ordered graphs.
The numbers 1-4 represent the momentum, spin and
charge states 
of the nucleons. Here $k^+_1>k^+_3$. (a) meson propagation terms 
(b) instantaneous  vector meson exchange of $v_3$, Eq.~(2.48)}
\label{fig:nn}
\end{center}
\end{figure}

It is worthwhile to begin by reviewing\cite{jerry,rmgm98}
how using  the light-front Hamiltonian
of Eqs.~(\ref{freef}-\ref{v3}) and (\ref{cv1}), (\ref{cv2})
leads to the one-boson exchange potential.
This derivation is  useful in understanding the full nuclear wave
function discussed in Sect.~X. 
Consider the scattering process $1+2\to 3+4$, of Fig.~\ref{fig:nn} .

The use of second-order perturbation theory shows that the
lowest-order contribution to the nucleon-nucleon scattering
amplitude is given by
\begin{equation}\langle 3,4|
 K|1,2\rangle=\langle 3,4| (v_1+v_1^\chi)g(P_{ij}^-)(v_1+v_1^\chi)
 +v_3|1,2\rangle,
 \label{kdef0}
 \end{equation}
 with
\begin{equation}
  g_0(P_{ij}^-)\equiv {1\over P_{ij}^--P_0^-} \; , \label{gij}
  \end{equation}
  where $P_{ij}^-$ is the negative component of the total initial or final
  momentum which are the same.
  In 
  constructing the NN potential one uses conservation of four-momentum between
  the initial and final NN states in
  constructing the NN potential.
  The  expression (\ref{kdef0}) yields a 
 one-boson exchange approximation to the nucleon-nucleon potential.

 It is worthwhile to discuss  the energy denominator $ P_{ij}^--P_0^-$
 in more detail. To be specific, suppose that $k_1^+>k_3^+$. Then
 the emitted meson of mass $\mu$
 has momentum $k$ with  $k^+=k_1^+-k_3^+, \bbox{k}_\perp=
 {\bbox{k}_1}_\perp-{\bbox{k}_3}_\perp$ and $k^-={k_\perp^2+\mu^2\over k^+}$.
 Then
 \begin{equation}
    P_{ij}^--P_0^- =  P_{12}^--P_0^- = P_{34}^--P_0^- =(k^-_1-k_3^-)
    -{{k_\perp}^2+\mu^2\over k_1^+-k^+_3}.
 \label{pdiff}   \end{equation}
    The interaction $K$ also contains a factor of $k^+$ in the denominator, so
    that the relevant denominator is
    \be D\equiv k^+ ( P_{ij}^--P_0^- )=
    (k_1^+-k_3^+) (k_1^--k_3^-)-k_\perp^2-\mu^2=q^2-\mu^2\label{det},\ee
    with \be q\equiv k_1-k_3.\ee
    This last familiar
    form involves the four-momentum transfer between nucleons 1 and 3
     leads to the  Yukawa-type potentials that include the effects of
     retardation. It is
    also useful to explore the form of the energy denominator using light-front
    variables by first defining the plus-component, $P^+$, of
    the initial and final total momentum. We
    may also
    define $k_1^+=xP^+$ and $k_3^+=x'P^+$ in which $x$ and $x'$  ($x>x'$),
   as ratios of plus-momenta,  are invariant under
    Lorentz transformations  in the three-direction. Then using 
    (\ref{pdiff}), we find
    \begin{equation}
      D=\left({{{k_1}_\perp}^2 +M^2\over x}-{{{k_3}_\perp}^2 +M^2\over
          x'}\right) (x-x') -
      k_\perp^2-\mu^2. \label{deq}
      \end{equation}
      This quantity is also    invariant under
    Lorentz transformations  in the three-direction. This expression is to be
    used only if $k^+_1>k^+_3$. If  $k^+_3>k^+_1$, then use a version
    of expression (\ref{deq}) in which $x$ and $x'$ are interchanged.

 A straightforward evaluation of Eq.~(\ref{kdef0})
 using Eqs.~(\ref{v1}-\ref{v3}) leads to the result 
\begin{equation}\langle 3,4|
 K|1,2\rangle=2\;\langle 3,4|V|1,2\rangle\quad
 {M^2\delta^{(2,+)}(P_i-P_f)\over \sqrt{k_1^+k_2^+k_3^+k_4^+}},\label{kdef}
\end{equation} 
where $\delta^{(2,+)}(P_i-P_f)\equiv
\delta^{(2)}(\bbox{P}_{i\perp}-\bbox{P}_{f\perp}) \delta(P_i^+-P_f^+)$ and $V$
is the standard expression for the sum of the $\pi,\phi $ and vector meson
exchange potentials:
\begin{equation}\langle 3,4|V|1,2\rangle=\langle 3,4|V(\phi) +V(\bbox{\pi})+
  V(V)|1,2\rangle. 
\end{equation}
The operator $K$ is twice the usual two-nucleon potential times a factor which
includes  the light front phase space factor and a momentum-conserving delta
function. 

For the exchange of
scalar and pseudoscalar mesons, only the term $(v_1+v_1^\chi)g_0(P_i^-)
(v_1+v_1^\chi)$
enters, and one finds
\begin{equation}
\langle 3,4|V(\phi,\pi)|1,2\rangle
 = {\bar u(4)\Gamma u(2)\; \bar u(3)\Gamma u(1) \over
4M^2 (2\pi)^3 \left(q^2-\mu^2\right)} \; . \label{sex}
\end{equation}

The notation is that $u(i)$ is the Dirac-spinor 
for a free nucleon of quantum numbers $i$, and $\Gamma$ is either of the
form $g_s$  or $i\;g_\pi\gamma_5{\bbox{\tau}}$. 
The derivation of the contribution of vector meson exchange proceeds by 
including the meson exchange $v_1g_0(P_i^-)v_1$
plus the meson instantaneous term $v_3$, and
the result 
takes the familiar form:
  \begin{equation}
\langle 3,4|
V(V)|1,2\rangle = -g_v^2
{\bar u(4)\gamma_\mu u(2) \bar u(3)\gamma^\mu u(1) \over
4M^2 (2\pi)^3 \left(q^2-m_v^2\right)}.\label{vex}
\end{equation}
The expressions  (\ref{sex}) and (\ref{vex}) represent
the usual \cite{rm,Mac89,Mac93,rm1,geb} expressions for  the chosen
one-boson exchange 
potentials, if no form 
factor effects are included.
The sum
of the amplitudes arising from each of the 
individual one-boson exchange terms
gives  the invariant amplitude to second order in each of the
coupling constants. 
The factors $1\over 4M^2$ in Eqs.~(\ref{sex})   and (\ref{vex})
can be thought of as re-normalizing the spinors so that $\bar{u}u=1$, and
the factors $\sqrt{M\over k^+}$ of Eq.~(\ref{kdef}) serve 
to further change the normalization to $u^\dagger u=1$

These amplitudes are strong, so computing the nucleon-nucleon scattering 
amplitude and phase shifts requires including higher order terms.
One may include  a sum which gives unitarity by including 
all iterations of the two particle irreducible
scattering operator $K$ 
through intermediate two-nucleon states. One first removes kinematic factors by
defining
 a $T$-Matrix $T$ using
\begin{equation}
  {\cal M}\equiv 2
  T {M^2\delta^{(2,+)}(P_i-P_f)\over \sqrt{k_1^+k_2^+k_3^+k_4^+}}, 
\end{equation} to   find that
\begin{equation} 
\langle3,4|T |1,2\rangle
=\langle 3,4|
V|1,2\rangle+
\sum_{\lambda_5,\lambda_6}\int  \langle 3,4|
V|5,6\rangle 
{2M^2\over k_5^+k_6^+}
{d^2k_{5\perp}dk^+_5\over P_i^--(k_5^-+k_6^-)+i\epsilon}
\langle5,6|T|1,2\rangle. \label{weinberg}
\end{equation}

One realizes that Eq.~(\ref{weinberg})
is of the form an equation, similar that of Weinberg\cite{We66},
derived for the
scattering of nucleons by Frankfurt and Strikman\cite{fs},  
by expressing the plus-momentum   variable in terms of a light-front 
momentum  fraction
$\alpha$ such that
\begin{equation}
p_5^+=\alpha P_i^+,\label{alphadef}
\end{equation}
and using the relative and total momentum variables:
\begin{eqnarray}
\bbox{k}_\perp\equiv (1-\alpha)\bbox{k_5}_\perp-\alpha
\bbox{k_6}_\perp \; , \nonumber\\
\bbox{P_i}_\perp=\bbox{k_5}_\perp+\bbox{k_6}_\perp \; . \label{alpha2}
\end{eqnarray}
Then,
\begin{equation} 
\langle3,4|
T|1,2\rangle
=\langle 3,4|
V|1,2\rangle+
\int\sum_{\lambda_5,\lambda_6} \langle 3,4|
V|5,6\rangle 
{2M^2\over \alpha(1-\alpha)}
{d^2k_\perp d\alpha\over P_i^2-{k_{\perp}^2+M^2\over\alpha(1-\alpha)}
+i\epsilon}
\langle5,6|T
|1,2\rangle, \label{419}
\end{equation}
where $P_i^2$ is the square of the total initial  four-momentum,
otherwise known as the invariant energy $s$ and 
${k_{\perp}^2+M^2\over\alpha(1-\alpha)}$
 is the corresponding quantity for the intermediate state.
 Because the kernel $V$ 
is itself  invariant under Lorentz transformations in the  three-direction
and the integral involves $k_\perp $ and $\alpha$ 
the procedure of solving this equation gives  $T$ with the same invariance. 
Note that we use the labels
$k_i$ to designate momenta in the intermediate state, and $k_i$ for the
initial and final
states.

Equation~(\ref{419}) can
in turn be re-expressed (in the center of mass frame)
as the Blankenbecler-Sugar (BbS) equation
\cite{BbS} 
by  using the variable transformation\cite{Te 76}:
\begin{equation}
\alpha={E(k)+k^3\over 2E(k)}, \label{alpha}
\end{equation}
with $E(k)\equiv\sqrt{\bbox{k}\cdot\bbox{k}+M^2}$, and $P_i^2=
4(\bbox{k}\cdot\bbox{k} +M^2)$
The result is:
\begin{equation}
\langle3,4|T
|1,2\rangle
=\langle3,4|V
|1,2\rangle+\int
\sum_{\lambda_5,\lambda_6} \langle 3,4|V
|5,6\rangle
{M^2\over E(p)}
{d^3p \over \bbox{k_i}^2-\bbox{k}^2
+i\epsilon}
\langle5,6|T
|1,2\rangle, \label{bsbs}
\end{equation}
which is the desired equation. Rotational invariance is manifestly obeyed. 
The three-dimensional propagator is exactly that of the BbS equation. 
There is,
one difference between Eq.~(\ref{bsbs}) and the standard BbS equation.
 Our one-boson exchange potentials depend on the 
square of the four momentum $q^2$ transferred when a meson is absorbed or
emitted by a nucleon. Thus the energy difference between the initial and final
on-shell nucleons is included and $q^0\ne 0$. This non-zero value is a
consequence of the invariance of $D$ of Eq.~(\ref{deq}) under Lorentz
transformations in the three-direction.  
 The usual derivation of the BbS equation from
the Bethe-Salpeter equation specifies that  $q^0=0$ is used 
in the meson propagator. Including $q^0 \ne 0$ instead of $q^0=0$ increases the
range of the potential relative to the usual treatment,
and its consequences are explored below.
One can convert Eq.~(\ref{bsbs}) into
the Lippmann-Schwinger equation of non-relativistic scattering theory
by removing the factor $M/E(p)$ with a simple transformation\cite{pl}.

\subsection{ Realistic One-Boson Exchange Potential}
So far we have reviewed how  the light front technique is used to
derive nucleon-nucleon potentials in the one-boson exchange (OBE)
approximation and use these in an appropriate wave equation (\ref{bsbs}).
It is also true that
 this  procedure 
yields potentials essentially identical to the Bonn OBEP potentials
\cite{Mac89,Mac93}  and
these potentials 
lead to a good description of the NN data\cite{rmgm98}.

The Bonn one-boson exchange  potentials employ six different mesons, namely,
$\pi,\eta, \omega,\rho, \sigma$ and the (isovector scalar)
$\delta/a_0$ meson.
The present formalism can account   for the
$\pi,\eta, \omega$ and $ \sigma$ in an approximately chiral invariant manner.
We wish to add in couplings $\bar{\psi}\bbox{\tau}\cdot\bbox{\delta}\psi$
  and $\bar{\psi}\bbox{\tau}\cdot\bbox{\rho}^\mu\gamma_\mu\psi$
 in a chiral invariant manner. Simply adding such terms to the Lagrangian
 of Eq.~(\ref{lagv}) would lead to a violation of the approximate symmetry of
 Eq.~(\ref{chiral}). 
 However, one can redefine the operator $U$ so that
 the symmetry remains. We replace the operator $\bar{\psi}'U\psi'$ in the
 Lagrangian (\ref{lagv}) by $\bar{\psi}'\tilde {U}\psi'$ :
 \begin{equation}
   \tilde{U}\equiv
   e^{{i\over 2 f_\rho}\bbox{\tau}\cdot\bbox{\rho}^\mu\gamma_\mu}
     e^{{i\over 2 f_\delta}\bbox{\tau}\cdot\bbox{\delta}}U
        e^{{i\over 2 f_\delta}\bbox{\tau}\cdot\bbox{\delta}}
          e^{{i\over 2 f_\rho}\bbox{\tau}\cdot\bbox{\rho}^\mu\gamma_\mu}.
            \end{equation}
Then the new  Lagrangian  is invariant under the transformation
\begin{eqnarray}
\psi^\prime\to e^{i \gamma_5 \bbox{\tau}\cdot\bbox{a}}\psi^\prime,\qquad
\tilde{U}\to e^{-i \gamma_5 \bbox{\tau}\cdot\bbox{ a}} \;U\; 
e^{-i \gamma_5 \bbox{\tau}\cdot \bbox{a}}.
\label{nchiral}\end{eqnarray} In the  applications 
 the exponential is expanded to first order in the meson fields.

The final term  to be included is the 
 tensor 
$\sigma_{\mu\nu}q^\nu$ part of the $\rho$-nucleon interaction. 
The presence of such a tensor interaction makes it difficult
to write the equation for $\psi_-$ as $\psi_-=1/p^+\cdots
\psi_+.$ This is a possible problem
because the standard value of the ratio of
the tensor to vector $\rho$-nucleon coupling $f_\rho/g_\rho$ is 6.1,
based upon Ref.~\cite{hp}.  Reproducing the
observed values of $\varepsilon_1$ and P-wave wave phase shifts
requires a large value 
$f_\rho/g_\rho$; see Ref.~\cite{brm}.  However our Lagrangian
generates  such a term via vertex correction diagrams
(which are the origin of the anomalous magnetic moment of the electron
in QED). 
Thus the procedure of Ref.~\cite{rmgm98} was to simply
add  the necessary tensor terms to the one boson exchange potential.   
Recently Cooke\cite{cooke} has provided the necessary light front
quantization of a Lagrangian containing the  tensor 
$\sigma_{\mu\nu}q^\nu$ part of the $\rho$-nucleon interaction. 

This brings us to the treatment of divergent terms
used in Ref.~\cite{rmgm98}.
The definition of any effective Lagrangian
requires the specification of such a procedure.  
Meson ($m$) nucleon form factors  $F_m(q^2)$ were introduced 
which reduce the strength of the  coupling for
large values of $-q^2$. This is also the procedure of 
Refs.~\cite{Mac89,Mac93,rm1}. In principle, calculating the
higher order terms using the correct Lagrangian can lead to consistent
calculations of these form factors.   We use a more
phenomenological approach here.

The net result is that the one-boson exchange treatment of the
nucleon-nucleon potential and the T-matrix resulting from its use in
the BbS equation is essentially the same as the one-boson exchange
procedure of Refs.~\cite{rm,Mac89,Mac93,rm1}. The only difference is
the keeping
of the retardation effects---the square of the four-vector momentum transfer
enters in our potentials.

\subsection{Results for the two-nucleon system}
Following established procedures~\cite{Mac89,Mac93}, the coupling constants
and cutoff masses of the six OBE amplitudes are varied within 
reasonable limits such as to reproduce the two-nucleon bound
state (deuteron) and the two-nucleon scattering data below the
inelastic threshold (about 300 MeV laboratory kinetic energy).

In Table~\ref{tab:parms}, we show the meson parameters for the  Light-Front (LF) OBEP
of Ref.~\cite{rmgm98} together with the predictions
for the deuteron as well as low-energy neutron-proton ($np$)
scattering.
For comparison, we also give the parameters from an OBEP that
was previously constructed and applied in the Dirac-Brueckner
approach to nuclear matter~\cite{Mac89,rm1}. The latter
uses the Thompson formalism~\cite{Tho70} which is very similar
to the BbS formalism.
Note that the Thompson OBEP uses $n_\alpha = 1$
also for vector meson form factors, which explains the differences
in the vector meson cutoff masses between the two OBEP.

\begin{table}
\caption{Potential parameters and predictions for the deuteron
and low-energy $np$ scattering.
For the deuteron, the binding energy $B_d$, the $D$-state probability $P_D$,
the quadrupole moment $Q_d$, and the asymptotic $D$-state over $S$-state
ratio $D/S$ are given. Low-energy $np$ scattering is parameterized in terms
of $a_{np}$ and $r_{np}$ in $^1S_0$ and $a_t$ and $r_t$ in $^3S_1$,
where $a$ denotes the scattering length and $r$ the effective range.
The nucleon mass is $M=938.919$ MeV.}
\begin{tabular}{lllllllll}
  &            & \multicolumn{2}{c}{\bf Light-Front OBEP}
&\hspace*{.1cm}& \multicolumn{2}{c}{{\bf Thompson OBEP}$^a$}
&\hspace*{.1cm}& \multicolumn{1}{c}{{\bf Empirical}$^b$} 
\\ \cline{3-4} \cline{6-7} \cline{9-9} \\
\multicolumn{9}{l}{\rm Meson Parameters:}\\
\\
  & $m_{\alpha}$(MeV)
     &  $g^{2}_{\alpha}/4\pi$ $[f/g]$
     &  $\Lambda_{\alpha}$(GeV)
     &&  $g^{2}_{\alpha}/4\pi$ $[f/g]$
     &  $\Lambda_{\alpha}$(GeV)
     &&  $g^{2}_{\alpha}/4\pi$ $[f/g]$ 
\\ \hline 

$\pi$ & 138.04  & 14.0 & 1.2 && 14.6 & 1.2 && 13.5 -- 14.6 \\

$\eta$ & 547.5  &  3   & 1.5 &&   5  & 1.5 && $\leq 5$ \\

$\rho$ & 769 & 0.9  [6.1] &1.85&& 0.95 [6.1]& 1.3 && 0.6(1) $[6.6\pm 1.0]$\\

$\omega$ & 782 & 24.5 [0.0] & 1.85 && 20.0 [0.0] & 1.5 && $24\pm 5\pm 7$ \\

$a_0$ & 983  & 2.0723 & 2.0 && 3.1155 & 1.5 &&              \\

$\sigma$ & 550  & 8.9602 & 2.0 &&  8.0769  &  2.0 &&  
\\ \hline \\
\multicolumn{9}{c}{\rm Deuteron}\\
\\
 $B_{d}$ (MeV)&&  2.2245 &&& 2.2247 &&& 2.224575(9)\\
 $P_{D}$ (\%) && 4.53 &&& 5.10 &&&  ---\\
 $Q_{d}$ (fm$^{2}$) && 0.270$^{c}$ &&& 0.278$^{c}$ &&&  0.2860(15)\\
 $D/S$ &&   0.0250 &&&   0.0257  &&&  0.0256(4)\\ 
\\
\multicolumn{9}{c}{\rm Low-Energy $np$ Scattering}\\
\\

$a_{np}$ (fm) && --23.745  &&& --23.747 &&& --23.748(10)\\
 $r_{np}$ (fm) && 2.671    &&&  2.664   &&& 2.75(5)\\
 $a_{t}$ (fm)  && 5.494    &&&  5.475   &&& 5.424(4)\\
 $r_{t}$ (fm)  && 1.856    &&&  1.828   &&& 1.759(5)
\end{tabular}
$^{a}$Potential B of Brockmann and Machleidt~\cite{rm1}.\\
$^{b}$For more comprehensive information on the empirical data and references,
see Table 4.1 and 4.2 of Ref.~\cite{Mac89}.\\
$^{c}$Meson-exchange current contributions not included.
\label{tab:parms}
\end{table}

\begin{figure}
\unitlength1.cm
\begin{picture}(15,9)(-3,1)
\includegraphics{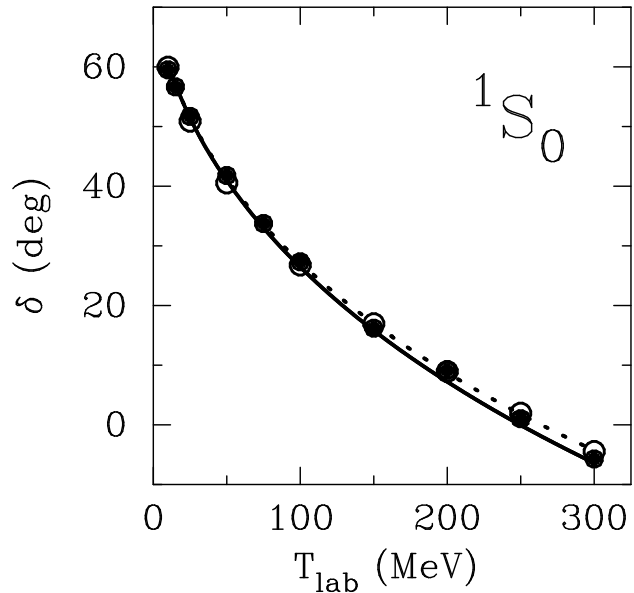}
\end{picture}
\caption{$^1S_0$ phase shifts-solid light front, dashed Thompson equation. }
\label{fig:f1s0}
\end{figure}

Phase shifts for $np$ scattering are shown in Fig.~1 of the  long paper of
Ref.~\cite{rmgm98} for
all partial waves with $J\leq 2$. A typical result is shown here in
Fig.~\ref{fig:f1s0}.
Over-all, the reproduction of the $NN$ data\cite{Sto93,VPI97}
by our LF OBEP is quite
satisfactory and certainly as good as by
OBEP constructed within alternative relativistic
frameworks.
Based upon these results, we  applied 
this OBEP to the relativistic nuclear many-body problem.

\section{Correlated Nucleons in Infinite Nuclear Matter }
I begin with an outline of the procedure and then review the formalism of
Ref.~\cite{rmgm98}. 
The first step is to derive a light front version of the
nucleon-nucleon interaction. 
This is most easily done within the framework of the one boson exchange
approximation. 
The resulting nucleon-nucleon
potential $V(NN)$ describes phase shifts reasonably well, as discussed in
Sect.IX.
The corresponding
interaction density is
 ${\cal V}(NN)$. The basic Lagrangian density
 contains a free nucleon term ${\cal L}_0(N)$,
 a free meson term ${\cal L}_0({\rm mesons})$ and an interaction term
 ${\cal L}_I(N,{\rm mesons})$  but does not contain ${\cal V}(NN)$. Thus 
 one adds this term and subtracts it:
\bea{\cal L}&=&{\cal L}_0(N)-{\cal V}(NN) + {\cal L}_{\rm m}\\
 {\cal L}_{\rm m}& =&{\cal L}_I(N,{\rm mesons})+{\cal L}_0({\rm mesons})
+{\cal V}(NN).\eea
We  use
the term ${\cal L}_0(N)-{\cal V}(NN)$ to obtain a first solution
$\mid\Phi\rangle$ to the
many-body problem. The term 
${\cal L}_{\rm m} $
accounts for mesonic content of Fock space, and we
present\cite{rmgm98} a scheme to incorporate the 
effects of ${\cal L}_{\rm m} $ and calculate the full wave function
$\mid\Psi\rangle$. Our procedure allows us to assess whether or not
${\cal V}(NN)$ has been chosen well. If it has, the effects of 
${\cal L}_{\rm m} $ can be treated perturbatively.

Solving for $\mid\Phi\rangle$ is no easy task --it demands a special 
non-perturbative treatment. One introduces a mean field 
$U_{MF}$ which acts
on single nucleons.
\be{\cal L}_0(N)-{\cal V}(NN) ={\cal L}_0(N)-U_{MF} +
\left(U_{MF}-{\cal V}(NN)\right).\ee
The operator 
 $U_{MF}$ is chosen to minimize the effects of  $\langle \Psi|U_{MF}-{\cal
V}(NN)|\Psi \rangle.$
There is a well-known procedure, called Brueckner theory, which is used to
determine $U_{MF}$. In schematic terms: 
\be  U_{MF}\sim G  \times \rho,\ee
in which $G$ is a nucleon-nucleon scattering matrix, as modified by the Pauli
principle, $\rho$ is the nuclear density, and the $\times$ represents a
convolution.

The result \cite{rmgm98} is a rather complete theory in which the 
full wave function $|\Psi\rangle$ includes the
effects of both NN correlations and explicit
mesons. This theory is reviewed next.

\subsection{Nucleonic Truncation For The Many-Body Problem}
One starts with a Lagrangian and can   derive a light front
Brueckner
theory  from first principles\cite{rmgm98}.  
The nuclear wave function for the ground state of infinite nuclear matter 
at rest is
 defined as $|\Psi\rangle$, and we wish to solve the equation
\begin {equation}
P^-|\Psi\rangle=M_A|\Psi\rangle, \label{sc}
\end{equation} 
in which 
$P^-$ is the light-front Hamiltonian
discussed above. For a nuclear system at rest we must have also
\begin {equation}
P^+|\Psi\rangle=M_A|\Psi\rangle. \label{scp}
\end{equation}
The operator $P^-$ can be separated as follows:
\begin{equation}
 P^- = P^-_0(N) + J \label{op}
\end{equation}
in which $P^-_0(N)$ is the kinetic contribution to the $P^-$ operator,
 giving ${p_\perp^2 + m^2}\over{p^+}$ for the minus-momentum of free
 fermions.
The operator $J$ is the sum of the  terms of Eqs.~(\ref{v1})-(\ref{v3}),
(\ref{cv1})
and (\ref{cv2}):
\begin{eqnarray}
J \equiv v_1 + v_2 + v_3 +v_1^\chi+v_2^\chi 
\end{eqnarray}

An approximate solution of Eq.~(\ref{sc}) is obtained in two
stages. First
 the nucleons-only part of the Hilbert space is constructed. This
involves
the assumption that using  a nucleon-nucleon
interaction  $K$ accounts for the meson-nucleon dynamics.
This assumption in then relaxed, and the formalism
necessary  to
construct the best possible potential and to include meson degrees
of freedom  in the wave function is reviewed.
Our  Hamiltonian ($P^-$) contains no important  terms in which the
vacuum can spontaneously
emit particles.  (Some of the meson self interaction terms do involve non-zero
creation from the vacuum\cite{mb:adv}, but their influence on the physics would
enter only through a medium modifications of the  nucleon mass which are not
part
of standard relativistic Brueckner theory.) 
This simplifying feature  of the dominant terms causes the
 derivations  to look
very similar to those of non-relativistic theory,
even though the treatment is
relativistic.

 All of the interactions in the Lagrangian are expressed in terms of the
meson-nucleon vertex functions and contact terms represented by the
operator $J$.
The
traditional approach is to introduce  a two-nucleon potential and temporarily
eliminate the meson degrees of freedom.  This accomplished by  
subtracting and adding  the two-nucleon potential
to the  Lagrangian.
It is worthwhile to  define the 
P-minus operator so  obtained 
as $P^-_0$ with: 
\begin{eqnarray}
P^-_0 \equiv  P^-_0(N) +K.
\label{p0m}\end{eqnarray}                                
The operator $K$ is given in Eq.~(\ref{kdef0}).
The complete P-minus operator is given by 
\begin{eqnarray}P^- &=& P^-_0 + H_1,
 \end{eqnarray} with
 \begin{eqnarray}
 H_1 &\equiv& J-K +P^-_0(m)
\end{eqnarray}
where $P^-_0(m)$ accounts for the contributions to $P^{-}$ of  mesons
which do not involve
interactions with the nucleons. 
The formal problem of choosing  the best $ {K}$
by minimizing the effects of
$H_1=J-K$ is discussed below in sub-section X.E.

The purely nucleonic   part of the full wave function is defined as 
$\mid \Phi  \rangle$, and is
the solution of the light front Schroedinger equation
\begin{equation} P^-_0 \mid \Phi \rangle
  =\left( P^-_0(N) +K\right)\mid \Phi \rangle=
M_0 \mid \Phi \rangle  .\label{big}
\end{equation}
The eigenvalue problem stated above is considerably simpler than the initial
one, but does contain the full complications of the
nuclear many-body problem.
We shall next review  the derivation of the light front 
Brueckner Hartree-Fock \cite{bhf} approximations.

\subsection{Light Front Brueckner  Hartree-Fock Approximation}
The interaction $K$ that appears  in Eq.~(\ref{big})  is  strong and  
the scattering amplitude is obtained, as discussed in Section 9,
by solving the
light front version of the Lippmann-Schwinger equation which 
treats the interaction between two nucleons to all orders in $K$. 
One needs to find an all-orders treatment for the ground state. This can be
done, and one can 
also  find the Slater determinant
$\mid\phi\rangle$, recall Eq.~(\ref{phi}),
that leads to the best approximation for the energy $M_0$ of
the full nucleonic wave function
 $\mid \Phi\rangle$:  
\begin{equation} P^-_{MF}\mid\phi\rangle
  =P_0^-(N)\mid\phi\rangle
= m_0\mid\phi\rangle.\label{phi}
\end{equation}
The operator $P_0^-(N)$ is expressed in terms of nucleon field operators
defined by the single particle basis:
\bea
k^-|i\rangle&=&k^-_i|i,\rangle\\
k^-_i
&=& {{\bbox{k}_i}_\perp^2 +(M+U_S)^2\over k^+_i} +U_V .\label{pim}
\end{eqnarray}
This is a generalization of 
Eq.~(\ref{sd}) in which self-consistent potentials, $U_S$ and $U_V$
to be defined below,
replace the meson field terms, $g_S\phi$ and $g_V\bar{V}^-$
in the equation defining the nucleon mode functions.
Ths details of the related spinors are given in Ref.~\cite{rmgm98}.

Both of the states 
$\mid\phi\rangle$ and  $\mid \Phi\rangle$ 
 are eigenstates of a P-minus operator, and both are eigenstates
of the operator $P_0^+(N)$. 
   We shall use standard techniques to derive a perturbation theory which
   employs the
   appropriate transition matrix generated by 
   $K $ 
   to obtain an expression for  the state $\mid\Phi\rangle$
   in terms of  $\mid\phi\rangle$.
Thus we write  
 \begin{eqnarray}
\mid \Phi \rangle &=& \mid \phi \rangle  + \Lambda \mid \Phi \rangle 
 \label{decomp}
\end{eqnarray}
with
\begin{eqnarray}
\Lambda &\equiv& 1 - \mid \phi \rangle \langle \phi \mid .
\end{eqnarray}
The use of algebra leads to the result
\begin{eqnarray}
\mid \Phi \rangle  &=& \mid \phi \rangle 
 + {1\over M_0 - \Lambda ( P_0^-(N)+K) \Lambda} 
\Lambda K 
\mid \phi \rangle . \label{Phi}
\end{eqnarray}
We can obtain a useful expression for $M_0$ by acting with the operator
$ \langle\phi\mid \left(P_0^-(N)+K\right)$ 
on the left of  Eq.~(\ref{Phi}) and using
 $\langle \phi \mid \Phi \rangle  = 1,$.
Then we find
\begin{eqnarray}
M_0 = \langle \phi \mid P^-_0(N) +K\mid \phi \rangle 
+ \langle\phi \mid K
\Lambda {1\over M_0-\Lambda ( P_0^-(N)+K)\Lambda}
\Lambda K
\mid \phi \rangle ,\label{M0}
\end{eqnarray}
and with 
 Eq.~(\ref{M0})  
\begin{eqnarray}
M_0 - m_0 = \langle \phi \mid K\mid \phi\rangle 
+ \langle \phi \mid K
\Lambda {1 \over M_0- \Lambda ( P_0^-(N)+K)\Lambda}
\Lambda K 
\mid \phi \rangle  . \label{mass}
\end{eqnarray} Thus
\begin{eqnarray}
M_0 - m_0 = \langle \phi \mid X\mid \phi\rangle,\label{mmm}
\end{eqnarray}
with 
\begin{eqnarray}
  X=K + K\Lambda{1\over M_0-\Lambda P_0^-(N)\Lambda}\Lambda X.
\end{eqnarray}
The operator $X$ is a many-body operator acting 
on all nucleons via the iterations
of the two-nucleon  interaction $K$. We shall make the independent pair
approximation of including only pair-wise interactions. Thus we approximate 
\begin{eqnarray}\langle \phi \mid X\mid \phi\rangle\approx
\langle \phi\mid\sum_{i<j} \Gamma_{i,j}(P^-_{ij})\mid \phi\rangle
\equiv\langle \phi \mid  \Gamma\mid \phi\rangle, \label{pgp}
\end{eqnarray}
where $\Gamma_{i,j}$ is a two-nucleon operator which is a solution of the
integral equation
\begin{eqnarray}
\Gamma_{i,j}(P^-_{ij}) &=& K_{ij} + K_{ij} \,\,{\Lambda\over
 P^-_{ij}-\Lambda P_0^-(N)\Lambda} \,\, \Gamma_{i,j}(P^-_{ij}).
\end{eqnarray}
The notation $i,j$ refers to a pair of particles. The relevant matrix element
is expressed using 
 the  eigenstates of Eq.~(\ref{sd})
as
\begin{equation} 
\langle3,4|\Gamma(P^-_{1,2})|1,2\rangle
=\langle 3,4|
K|1,2\rangle+
\sum_{\lambda_5,\lambda_6}\int  \langle 3,4|
K|5,6\rangle 
{2{M^*}^2\over k_5^+k_6^+}
{d^2k_{5\perp}dk^+_5\; Q\over P^-_{1,2}-(k_5^-+k_6^-)+i\epsilon}
\langle5,6|\Gamma|1,2\rangle, \label{bweinberg}
\end{equation}
in which we define
\begin{equation}
  M^*\equiv M+U_S. \label{mstar}
  \end{equation}
The operator $Q$ 
is the two-body version of
$\Lambda$ and
projects the momenta $k_5$ and $k_6$ above the
Fermi sea. 
The factor $ {{M^*}^2\delta^{(2,+)}(P_i-P_f)\over \sqrt{k_1^+k_2^+k_3^+k_4^+}}$
appears in each of the terms of Eq.~(\ref{bweinberg}), so it is worthwhile
to define a Bruckner $G$-Matrix $G$ using
\begin{equation}
  \Gamma\equiv 2
  G {{M^*}^2\delta^{(2,+)}(P_i-P_f)\over \sqrt{k_1^+k_2^+k_3^+k_4^+}}.
\end{equation}

To follow the steps of Sect. IX in converting Eq.~(\ref{bweinberg}) into
one of a more familiar  form, in which rotational invariance is manifest,
one needs to know the values of
\begin{equation}P_{1,2}^-=k_1^-+k_2^-,
  \end{equation}
  which for the case of
relevance here in computing the nuclear expectation value in the independent
pair approximation, is the
same as $k_3^-+k_4^-$.
The single-particle minus-momentum 
eigenvalues are  given according to the light front
Eq.~(\ref{pim}).
Our approximation is that $U_V$ is independent of orbital $i$. Thus this
potential cancels out in computing the difference $ P_i^--(k_5^-+k_6^-)$
and the energy denominator is as in the free space considerations of Sect.~III,
except that the mass of the nucleon is replaced by $M+U_S$. Thus the previous
derivation of an equivalent three-dimensional integral equation that is
manifestly covariant and rotationally invariant proceeds as before.

One expresses  the plus-momentum   variable in terms of a light-front 
momentum  fraction
$\alpha$ of Eqs.~(\ref{alpha}) and (\ref{alpha2}) so that one obtains
\begin{equation} 
\langle3,4|
G|1,2\rangle
=\langle 3,4|
V|1,2\rangle+
\int\sum_{\lambda_5,\lambda_6} \langle 3,4|
V|5,6\rangle  
{2{M^*}^2\over \alpha(1-\alpha)}
{d^2k_\perp d\alpha\;Q\over P_i^2-{k_{\perp}^2+{M^*}\;^2\over\alpha(1-\alpha)}
}
\langle5,6|G
|1,2\rangle, \label{519}
\end{equation}
where $P_i^2$ is square of the total initial  four-momentum, computed using
$M+U_S$ for the nucleon mass.

Equation~(\ref{519}) can
in turn be re-expressed as a medium-modified
Blankenbecler-Sugar (BbS) equation
\cite{BbS} 
by  using the  medium-modified version of the
variable transformation\cite{Te 76}:
\begin{equation}
\alpha={E^*_{\bbox{k}}+k^3\over 2E^*_{\bbox{k}}}, \label{balpha}
\end{equation}
with $E^*_{\bbox{k}}=\sqrt{k^2+{M^*}^2}$.
 The result is:
\begin{equation}
\langle3,4|G
|1,2\rangle
=\langle3,4|V
|1,2\rangle+\int
\sum_{\lambda_5,\lambda_6} \langle 3,4|V
|5,6\rangle
{{M^*}^2\over E^*_{\bbox{k}}}
{d^3k\; Q \over \bbox{k}_i^2-\bbox{k}^2}
\langle5,6|G
|1,2\rangle, \label{bbsbs}
\end{equation}
which is the desired equation.

 The Brueckner light front Hartree-Fock (BHF)
 approximation is defined by taking
the mean field to be calculated from the average G-matrix according to
\begin{eqnarray}
   U_l=
 \langle\bar{l}\mid (U_S+\gamma^0U_V^0)^{BHF}\mid l\rangle&=&
 \sum_{m<F}
   \langle\bar{l}\;\bar{m}\mid G
   \mid l\;m \rangle_a\;. 
\end{eqnarray}
The sum over occupied orbitals $l,m$ gives
\begin{eqnarray}\sum_{l<F} U_l
  &=& 2\langle\phi\mid G
\mid\phi\rangle.
\label{ubhf}
\end{eqnarray}
We use  this BHF mean-field to determine the value of $m_0$ via 
Eq.~(\ref{phi}). Then the use of Eq.~(\ref{pgp}) in Eq.~(\ref{mmm})
determines the value of $M_0$.  as the eigenvalue of $P^-_0$. 
But $M_0$ is also the eigenvalue (or in this case the expectation value)
of  $P^+_0. $ 
The  minimization  of $P^-_0$ subject to the
constraint that the expectation value of $P^+_0$ is the value
of $P^-_{0}$ leads to 
(\ref{pplus}):
\begin{eqnarray}
{P^{-}_{0}\over \Omega}&=&
{4\over (2\pi)^3}\int_F d^2k_\perp dk^+\left\{ {\bbox{k}_\perp^2+ 
(M+U_S)^2\over k^+}-2\;
{1\over 2}\sum_\lambda\;
{\bar{u}(k,\lambda)\over \sqrt{2 k^+}}U_S
{u(k,\lambda)\over\sqrt{2 k^+}}\right\}\;
+\langle\phi\mid{\Gamma}\mid\phi\rangle
\label{pbminus}\\
{P^{+}_{0}\over \Omega}&=&
{4\over (2\pi)^3}\int_F d^2k_\perp dk^+ k^+.\label{pbplus}
\end {eqnarray}
Note that the quantity $k^+$ is defined as 
\be
k^+\equiv E^*_{\bbox{k}}+k^3
.\ee
Taking the average of  equations (\ref{pbminus})  and~(\ref{pbplus}), and
using the basis of Eq.~(\ref{pim}) leads to our result for the BHF version
of the nuclear mass:
\begin{equation}M_0=\sum_{l<F} \epsilon_l 
   -{1\over 2}\sum_{l,m<F}
   \langle\bar{l}\bar{m}\mid G
   \mid l\;m\rangle_a, \label{main}
 \end{equation}
 in which $\epsilon_l $ is the eigenvalue of the equal time self-consistent
 Dirac equation for
 the nuclear modes.   
This is equivalent to the usual expression of
Ref.~\cite{Mac89,rm1}. 

\subsection{Results for Energy versus Density}
\begin{figure}
\unitlength1.cm
\begin{picture}(15,9)(-3,1)
\includegraphics{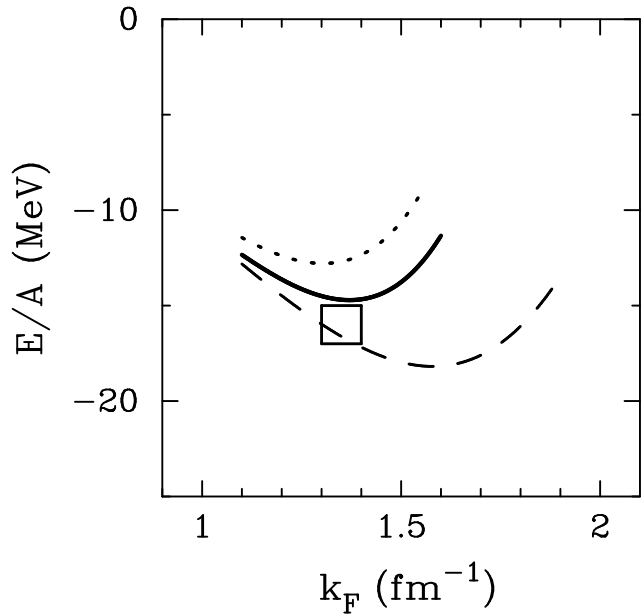}
\end{picture}
\caption{Binding energy vs density of nuclear matter}
\label{fig:nm}
\end{figure}

The formalism of the previous section is used to  calculate
the energy per nucleon in nuclear matter as a function of density,
Eq.~(\ref{main}). Our result is plotted in Fig.~\ref{fig:nm} by the solid line.
The curve has a minimum value of  ${\cal E}/A = -14.71$ MeV at $k_F = 1.37$
fm$^{-1}$, and gives  an incompressibility of $K=180$ MeV at the minimum.
These predictions agree well with the
empirical values 
${\cal E}/A=-16\pm 1$ MeV,
$k_F=1.35\pm 0.05$ fm$^{-1}$, 
and $K=210\pm 30$ MeV~\cite{Bla80}.

To  better understand  our predictions, it is useful to
compare with the results from alternative relativistic approaches.
Brockmann and Machleidt~\cite{rm1} predict
${\cal E}/A=-13.6$ MeV,
$k_F=1.37$ fm$^{-1}$, 
and $K=250$ MeV at saturation, using the equal-time formalism
and their `Potential B'.
The greatest difference occurs for the incompressibility
which is predicted smaller by
the light front  Brueckner theory implying a softer
equation of state. This can be partially attributed to the 
medium effect that comes
from the
meson propagators in the light front  approach and that is absent in the 
equal
time (ET) approach.
In the ET approach the meson propagators are
\begin{equation}
\frac{i}{-{\bbox q}^2-m_\alpha^2}=
\frac{i}{-({\bbox k'}-{\bbox k})^2-m_\alpha^2} \; .
\label{PropET}
\end{equation}
In nuclear matter, the free-space LF meson propagators 
are replaced by
\begin{equation}
\frac{i}{(E'^*-E^*)^2-({\bbox k'}-{\bbox k})^2-m_\alpha^2} \; ,
\label{PropLFinmed}
\end{equation}
while the ET propagators undergo no changes.
The medium effect on the LF meson propagators enhances them
off-shell which leads to more binding energy.
This is demonstrated
in Fig.~\ref{fig:nm} where the difference between the dotted and solid curve 
is generated by the medium effect on the meson propagators.

There is another difference that arises from a technical issue in the
solution of the transcendental equation for the
G-matrix. We obtain new values of the mean fields $M^*(k_F)=M_N+U_S=
718 $ MeV and $U_V=165$ MeV. The mean field potentials  obtained here
from the G-matrix 
are considerably 
smaller than those of mean field theory in which the potential is used.
The implications for nuclear deep inelastic scattering
 are discussed in Sec.~10.G.

The most important medium effect in
relativistic approaches to nuclear matter comes from the
use of in-medium Dirac spinors representing the nucleons
in nuclear matter (`Dirac effect').  This effect (and the medium effect on
meson propagators) is absent in the conventional
(non-relativistic) Brueckner calculation which yields the dashed
curve in Fig.~\ref{fig:nm}. 
A characteristic for all non-relativistic  Brueckner theory  calculations are
the results 
that the saturation density is too high, or the binding energy is
too low.

\subsection{Meson degrees of Freedom}

The nucleonic wave function  $\mid \Phi \rangle $ of 
Eq.~(\ref{Phi}) gives us the purely nucleonic part of the Fock space,
in which the effects of the mesons have been replaced by the
two-nucleon interaction $K$.
However, the full wave function is 
 $\mid \Psi \rangle $ of Eq.~(\ref{sc}). We need to assess whether
 $\mid \Phi\rangle$
 is a good approximation to $\mid \Psi \rangle $ and 
 to determine the mesonic plus-momentum distributions.
 
To obtain $|\Psi\rangle$
recall the  relation between the full $P^-$ operator and the one of
Eq.~(\ref{big}) ($P_0^-=P_0^-(N)+K)$  
corresponding to the nucleonic wave function $\mid \Phi \rangle$:
\begin{equation}
P^- = P^-_0 + J-K+P^-_0(m).
\end{equation}
Using this in Eqs.~(\ref{sc}) and (\ref{big}) yields 
\begin{equation}
 \mid \Psi \rangle = \mid \Phi \rangle
 + {1 \over M_A- \Lambda_\Phi P^- \Lambda_\Phi}
 \Lambda_\Phi 
(J-K) \mid \Phi\rangle , \label{psi}
\end{equation}
where $\Lambda_\Phi=1-\mid\Phi\rangle\langle\Phi\mid$.

An expression for the nuclear mass $M_A$ can be obtained by
multiplying Eq.~(\ref{psi}) by
  $\langle \Phi \mid P^-$ 
using 
  $\langle \Phi\mid \Psi\rangle=1$ to obtain the result:
\begin{eqnarray}
M_A = \langle  \Phi \mid P^- \mid \Phi\rangle  + \langle \Phi \mid (J-K)
\Lambda_\Phi 
{1 \over M_A- \Lambda_\Phi  P^- \Lambda_\Phi } \Lambda_\Phi  (J-K) \mid \Phi
\rangle .
\end{eqnarray}
For the purely nucleonic wave function  $\mid \Phi\rangle $ we have
\begin{eqnarray}
\langle  \Phi\mid P^- \mid \Phi\rangle  =
M_0 + \langle \Phi \mid J-K \mid \Phi \rangle,
\end{eqnarray}
so that
\begin{eqnarray}
M_A =M_0+\langle \Phi \mid J-K \mid \Phi\rangle  + \langle \Phi \mid (J-K)
\Lambda_\Phi 
{1 \over M_A- \Lambda_\Phi  P^- \Lambda_\Phi } \Lambda_\Phi  (J-K) \mid \Phi
\rangle .
\end{eqnarray}
The difference between $M_A$ and $M_0$ is the expectation value of the
 operator $O$,
which 
satisfies the integral equation
\begin{eqnarray}
  O = J-K + (J-K) \Lambda_\Phi G_0 (M_A)\Lambda_\Phi O,\label{oeq}
\end{eqnarray}
with
\begin{eqnarray}
G_0(M_A) \equiv  \Lambda_\Phi {1\over M_A-\Lambda_\Phi P^-_0\Lambda_\Phi}
\Lambda_\Phi .
\end{eqnarray}
The lowest relevant order of Eq.~(\ref{oeq}) is given by 
\begin{eqnarray}
O \approx  J-K + (J-K) \Lambda_\Phi G_0 (M_A) (J-K) \nonumber.
\end{eqnarray}

The one-boson exchange interaction   $K$ is given by Eq.~(\ref{kdef0})
(which now includes also the effects of Sect.~9.B) and 
we may determine if the expectation value
$ \langle \Phi \mid O \mid \Phi \rangle$
is reasonably  small. If  this is true,
 the 
quantity $M_0$ would be
 a good approximation to the true eigenvalue of the $P^-$
operator $M_A$.
In the one-boson exchange approximation
\begin{eqnarray}
 \langle \Phi \mid O \mid \Phi  \rangle  &\approx&
 \langle  \Phi \mid v_3-K \mid \Phi \rangle 
+  \langle  \Phi \mid J \,\, G_0 (M_A) J \mid \Phi \rangle \\
&=&  \langle  \Phi \mid  v_3-K \mid \Phi\rangle 
+  \langle  \Phi \mid (v_1+v_1^\chi) G_0 (M_A) (v_1+v_1^\chi)  \mid \Phi \rangle. 
\end{eqnarray}
The use of Eq.~(\ref{kdef0}) yields
\begin{eqnarray}
 \langle \Phi \mid O \mid \Phi\rangle &\approx& 
 \langle \Phi \mid (v_1+v_1^\chi) \left( G_0 (M_A)-g_0(P^-_{ij})\right)
 (v_1+v_1^\chi)
 \mid \Phi
 \rangle ,
\end{eqnarray}
where the term $P^-_{i,j}$ is specified in Eqs.~(\ref{bweinberg})-(\ref{pim}).
But within the independent pair approximation (in which one includes only the
energy (minus-momentum)  differences for a chosen pair of nucleons)
\begin{eqnarray} G_0 (M_A)\equiv g_0(P^-_{ij}),\label{gg}
  \end{eqnarray}
  and 
the expectation value of $O$ vanishes. 

Thus within our approximations, it is consistent to say that
the exact nuclear mass $M_A$ is well
approximated
by $M_0$. This means that we have shown that it is acceptable to remove the
explicit mesons for calculations of the nuclear mass.
The simplicity of the derivation of this result is made possible by the
dynamical simplicity of the vacuum, which is one of the defining features of
light front field theory.

\subsection{Momentum distributions} 
We  now indicate how to compute the + components of momentum.
Look at $T^{++}$ as given by Eq.~(\ref{tpp}) 
The plus momentum carried by the scalar meson is given by
\begin{eqnarray}
 P^+ (\phi) = \int d^2 k_\perp dk^+ k^+ a^\dagger(k) a(k),
\label{p+phi}\end{eqnarray}
while that of the pion is given by
\begin{eqnarray}
 P^+ (\pi) = \int d^2 k_\perp dk^+ k^+ \bbox{a}^\dagger(k)\cdot\bbox{a}(k),
\label{ppi}\end{eqnarray}
and  that of the vector meson is given by
\begin{eqnarray}
 P^+ (\omega) = \sum_{\omega=1,3} \int d^2 k_\perp dk^+ k^+ a^\dagger 
({\bbox k}, \omega) a ({\bbox k}, \omega) . \end{eqnarray}

The procedure\cite{rmgm98}. is to 
evaluate the expectation value of
the number operators  in the wave function $|\Psi\rangle$ of Eq.(\ref{psi}). The meson
destruction operator annihilates the first term of that equation but finds a
non-zero result when acting on the second term. Here only the scalar term is
considered.
 The evaluation of (\ref{p+phi}) using 
(\ref{psi}) leads to 
\begin{eqnarray}
 P^+(\phi)& =& \int d^2 k_\perp dk^+ k^+ \langle\Phi \mid J^\dagger
 ( {\bbox k} 
) {1\over (M_A-\Lambda P^- \Lambda)^2} J  ({\bbox k})
\mid \Phi\rangle\nonumber\\
&\approx &\int d^2 k_\perp dk^+ k^+ \langle\Phi \mid J^\dagger ( {\bbox k} 
) {1\over (M_A-\Lambda P^-_0 \Lambda)^2} J ({\bbox k}) \mid \Phi\rangle,
\label{ppp}
\end{eqnarray}
in which the approximation is motivated by the near equality of
$M_A$ and $M_0$ and the  expectation that the impulse approximation
evaluation of the meson exchange potential is valid.
The term $J(k)$ is defined in Eq.~(\ref{jdef}).

The operator $J^\dagger \cdots J$ to be
evaluated has one and two body pieces. The one-body
terms are related to a shift in the self energy of the nucleon caused by
the medium. In
infinite nuclear matter the ratio of pairs to single nucleons is infinite
so that the number density is well approximated by the two nucleon terms of
Eq.~(\ref{ppp}). The evaluation is simplified by the use of Eq.~(\ref{gg}) and
noting that the relevant matrix element is the same as occurring in the
one-boson exchange operator $K$ except that the denominator is squared.
Thus the momentum density $n_\phi({\bbox k})$, defined by
\begin{eqnarray}
 P^+ (\phi) \equiv \int d^2 k_\perp dk^+ k^+n_\phi({\bbox k}),
\end{eqnarray}
is given as a derivative of the scalar-meson exchange
contribution to the nucleon-nucleon potential:
\begin{eqnarray}
n_\phi({\bbox k}) 
\approx 
 -2\langle\Phi \mid
 {\partial V_\phi(P^-_{ij},\bbox{k}) \over \partial P^-_{ij} }
 \mid \Phi \rangle 
\end{eqnarray}
with
\begin{eqnarray}
  {\partial V_\phi(P^-_{ij},\bbox{k}) \over \partial P^-_{ij} }
\equiv \left[ J_i^\dagger ( {\bbox k} 
) {1\over (M_A-\Lambda P^- \Lambda)^2} J^\dagger_j ({\bbox k})\right] 
\end{eqnarray}
  in which the notation $i,j$ specifies that only two-nucleon contributions are
  included. Note that 
$\mid \Phi >$ is the correlated ground state. Note that the expectation value
is taken using the single particle basis specified by Eq.~(\ref{pim}).
Note that\cite{rmgm98}
\begin{equation}
  -{\partial \over \partial P_{ij}^-} {1\over (P_{ij}^--P_0^-)}=
   { k^+\over(q^2-m_\phi^2)^2}.
    \end{equation}
    This means that evaluating the  plus-momentum  distribution for scalar
    mesons is the same as evaluating the expression for the scalar meson
    contribution to the nuclear potential energy, except that the
    potential $V_\phi({\bbox k})$ (the dependence on $P^-_{i,j}$ is suppressed)
    is multiplied by the factor $ -k^+/(q^2-m_\phi^2)$.
    The net result is that
    \begin{eqnarray}
     n_\phi({\bbox k})=\sum_{l,m<F} \langle l\;m \mid
     \Omega^\dagger_{l\;m}V_\phi({\bbox k}){(- k^+)\over(q^2-m_\phi^2)}
    \Omega_{l\;m} \mid l\;m\rangle_a, \label{nphi}
     \end{eqnarray}
     where $\Omega_{l\;m}$ is the Moeller scattering operator for the
     two-nucleon state $ l\;m $. One finds a similar expression for the
     pionic
     terms.

     We can get the total number of each kind of meson
(except the $\omega)$ using a sum rule.
Consider the schematic form of the equation for the G-matrix
\begin{equation}
  G(P_{ij}^-)=V(P_{ij}^-) +V(P_{ij}^-){Q\over \Delta E} G(P_{ij}^-),
  \label{gmats}
  \end{equation}
 where
 ${Q\over \Delta E} $ is a schematic
 representation of the propagator of Eq.~(\ref{bweinberg}).
   Differentiating with respect to the denominator
   $P_{ij}^-$ appearing in  a given meson
   $m$
   exchange potential 
  yields the result:
\begin{equation}
  {\partial G^m(P_{ij}^-)\over \partial P_{ij}^-}\equiv
  (1+G{Q\over \Delta E})  {\partial V^m(P_{ij}^-)\over \partial P_{ij}^-}
(1 +{Q\over \Delta E}G), 
\end{equation}
in which the label $m$ refers to the type of meson. 
The potential $V^m$ appearing in Eq.~(\ref{gmats}) is simply
the Fourier transform
of the potential $V^m(\bbox{q})$. Thus  
an examination of Eq.~ (\ref{nphi}) and its pionic generalization
 that, considering the pion
for example, 
\begin{equation}
  N_\pi\equiv \int d^3q\; n_\pi(\bbox{q})=\sum_{\alpha,\beta<F}
  \langle l\;m\mid  {\partial G^\pi(P_{ij}^-)\over \partial P_{ij}^-}
  \mid l\;m\rangle_A. \label{npi}
  \end{equation}
  Numerical evaluation of Eq.~(\ref{npi})
  leads to the result that $N_\pi/A= 0.05.$
  This is smaller than the 18\% of Friman et al\cite{bf}
because we use scalar mesons  instead of intermediate $\Delta$ states to 
 provide the bulk of the attractive 
force.

   \subsection{Apparent Puzzle Resolved}

The values of $U_S$, $U_V^-$ and $N_\pi$ allow us to assess the 
deep inelastic scattering of leptons from our version of the ground state of
nuclear matter. Using $M^*(k_F)=718$ MeV  and 
neglecting the influence of
two-particle-two-hole states
to approximate
$f(k^+)$ using Eq.~(\ref{ybar}) 
shows that nucleons carry 83\% (as opposed to the 65\% 
of mean field theory\cite{jerry}) of  the nuclear plus momentum. 
This represents a vast improvement in the description of nuclear deep inelastic
scattering. The 
minimum value of the ratio $F_{2A}/F_{2N}$, obtained from 
the convolution formula (\ref{deep}) 
 is increased by a factor of twenty   towards 
the data 
as  
extrapolated in Ref.~\cite{sdm}. But this calculation provides only  a
lower limit of the nucleon contribution because of the neglect of 
effects of the two-particle-two hole states. We estimate  that
nucleons with momentum greater than $k_F$ would substantially increase the
computed ratio $F_{2A}/F_{2N}$ because $F_{2N}(x)$ decreases very rapidly
with increasing values of $x$ and because $M^*$ would increase at high momenta.

Turn now to the experimental information about the nuclear pionic content.
The Drell-Yan experiment on nuclear targets\cite{dyexp}
showed no enhancement of nuclear pions within an error of about 5\%-10\% for
their heaviest target. No substantial  pionic enhancement 
is found in (p,n) reactions. 
Understanding this result is 
an important challenge to the 
understanding of nuclear dynamics~\cite{missing}. 
Here we have a good description of nuclear dynamics, 
and our 5\%  enhancement is roughly consistent\cite{jm},
within errors, with the Drell-Yan
data.

\section{Coming attractions} 

The preceding sections have been concerned with reviewing
published work devoted to applications of  
light front dynamics with hadronic Lagrangians 
to   heavy nuclear systems. Only the first studies of mean field theory and
Brueckner theory have been completed. Many more detailed elaborations of these
studies are
possible and will be needed.
Much work in applying light front dynamics to nuclear physics
remains to be done.
The purpose of the present section is to briefly discuss 
a few new directions of our
research.

\subsection{Quark-composite nucleons in the nucleus}
The first new direction is to consider the nucleons as composite systems of 
quarks.  So far nuclear 
effects enter in deep inelastic scattering only through the influence of
the nucleon distribution function $f(y)$, recall Eq.~(\ref{deep}) 
However the 
internal wave functions of the nucleons 
and therefore 
structure functions $F_{1,2}$ could be modified by the presence of the 
nuclear medium. One model which includes such effects is the quark meson
coupling model (QMC)\cite{qmc},
 initiated by Guichon and developed by Thomas, Saito and
collaborators. 
Here the nucleons are treated as three quarks confined in a bag or under the
influence of a confining potential\cite{qmcme}. The nucleus is bound
by the exchange of mesons between quarks in different nucleons. 
This model has the nice property that the scalar and vector potentials are
much weaker than in the usual QHD theory. This means that nucleons carry a
large (but not unity) fraction of the nuclear momentum, so that one can
describe
the  nuclear deep inelastic scattering data. We
\cite{gk} have been 
developing a light front 
version of the theory. In this model, the light quarks move relativistically
within nucleons which move  relativistically within the nucleus. The second
part can be handled by using the techniques discussed in the present review. 
To proceed further, one needs to obtain a light front description of
the  three-quark system.
The binding of two particles is well-described within light front dynamics,
 see the reviews and our Sect.~3.
Therefore, as a first step,
we consider the nucleon as a  bound state of a quark
and a di-quark. In a nucleus the quark moves under the influence of the
interactions with other nucleons as well as under the confining effects of the
di-quark. The preliminary results are that the saturation properties of the
system as about the same as in the usual QMC  model. This means that we will be
able to
evaluate nuclear deep inelastic scattering with nuclear modified versions of
the
nucleon structure function.

\subsection{Few Body Problems}
Applications of light front 
dynamics to nuclear physics began with studies of the 
deuteron\cite{fs,coester,Ka 88}.
The existence of  many previous extensive applications,
was one reason behind my decision to concentrate on heavy nuclei.
However, there are many planned experiments on the deuteron.
The Lagrangian  approach of this review could be applied
to the deuteron and also to bound
states of three particles.

We have begun by doing some toy model calculations using  a version of
the 
Wick-Cutkosky model in  
which two heavy scalar particles are bound by the exchange
of a light scalar particle\cite{cmp}. This model had already been applied\cite{blb1} in
light front calculations, with
considerable success in understanding rotational invariance in scattering
problems.
The first step of Ref.~ \cite{cmp}was to study the
breaking of manifest rotational invariance, present in light front dynamics.
There can be many bound states in this
model, so a test of rotational invariance is to reproduce the degeneracies of
the eigenvalues.
We found that 
the use of 
an effective Hamiltonian
that takes into account two
       meson exchanges, leads to significant improvement in the degree to
       which the computed spectra have the correct degeneracies. 
       This is much improved compared to results obtained
       when only one meson exchange is included in the effective Hamiltonian.
       The largest 
       improvement occurs when the states are weakly bound.

A more detailed study of the  ground state using carefully computed 
one and two boson exchange potentials\cite{cm} shows that the light front 
techniques reproduce the results of  exact  model calculations\cite{tjon},
provided
the binding energy of the two particle system is small compared to the mass of
the heavy particles.

The next step is to use our Lagrangian to describe  deuteron
form factors measured in electron and neutrino scattering.
Much of the previous work has been devoted to using the impulse approximation
to compute form factors measurable in electron- and neutrino- interactions
with
the deuteron. In the light front this means that one evaluates the matrix
elements of the so-called
good or $J^+,A^+$ components of the electromagnetic and axial-vector
currents. 
We studied this some time ago\cite{fhm} and
found that the computed axial form factor is
very sensitive to the choice of matrix element of $A^+$ used to extract the
form factor $F_A(Q^2)$. This is because of an  inherent ambiguity
arising from the spin-one nature of the deuteron:
there are two independent matrix
elements, 
but only one operator $A^+$ is used in the impulse approximation.
The D-state of the deuteron was responsible for most of the numerical effect.
Keister\cite{keister} has 
studied relativistic effects in the deuteron axial form factor and
confirms the dominant role of the  D-state in observable effects beyond the
non-relativistic limit.   
The present formalism, based on the use of a Lagrangian should allow us to
compute 
the matrix elements of operators other
than $J^+,A^+$. In practice, this means that including
two-body exchange currents 
effects of the $|\pi \;NN\rangle$ Fock-state components will be necessary.

A needed important effort would be
to apply light front dynamics to the three-nucleon and
three-quark systems.
Studies of this problem are in their infancy\cite{coester3,fs3,bakker3}.
One can use the light front version of relative momentum variables and write
done the light front version of
the Fadeev equations. These should be solvable using 
 modern high-speed  computers.
 
\subsection{Possible experimental signature-HERMES Effect}

So far we have shown that light front dynamics can be applied to the physics
of heavy nuclei. The standard good results for  binding energy and density have
been reproduced (with some improvement in the computed value of the
compressibility of nuclear matter\cite{rmgm98}). The salient new finding is
that vector mesons carry a significant fraction of the nuclear plus-momentum,
and that the meson fields can be treated in a
quantum fashion that enables other
matrix elements to be computed. The use of the simplest field theories leads
to very strong meson fields, which account for
an experimentally unacceptably large  plus-momenta. This problem is
ameliorated by the use of more complicated Lagrangians or by the use of
Brueckner theory. 

Thus it seems that nuclei can be well-described using the light front approach.
But it would be better to go further. In particular, 
it would be highly desirable to find specific  experimental evidence
for a previously
unappreciated feature of nuclei that is most conveniently computed within the
framework of the 
light front approach.
The HERMES collaboration discovery in high energy, low momentum transfer
($0.5\;
{\rm GeV}^2< Q^2< 2 \;{\rm GeV}^2$ positron-nucleus scattering that the value
of  
 $R\equiv{\sigma_L\over \sigma_T}$ is enhanced (by a factor as big as 5)
 in nuclei\cite{hermes} may provide the experimental evidence we desire.

 The large nuclear value of  $R$ occurs as the result of two significant
 effects: 
 $\sigma_L$ is enhanced  by a factor of two or more
 and $\sigma_T$ is depleted by as much as 50\%.
Since a small value of $R$ is the signature of spin 1/2 partons\cite{cagr}
it
is natural to attempt an explanation of this unusual finding in terms of the
mesonic field of nuclei. We found\cite{3m} we find that it
is possible to reproduce the qualitative features of the data using the
nuclear $\omega$ and $\sigma $ fields which are responsible for binding
nuclei. A mechanism in which
a virtual photon strikes a nuclear $\omega$ meson
turning it into a $\sigma $ meson gives a large enhancement to $\sigma_L$.
To reproduce the small value of $\sigma_T$ it is necessary to find a mechanism
which interferes destructively with the dominant process of $\gamma^*$
conversion to a vector meson which is scattered onto its mass shell by the
target. The process of a virtual photon striking a nuclear $\sigma$ meson and
converting  it into a vector meson can provide the necessary destructive
interference.  Both of these mechanisms depend on previously unmeasured
photon-meson coupling constants and form factors, so that reproducing the
HERMES data is not a necessary consequence of our approach. 
However, we are able
to argue that the data do not violate laws of physics or or disagree with other
experiments. The model of Ref.~\cite{3m} contains many specific predictions
which should be easy to test.
\section {Summary}

This  paper begins  with a discussion of the motivation for employing
light front  dynamics to compute the wave functions of nuclei, Sect.~1.
 The main reason
is that high energy experiments are best analyzed in terms  of the plus-momenta
of the particles involved. Sect.~2 contains a
brief description of  light front dynamics
 which stresses the importance of Eq.~(\ref{ONE}) and the use of a Lagrangian
 to define the total momenta $P^+$ and $P^-$ which are the light front
 plus-momentum and $\tau$ development operators.

 Since light front technology can appear cumbersome and recondite to a new
 reader,  Sect.~3 is devoted to explaining how ordinary potentials\cite{toy}
 are
 handled using light front technology. The influence of relativistic
 corrections to means square radii of heavy
 nuclei is examined and found to vary as $A^{-5/3}$, Eq.~(\ref{relre}).
 Light front nuclear momentum distributions are examined using a toy Hulth\'en
 model and the separate $p^+$ and $\bbox{p_\perp}$ dependences are interesting,
 Fig.~3.1.

 Light front quantization of hadronic Lagrangians (without pions and chiral
 symmetry) is presented in Sect.~4. An important element is the
 transformation (\ref{sytran})
 of Soper\cite{des71} and Yan\cite{yan34} of the Fermion fields
 used to separate the dependent and independent fermion degrees of freedom.

 The formalism is applied to infinite nuclear matter under the mean field
 approximation, Sect.~5,\cite{jerry}.
 Standard results for saturation properties are
 reproduced. The special feature of the light front formalism the 
use of the plus momentum as one of the canonical variables. This enables
a close contact with the experimental variables used to analyse 
deep inelastic scattering and any experiment in which there is one large
momentum. This feature is exploited here in the derivation 
(within the mean field approximation) of the 
nucleonic and mesonic distribution functions for infinite nuclear
matter, Eqs.~(\ref{ndist})-(\ref{v10}). The vector mesons are shown to
carry a significant fraction of the nuclear plus momentum, but only 
a zero plus-momentum, and therefore do not
participate in nuclear deep inelastic scattering or Drell-Yan experiments.

This restriction to zero plus momentum is investigated in Sect.~6, using a
simple static model\cite{bm98}. 
The feature that a static source in the usual coordinates corresponds
to a source moving with a constant velocity in light front coordinates is
exploited. 
The dependence of the vector meson distribution function on the nuclear radius
is studied, with the key result reproduced  in Fig.~6.1. The plus momenta
are no longer restricted to zero, except in the limiting case of infinite
nuclear matter. For finite nuclei, of radius $R_A$ the plus-momentum
distributions are spread, with
$k^+\sim1/R_A.$ Thus experimental verification or disproof is possible.

Sect.~7 is devoted to reviewing the application of light front dynamics to
finite nuclei\cite{bbm99}. The necessary technique is to minimize expectation
value of the sum $P^-+P^+$. This leads to a new set of coupled
equations (\ref{nnplus}) and (\ref{nnminus}) for the single nucleon
modes. These depend on the meson fields of Eqs.~(\ref{phieq}) and
(\ref{veq}).
The most qualitatively startling feature emerging from the derivation
is that the meson field equations (\ref{phieq}) and (\ref{veq}) are the
same as that of the usual theory, except that $z$ of the equal-time
theory translates to $-x^-/2$ of the light-front version. This can be
understood in a simple manner by noting that light-front quantization
The  general argument is that this result emerges from 
the feature that a static source in the usual coordinates corresponds
to a source moving with a constant velocity in light front coordinates, as in
Sect.~6.

Even though the meson field equations of the light-front and equal-time
theories are the same, there are substantial and significant
differences between the two theories. In our treatment, the mesonic
fields are treated as quantum field operators. The mean field
approximation is developed by replacing these operators by their
expectation values in the complete ground state nuclear wave function.
This means that the ground state wave function contains Fock terms with
mesonic degrees of freedom. We can therefore compute expectation
values other than that of the field and  
obtain the mesonic momentum distributions, Table~7.2 and Fig.~7.2.
This feature has
been absent in standard approaches.

We obtain an approximate solution (\ref{emcv}) of our nucleon mode
equation. Our nucleon mode functions are approximately a phase factor
times the usual equal-time mode functions (evaluated at $x^-=-2z$).
This shows that the energy eigenvalues of the two theories should have
very similar values. But the wave functions are different-- the
presence of the phase factor explicitly shows that the nucleons give up
substantial amounts of plus momentum to the vector mesons.

A new numerical technique,
is introduced in Ref.~\cite{bbm99}
to solve the coupled nucleon and meson field equations.
Our results, Table 7.1,  display the expected $2j_\alpha +1$ degeneracy of the
single nucleons levels, and the resulting binding energies are
essentially the same as for the usual equal-time formulation. This
indicates that the approximation (\ref{emcv}) is valid.

The present results for finite nuclei, which use the original Walecka model,
are not consistent with experiments on
lepton-nucleus deep inelastic scattering  and $(e,e')$
reactions.  This is
because, in $^{40}$Ca for example, the nucleons carry only 72\% of the
plus momentum. This is a result of the quantity $M +g_s\phi$, which
acts as a nucleon effective mass, is very small, about 670 MeV. The use
of a small effective mass and a large vector potential enables a simple
reproduction of the nuclear spin orbit force \cite{bsjdw,hs}. However,
the use of a
different Lagrangian, including  non-linear couplings between scalar
mesons  should provide significant improvement as indicated by the light front
results for infinite nuclear matter, Eq.~(\ref{v10}).
Another interesting possibility would be to obtain a light-front
version of the quark-meson coupling model \cite{qmc,qmcme,gk},
in which confined
quarks interact by exchanging mesons with quarks in other nucleons.
This model, also has smaller magnitudes of the scalar and vector
potentials.

In any case, these kinds of nuclear physics calculations can be done in
a manner in which modern nuclear dynamics is respected, boost
invariance in the $z$-direction is preserved, and in which the
rotational invariance so necessary to understanding the basic features
of nuclei is maintained.

Sect.~8 reviews  how the light front quantization of a chiral
Lagrangian can be accomplished.  The resulting formalism can be
applied to many problems of interest to nuclear physics. 
  Soft pion theorems for pion-nucleon scattering,
Eq.(\ref{wtt}) are
reproduced.

Sect.~9 reviews the our light front studies\cite{jerry,rmgm98}
of  nucleon-nucleon scattering. The T-matrix is shown to
be manifestly covariant, using  the one boson exchange approximation to the
nucleon-nucleon potential. An essential feature is the the meson propagators
of Eqs.~(\ref{det}) and (\ref{det})
contain the influence of retardation.
 Light front quantization is
used to obtain a  nucleon-nucleon potential which yields phase shifts in good
agreement with data. See, for example, Fig.~9.2.

Sect.~10 reviews our  derivation\cite{rmgm98}
of  the
 Brueckner Hartree-Fock equations. 
Applying our light front OBEP, the nuclear matter
saturation properties
are reasonably well reproduced. The binding energy per nucleon
is 14.7 MeV and $k_F=1.37 $fm$^{-1}$. A  value of 
the compressibility, 180 MeV,
that is  smaller than that of
alternative relativistic approaches is obtained. This is largely due to the
light front requirement that retardation terms be kept.
The  derivation starts with a  meson-baryon
Lagrangian, so we are able to show that replacing the meson degrees of freedom
by a NN interaction is a consistent   approximation, and the
formalism allows one to calculate corrections to this approximation in a
well-organized manner.  The simplicity of the vacuum
in our light front approach
is an important feature in allowing the derivations to proceed.
The mesonic Fock space components of the nuclear wave
function are obtained also, and aspects of the
meson and nucleon  plus-momentum distribution functions
are computed. We find that   there are about 0.05 excess pions
       per nucleon. 

       There are promising
       implications of the Brueckner theory results for studies of
       nuclear deep inelastic scattering and Drell-Yan reactions.
      The nucleons probably carry about 90 \% of the nuclear plus momentum,
      enough for rough accord\cite{sdm} with data. Furthermore having 0.05
      excess pions  is consistent\cite{jm},
within experimental  errors, with the Drell-Yan
data.

A few future directions are discussed in Sect.~11.
Despite the extreme length of this review, 
much  work in applying light front dynamics to nuclear physics
remains to be done. For example,
only the first studies of mean field theory and
Brueckner theory have been completed. Many more detailed elaborations for heavy
nuclei, few-body systems and composite nucleons are
possible and will be needed. I hope that this review will stimulate the reader
to future research using light front dynamics.

\section*{Acknowledgments}
This review is based on  work performed in collaboration with 
P.G.~Blunden, J.R.~Cooke, M.~Burkardt, and R.~Machleidt, and D.R.~Phillips.
This work is partially supported by the USDOE. I thank S.J.~Brodsky, L.
Frankfurt
and
M. Strikman for introducing me to this subject. 
My work   light front dynamics began during a sabbatical hosted by the national INT,
the SLAC theory group and the CSSM in Adelaide.

\end{document}